%% file: main.tex
\documentclass[acmtog]{acmart}

\input{config/packages_commands}
\input{config/authors}
\input{config/meta}

\begin{document}

\title{Embodied Hands: Modeling and Capturing Hands and Bodies Together}

\input{tex_FIG/FIG_teaser}

\input{tex_main/00_abstract}

\maketitle

\input{tex_main/01_intro}
\input{tex_main/02_previous}
\input{tex_main/03_method}
\input{tex_main/04_handsBody_4d_alignment}
\input{tex_main/05_experiments}
\input{tex_main/06_discussion}
\input{tex_main/07_acknowledgements}

\balance
\bibliographystyle{ACM-Reference-Format}
\bibliography{bib/hands,bib/bodies}

\pagebreak

\input{tex_main/08_appendix}

\end{document}

%% file: config/packages_commands.tex
\usepackage{booktabs}
\usepackage{balance}
\usepackage{subfig}
\usepackage{multirow}
\usepackage{hhline}
\usepackage{tabularx}
\usepackage{rotating}
\usepackage{xfrac}

\newcommand{\mypm}{\mathbin{\smash{\raisebox{0.35ex}{$\underset{\raisebox{0.5ex}{$\smash -$}}{\smash+}$}}}}
\newcommand{\mycpm}{\mathbin{\smash{\raisebox{0.35ex}{$\underset{\raisebox{0.5ex}{\textcolor{RoyalBlue}{$\smash -$}}}{\textcolor{ForestGreen}{\smash+}}$}}}}
\definecolor{ForestGreen}{RGB}{34,139,34}
\definecolor{RoyalBlue}{RGB}{65, 105, 225}

\newcommand{\lotsOfHandReferences}{\cite{leapMotion, Hamer_Hand_Manipulating, Tzionas:IJCV:2016, JavierHandsInActionJournal, Melax:2013:handPhysics, Lepetit:ICCV:2015:handCnnLoop, OikonomidisBMVC, Oikonomidis_2hands, DART-Schmidt-RSS-14, srinath_cvpr2015, htrack_sgp15, NYU_tracker_tompson14tog, Tang:PAMI:2017, schroder2014real, sg_IJCV16_XuEtAl, MSR_2016_Siggraph_handTrack, wang2013handObject}}

\newcommand{\mano}{\mbox{MANO}\xspace}
\newcommand{\smplH}{\mbox{SMPL+H}\xspace}

\newcommand{\template}{\boldsymbol{\bar{\mathrm{T}}}}
\newcommand{\pose}{\vec{\theta}}
\newcommand{\shape}{\vec{\beta}}
\newcommand{\bweights}{\mathcal{W}}
\newcommand{\algn}{\boldsymbol{\mathrm{V}}}
\newcommand{\algnsurf}{\mathcal{V}}

\newcommand{\valgnsurf}{v}

\newcommand{\scanvert}{s}
\newcommand{\scanverts}{\boldsymbol{\mathrm{S}}}

\usepackage[ruled]{algorithm2e}

\SetAlFnt{\small}
\SetAlCapFnt{\small}
\SetAlCapNameFnt{\small}
\SetAlCapHSkip{0pt}
\IncMargin{-\parindent}

\newcommand{\thetab}{\theta_b}
\newcommand{\thetah}{\theta_h}
\newcommand{\poseb}{\pose_b}
\newcommand{\poseh}{\pose_h}
\newcommand{\anneal}{\lambda_a}
\newcommand{\modelsurf}{\mathcal{M}}
\newcommand{\gmmean}{\vec{\mu}}

\usepackage[toc,page]{appendix}

%% file: config/authors.tex
\author{Javier Romero}
\authornote{Both authors contributed equally to the paper}
\authornote{This research was performed by JR at the MPI for Intelligent Systems.}
\affiliation{
  \institution{Body Labs Inc.}
  \city{New York} 
  \state{NY} 
}	
\email{javier.romero@bodylabs.com}

\author{Dimitrios Tzionas}
\authornotemark[1]
\affiliation{
  \institution{Max Planck Institute for Intelligent Systems}
  \city{T\"{u}bingen} 
  \state{Germany} 
}
\email{author@tuebingen.mpg.de}

\author{Michael J. Black}
\affiliation{
  \institution{Max Planck Institute for Intelligent Systems}
  \city{T\"{u}bingen} 
  \state{Germany} 
}
\email{black@tuebingen.mpg.de}
\authorsaddresses{Authors' email addresses: javier.romero@bodylabs.com; \{dtzionas, black\}@tue.mpg.de}

\renewcommand{\shortauthors}{Romero et. al.}

%% file: config/meta.tex
\setcopyright{rightsretained} 
\acmJournal{TOG}
\acmYear{2017}\acmVolume{36}\acmNumber{6}\acmArticle{245}\acmMonth{11} \acmDOI{10.1145/3130800.3130883}

\ccsdesc[500]{Computing methodologies~Motion capture}

\keywords{Hands, Human body shape, 3D shape, Learning, Performance capture, Motion capture}

\setcitestyle{square}

\hypersetup{final}

%% file: tex_FIG/FIG_teaser.tex
\begin{teaserfigure}
   \includegraphics[width=\textwidth]{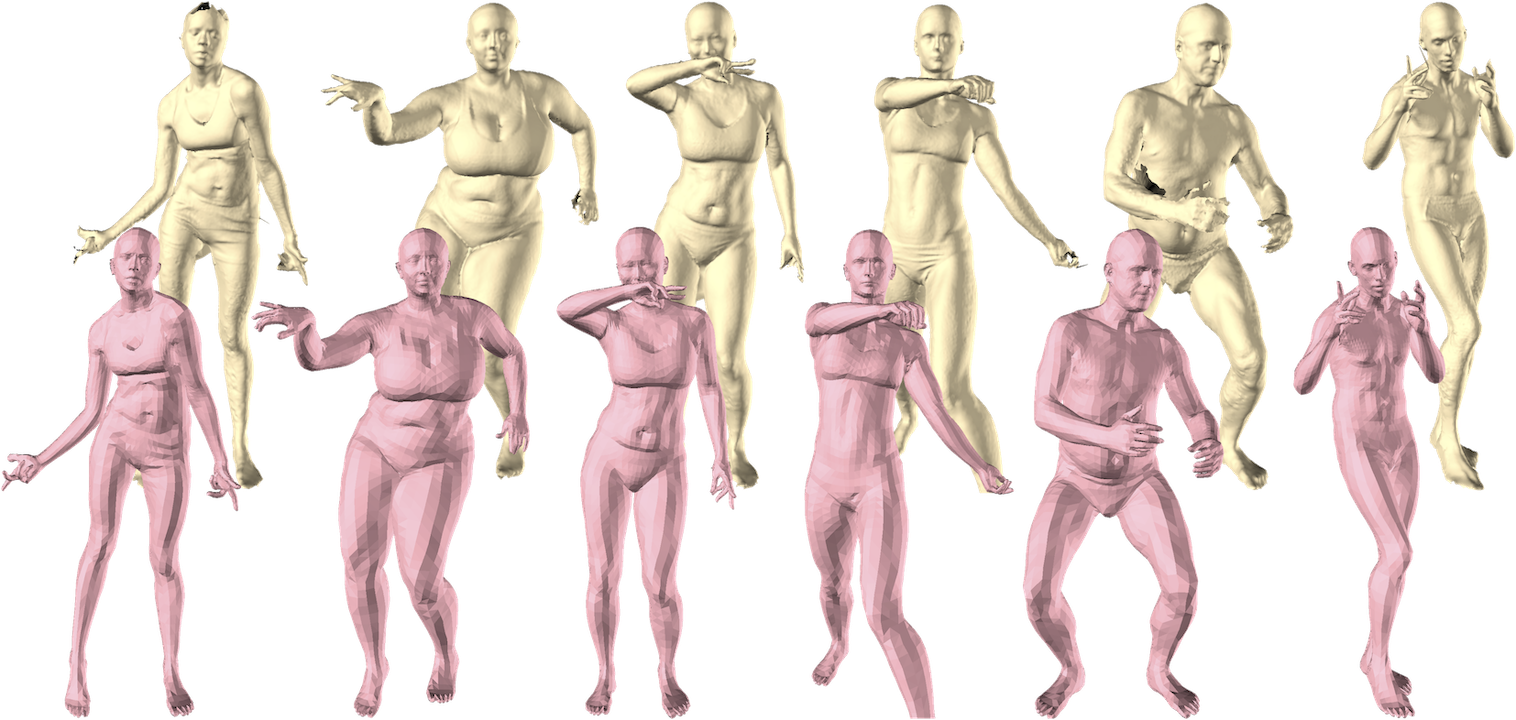}
   \caption{
		{\bf Capturing hands and bodies together.}
		Beige: raw 4D scan.  
		Pink: recovered shape and pose using our model of \emph{hands and bodies interacting together (\smplH)}.
		To deal with noise and missing data, we develop a new model of hands called  \emph{\mano} (\emph{hand Model with Articulated and Non-rigid defOrmations}) and combine it with a parameterized  3D body model (SMPL). 
		This combination enables natural performance capture even under severely noisy measurements.
	}
	\label{fig:teaser}
\end{teaserfigure}

%% file: tex_main/00_abstract.tex
\begin{abstract}
Humans move their hands and bodies together to communicate and solve tasks.
Capturing and replicating such coordinated activity is critical for virtual characters that behave realistically.
Surprisingly, most methods treat the 3D modeling and tracking of bodies and hands separately.
Here we formulate a model of {\em  hands and bodies interacting together} and fit it to full-body 4D sequences.
When scanning or capturing the full body in 3D, hands are small and often partially occluded, making their shape and pose hard to recover.
To cope with low-resolution, occlusion, and noise, we develop a new model called
\emph{\mano} (\emph{hand Model with Articulated and Non-rigid defOrmations}).
\mano is learned from around 1000 high-resolution 3D scans of hands of 31 subjects in a wide variety of hand poses.
The model is realistic, low-dimensional, captures non-rigid shape changes with pose, is compatible with standard graphics packages, and can fit any human hand.
\mano provides a compact mapping from hand poses to pose blend shape corrections and a linear manifold of pose synergies.
We attach \mano to a standard parameterized 3D body shape model (SMPL), resulting in a fully articulated body and hand model (\smplH).
We illustrate \smplH by fitting complex, natural, activities of subjects captured with a 4D scanner.
The fitting is fully automatic and results in full body models that move naturally with detailed hand motions and a realism not seen before in full body performance capture.
The models and data are freely available for research purposes at \mbox{\url{http://mano.is.tue.mpg.de}}.
\end{abstract}

%% file: tex_main/01_intro.tex
\section{Introduction}

Bodies and hands are literally inseparable.
Yet, despite this, research on modeling bodies and hands has progressed separately.
Significant advances have been made on learning realistic 3D statistical shape models of full bodies but these models typically have limited, or no, hand articulation.
Similarly there is significant work on tracking hands using depth sensors and video sequences but these hands are modeled and tracked in isolation from the body.
We argue that the hands and body together are important for communication and that a complete picture of our actions, emotions, and intentions is not possible without the joint analysis of hands and bodies.
The growth of interest in virtual and augmented reality has increased this need for
characters and avatars that combine realistic bodies and hands in motion.
Here we develop a new approach to capture the 4D motion of the hands and body together.

Several factors have led to the separation of hands from bodies.
Full body models such as SCAPE~\cite{anguelov2005scape} are learned from subjects making a tight fist, while more recent models like SMPL~\cite{Loper:SIGASIA:2015} assume an open, rigid, hand.
Neither looks realistic when animated.
Such 3D body models are created from 3D scans of the whole body.
At the resolution of most body scanners, the hands are small and the fingers are hard to resolve resulting in noise and ``webbing'' between the fingers.
Additionally, occlusion of the hand by the body and by itself results often in significant missing data.
Consequently, to get sufficient resolution and unconstrained movement, most visual hand tracking work focuses on capturing hands alone, possibly together with the forearms, using RGB-D sequences.

We make several contributions, which can be roughly divided into two
categories: learning a new model of the hand, and tracking hands and
bodies together.

First we collect a new database of detailed hand scans of 31 subjects in up to 51 poses.
We capture both left and right hands of men and women, use a wide range of poses, and capture hands interacting with objects.

Second, we use this data to build a statistical hand model similar to the SMPL body model \cite{Loper:SIGASIA:2015}, which we call \textit{\mano} for \emph{hand Model with Articulated and Non-rigid defOrmations}. 
Like SMPL, the model factors geometric changes into those inherent to the identity of the subject and those caused by pose. 
The model is trained to minimize the vertex error in the training set.
The pose space uses linear blend skinning for simplicity, with corrective blend shapes that are automatically learned from the scans.
\mano is created from the SMPL hand topology, and has analogous components to those in SMPL: a template shape, kinematic tree, shape and pose blend shapes, blend weights and a joint regressor. 

Hand articulation is very different, however, from the full body; the hand contains a large number of joints with restricted articulation. 
Consequently, \mano differs from SMPL in several ways.
Like SMPL, \mano uses corrective blend shapes that are a function of pose.
Unlike SMPL, following the example of \cite{mohr2003building} that introduced localized skinning blend weights, we encourage the corrective pose blend shapes to be local by penalizing the dependency of corrections on joints which are far away in geodesic terms.
Additionally, the high dimensionality of the full hand space makes fitting to noisy, low-resolution, body scans computationally expensive and prone to local minima.
Consequently we reduce the dimensionality of the pose space by computing a linear embedding of the pose parameters from our dataset.
We believe this is the first such analysis based on such high-quality hand data.
The resulting \mano model is lightweight, simple to animate, and compatible with existing graphics software.

Third, we combine \mano with the SMPL body model to give a new  combined model of {\em  hands and bodies interacting together (\smplH)}.
Here we take the shape space of the hand from the SMPL model, which captures correlations between hand and body morphology.
To this we add the \mano joints, kinematic tree, blend weights, and pose blend shapes.

Fourth, we address the problem of capturing full bodies and hands in motion.
To that end, we employ a 4D body scanning system that captures full 3D body shape at 60 frames per second.
At the resolution of the scanner, the hands can be extremely noisy and low resolution, sometimes disappearing entirely.
To recover hand pose, we modify \emph{4Cap}, the temporal mesh registration algorithm used in DYNA~\cite{Dyna:SIGGRAPH:2015}, to include a simple velocity prior that prevents sudden hand motion in the complete absence of data.
With this we fit \smplH to full-body 4D sequences to recover the body intrinsic shape and its changing pose, including finger articulation.

To illustrate the models and methods, we fit \smplH to a wide variety of complex sequences with fast motions.
Figure~\ref{fig:teaser} shows a few examples and many more can be found in the {\bf supplemental video}.
In addition to recovering the model parameters, we further allow the vertices of alignments to move to better fit the scan (with a penalty for deviating from the model surface) increasing their realism and level of detail.
The resulting models and alignments look much more natural than those using SMPL. 

In summary we propose a new model of hand shape and pose that is learned from data, compatible with existing graphics systems, low-dimensional, realistic and compatible with the SMPL body model.
By combing \mano with SMPL we are able to jointly capture bodies and hands in motion with high realism and deal with missing and noisy data.
This opens up the study of correlated body and hand motion at a level of detail not previously possible.
The \mano model and \smplH are \textbf{available for research purposes} \cite{dataset_SMPLH_PSHAND} along with the aligned meshes needed to train \mano and the test-datasets for evaluation.

%% file: tex_main/02_previous.tex
\section{Related Work}

\input{tex_FIG/FIG_previous_handModels}

The body and the hands have often been studied separately.
For example, models of neural control have long embraced this divisional approach \cite{Penfield:1937}.
Here we argue that modeling and tracking hands and bodies together is important for vision, graphics, and virtual reality.

\paragraph{Hand capture.}
Capturing the 3D motion of human hands has been studied for decades \cite{HoggHand96,Review_Erol_HandPose} due to its applications in computer graphics, animation, human computer interaction, rehabilitation and robotics to name a few.
This interest is increasing~\cite{survey13,Supancic:ICCV:2015} in light of the recent advancements in consumer RGB-D sensors and virtual reality \lotsOfHandReferences.

Despite this interest, key problems remain and hands have proven difficult to detect and track.
Hands are dexterous, with many degrees of freedom, and the fingers all look similar in color and shape.
They also move quickly, resulting in motion blur and making tracking difficult.
The 3D shape of the hand leads to self occlusion, which makes hand pose inference ambiguous.
The key problem is that {\em hands are small}; in images or scans of the full body, hands are often difficult to fully resolve.

Even traditional marker-based motion capture has limitations for capturing hands and bodies together.
Most capture protocols ignore the hands and most publicly available datasets do not contain any information about finger pose.
Hands require small markers which, in a large volume, require many mocap cameras to sufficiently resolve.
Markers also limit hand movement, can fall off, and are often occluded.
This limits the capture and study of hand and body movement together.

Several methods use additional instrumentation like wearable cameras \cite{Oikonomidis_MSR_Digits}, markers \cite{zhao2012markersHand}, data-gloves \cite{dipietro2008survey}, or colored-gloves \cite{glove_MIT} to simplify hand tracking.
Such approaches are intrusive, alter the hand shape, and hinder natural hand motion.
Furthermore, they need to be combined and synchronized with other systems to capture the full body and hands together.
Recent commercial solutions such as Leap Motion \cite{leapMotion} require no markers or gloves but only capture the hands without the body.

Most recent research systems use commodity RGB-D cameras \cite{survey13,Supancic:ICCV:2015} but these have limited resolution and existing methods only capture the hands and not the full body.
Multicamera systems \cite{LucaHands,Tzionas:IJCV:2016,srinath_iccv2013} on the other hand scale to large volumes but currently do not enable full body and hand tracking together.
	
\paragraph{Fitting and tracking.}
Beginning with Rehg and Kanade \shortcite{Kanade94}, there have been many generative hand models proposed and many methods for fitting them to data.
These include many  filtering approaches \cite{Isard2000,Cipolla_ModelBased,Huang_capturingNaturalHandArtic2001,vanGool_smartParticle}, belief-propagation \cite{Hamer_ObjectPrior,Hamer_Hand_Manipulating,SudderthNonparamBeliefPropag}, Particle Swarm Optimization (PSO)	\cite{OikonomidisBMVC}, sampling \cite{Oikonomidis14},  inverse-kinematics \cite{Oikonomidis_MSR_Digits,glove_MIT}, probabilistic line matching \cite{AthitsosCluttered2003}, reduced linear search sub-spaces \cite{HoggHand96,Huang_capturingNaturalHandArtic2001}, Bayesian filtering with Chamfer matching \cite{Cipolla_TORR}, and physics-based methods \cite{Melax:2013:handPhysics}.
\citet{DART-Schmidt-RSS-14} extend the Signed Distance Function representation to articulated objects, while \citet{MSR_ASIA_handTracking} combine a gradient based ICP approach with PSO.
\citet{srinath_iccv2013} explore the use of a Sum of Gaussians model for hand tracking in RGB images. 
None of these approaches address the tracking of full bodies and hands together.

\paragraph{Hand Models.}
Many hand models have been proposed and are summarized in Fig.~\ref{fig:HandsRelated:handModels}.
A popular approach is to approximate the hand with shape primitives (Fig.~\ref{fig:HandsRelated:handModels:shapePrimitivesFORTH}) which enable fast evaluation of distances \cite{OikonomidisBMVC,Oikonomidis_2hands,Oikonomidis14,MSR_ASIA_handTracking,htrack_sgp15,Kanade94}. 
An alternative is the Sum-of-Gaussians model \cite{srinath_iccv2013,srinath_cvpr2015} (Fig.~\ref{fig:HandsRelated:handModels:SoG}). 
The sphere-mesh model in Fig.~\ref{fig:HandsRelated:handModels:Tkach} \cite{Tkach:SIGGRAPH:2016} can be thought of as a generalization of the these models, while models based on shape-primitives are also used as underlying collision models \cite{Oikonomidis_1hand_object}. 
\citet{DART-Schmidt-RSS-14} voxelize each shape-primitive and compute a Signed Distance Function for the local coordinate frame (Fig.~\ref{fig:HandsRelated:handModels:articuTsdfDART}).
\citet{Melax:2013:handPhysics} use a union of convex bodies for hand tracking. 
However these approaches only roughly approximate hand shape. 
	
A triangulated mesh with Linear Blend Skinning (LBS) \cite{LBS_PoseSpace}  (Fig.~\ref{fig:HandsRelated:handModels:mesh}) is more realistic and better fits image data \cite{LucaHands,Tzionas:IJCV:2016}. 
Despite their fixed shape, meshes are useful for computing contact points during interaction \cite{Tzionas:ICCV:2015}; this computation can be accelerated by approximating the mesh as an ensemble of convex hulls \cite{Tzionas:IJCV:2016} (Fig. \ref{fig:HandsRelated:handModels:Tzionas}).  
A recent alternative~\cite{MSR_2016_Siggraph_handTrack} models the hand with smooth loop subdivision surfaces \cite{loop1987smooth} (Fig.~\ref{fig:HandsRelated:handModels:subdivision}), which facilitate efficient and accurate computation of derivatives.
\citet{ParagiosHandMonocular2011} define a triangulated hand model and introduce scaling terms for each bone, allowing the hand to change shape.
They use surface texture and shading to fit this to images in an analysis-by-synthesis approach. 

Recently deep learning is providing new options for estimating hands from images and depth maps.
Instead of using a single Convolutional Neural Network (CNN) to predict hand joints from the input image as \citet{NYU_tracker_tompson14tog}, \citet{Lepetit:ICCV:2015:handCnnLoop} train a second CNN to close the loop from output to input and refine the predicted pose. 
This second CNN (Fig.~\ref{fig:HandsRelated:handModels:cnnLoop}) is an image synthesizer inspired by \citet{dosovitskiy2015learning} that renders a hand given the hand pose predicted by the first CNN. 

All above approaches model hands in isolation from the rest of the body, with the exception of \cite{MSR_2016_Siggraph_handTrack, Melax:2013:handPhysics} that also model the forearm. 
However, as it is shown in our experiments section, having very noisy or even completely missing hand and forearm is not uncommon in real full-body data. 

\paragraph{Dimensionality Reduction.}
While hands have many degrees of freedom, a large number of them are not independently controllable and, in natural movements, hand poses are effectively low dimensional.
Various glove- \cite{Santello:Neuroscience:1998, Todorov:2004} or marker-based \cite{schroder2014real} capture methods have been used to study this with the effective dimensionality depending on the task and the capture system.
\citet{Santello:Neuroscience:1998} find that two principal components account for over $80\%$ of the variance in their data. 
\citet{schroder2014real} find that $3$-$6$ components account for $80-90\%$ of the variance. 
\citet{Todorov:2004} use a different protocol and find the effective dimensionality is about 6.5.  
While our capture method is different and more detailed, our results are broadly consistent with those of \citet{Todorov:2004}. 
	
The practical use of a low dimensional pose space for hands was shown by \citet{elkoura2003handrix}, who  used it in a data-driven approach to generate physically plausible poses for music generation. 
In addition, \citet{schroder2014real} employ it for efficient hand tracking with noisy RGB-D data. 
However they train on motion data acquired by an intrusive marker-based approach that alters the shape and pose space, and rely on a fixed shape model constituted by cylindrical shape primitives. 

\paragraph{Personalized hand models.}
Personalizing the hand model to the user improves tracking \cite{MSR_2016_CVPR_fitsGlove}. 
A personalized template mesh can be reconstructed offline with a multiview-stereo method \cite{LucaHands,Tzionas:IJCV:2016}, however this does not scale well with the number of users. 
\citet{ParagiosHandMonocular2011} define 56 scaling terms to allow bone lengths to vary but they do not learn how they correlate from data.
\citet{MSR_2014_CVPR_handShapeAdaptation} personalize a template hand mesh by adapting its shape to fit the depth data of a user's hand using an off-line calibration step for each user.

\citet{MSR_2015_CVPR_learnShapeModel} go further to learn a model of shape variation from scans of 50 people captured with a depth sensor.
They fit a linear blend skinned model to partial hand scan data and learn a low-dimensional PCA model of the deviations from this.
This is similar to \mano, except that they restrict the pose-dependent deformations to be modeled by LBS; we produce more realistic posed meshes by learning pose-dependent corrective blend shapes \cite{Loper:SIGASIA:2015}.
With a calibration step \cite{MSR_2015_CHI_sharp_track} their model can be used for realtime tracking \cite{MSR_2016_CVPR_fitsGlove,MSR_2016_Siggraph_handTrack}.

\paragraph{Body models.}
The modeling and tracking of bodies has a similar history to hands, paralleling the examples in Fig.~\ref{fig:HandsRelated:handModels}.
Early body models are based on geometric primitives, while more recent ones are learned from data.
Early methods capture the statistics of body shape but do not model how the geometry of the body changes with pose \cite{allen2003space,Seo:2003:SAB:846276.846292}.
The introduction of the SCAPE model \cite{anguelov2005scape} showed how to learn a model that combines shape and pose variations.
SCAPE achieved high realism and spawned many related approaches \cite{tenbo,Freifeld:ECCV:2012,hasler2009statistical,Hirshberg:ECCV:2012,Dyna:SIGGRAPH:2015}.
These models represent shape in terms of triangle deformations, which present technical challenges for optimization and fitting.
Vertex based models~\cite{Allen:2006:LCM,Hasler10,Loper:SIGASIA:2015} are preferable in this respect.
None of these models capture hand motion, adopting a fixed posture with either a fist or an open hand.

Here we base our hand model on SMPL \cite{Loper:SIGASIA:2015} and replace the hands in SMPL with our new learned \mano model.
The original SMPL hands include a small amount of bending and wrist rotation, but this is not sufficient to capture any significant hand pose variation.
The new \mano model can be used separately or together with the body (\smplH).  
When used separately, it should be able to replace most of the hand models described above easily since it is built on standard vertex-based modeling methods.
When combined with the body, we are able to estimate both hand poses and body movement together. 
We know of no previous models that attempt to model and capture the full body and detailed hands together.

%% file: tex_FIG/FIG_previous_handModels.tex
\newcommand{\ImgSquizModelsLiteratureWHHa}{+04.0mm}
\newcommand{\ImgSquizModelsLiteratureWHHb}{+03.0mm}
\newcommand{\ImgSquizModelsLiteratureWHHc}{+01.0mm}
\newcommand{\verticalspace}{-01.0mm}
\newcommand{\ImgSquizModelsLiteratureWSS}{0.110}
{
\begin{figure}[t]
		\subfloat[][]{	\includegraphics[trim=000mm 000mm 000mm 000mm, clip=true, height=\ImgSquizModelsLiteratureWSS \textwidth]{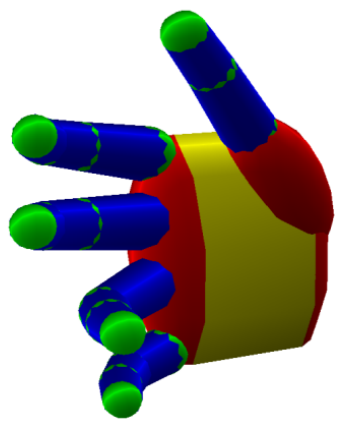}	\label{fig:HandsRelated:handModels:shapePrimitivesFORTH}	}	\hspace*{\ImgSquizModelsLiteratureWHHa}
		\subfloat[][]{	\includegraphics[trim=000mm 000mm 000mm 000mm, clip=true, height=\ImgSquizModelsLiteratureWSS \textwidth]{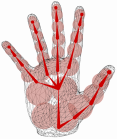}					\label{fig:HandsRelated:handModels:SoG}					}	\hspace*{\ImgSquizModelsLiteratureWHHa}
		\subfloat[][]{	\includegraphics[trim=000mm 000mm 000mm 000mm, clip=true, height=\ImgSquizModelsLiteratureWSS \textwidth]{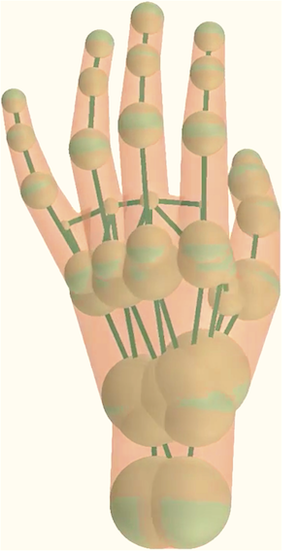}				\label{fig:HandsRelated:handModels:Tkach}				}	\hspace*{\ImgSquizModelsLiteratureWHHb}
		\subfloat[][]{	\includegraphics[trim=000mm 000mm 000mm 000mm, clip=true, height=\ImgSquizModelsLiteratureWSS \textwidth]{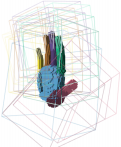}			\label{fig:HandsRelated:handModels:articuTsdfDART}		}\\
		\subfloat[][]{	\includegraphics[trim=000mm 000mm 000mm 000mm, clip=true, height=\ImgSquizModelsLiteratureWSS \textwidth]{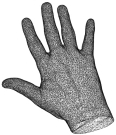}			\label{fig:HandsRelated:handModels:mesh}					}	\hspace*{\ImgSquizModelsLiteratureWHHb}
		\subfloat[][]{	\includegraphics[trim=000mm 000mm 000mm 000mm, clip=true, height=\ImgSquizModelsLiteratureWSS \textwidth]{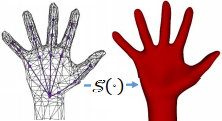}		\label{fig:HandsRelated:handModels:subdivision}			}	\hspace*{\ImgSquizModelsLiteratureWHHb}
		\subfloat[][]{	\includegraphics[trim=000mm 000mm 000mm 000mm, clip=true, height=\ImgSquizModelsLiteratureWSS \textwidth]{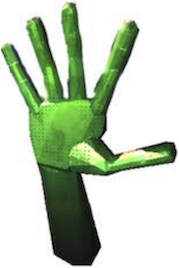}				\label{fig:HandsRelated:handModels:Melax}				}\\
		\subfloat[][]{	\includegraphics[trim=000mm 000mm 000mm 000mm, clip=true, height=\ImgSquizModelsLiteratureWSS \textwidth]{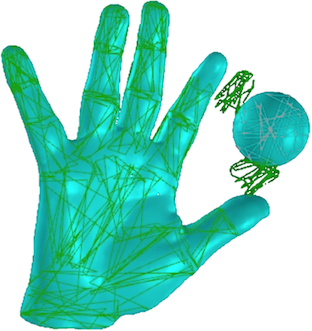}		\label{fig:HandsRelated:handModels:Tzionas}				}	\hspace*{\ImgSquizModelsLiteratureWHHc}
		\subfloat[][]{	\includegraphics[trim=000mm 000mm 000mm 000mm, clip=true, height=\ImgSquizModelsLiteratureWSS \textwidth]{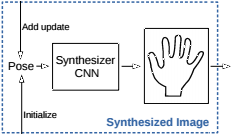}				\label{fig:HandsRelated:handModels:cnnLoop}				}	\hspace*{\ImgSquizModelsLiteratureWHHc}
		\subfloat[][]{	\includegraphics[trim=000mm 000mm 000mm 000mm, clip=true, height=\ImgSquizModelsLiteratureWSS \textwidth]{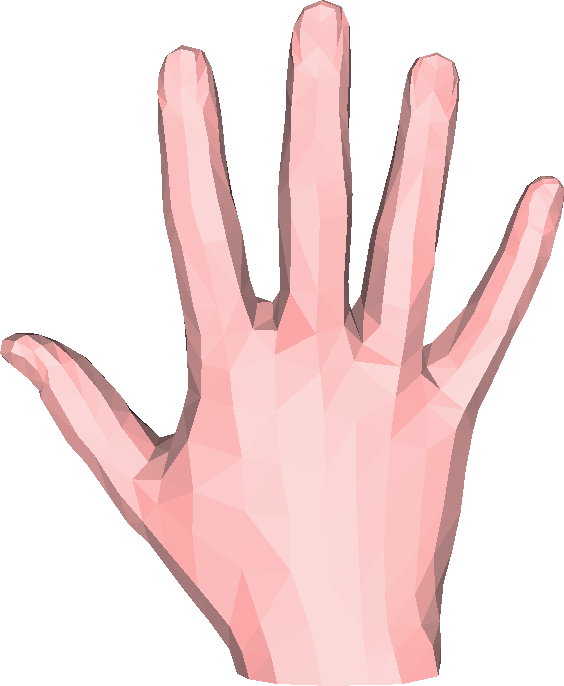}				\label{fig:HandsRelated:handModels:HAMPL}				}
	\caption[Hand models]{Hand models.
			(a) \emph{Primitives} approximation \cite{OikonomidisBMVC},
			(b) \emph{Sum-of-Gaussians} model \cite{srinath_iccv2013},
			(c) \emph{Sphere-Meshes} \cite{Tkach:SIGGRAPH:2016}  can be thought of as a generalization of the previous models, 
			(d) \emph{Articulated TSDF} for a voxelized shape-primitive hand model \cite{DART-Schmidt-RSS-14},
			(e) \emph{Triangular Mesh} \cite{LucaHands,Tzionas:IJCV:2016},
			(f) \emph{Loop Subdivision Surface} of a triangular control mesh \cite{MSR_2015_CVPR_learnShapeModel},
			(g) \emph{Convex Bodies} for tracking \cite{Melax:2013:handPhysics},
			(h) \emph{Convex Parts} of a triangular mesh for contact point detection \cite{Tzionas:IJCV:2016},
			(i) \emph{Learned Model} using a CNN to synthesize images of a given hand pose \cite{Lepetit:ICCV:2015:handCnnLoop},
			(j) Our proposed hand \emph{\mano} model.
			Images reproduced from the cited papers or their supplementary videos. 
	}
	\label{fig:HandsRelated:handModels}
\end{figure}
}

%% file: tex_main/03_method.tex
\section{Model}		\label{sec:model}

\begin{figure*}[t]
  \includegraphics[width=1.0\linewidth]{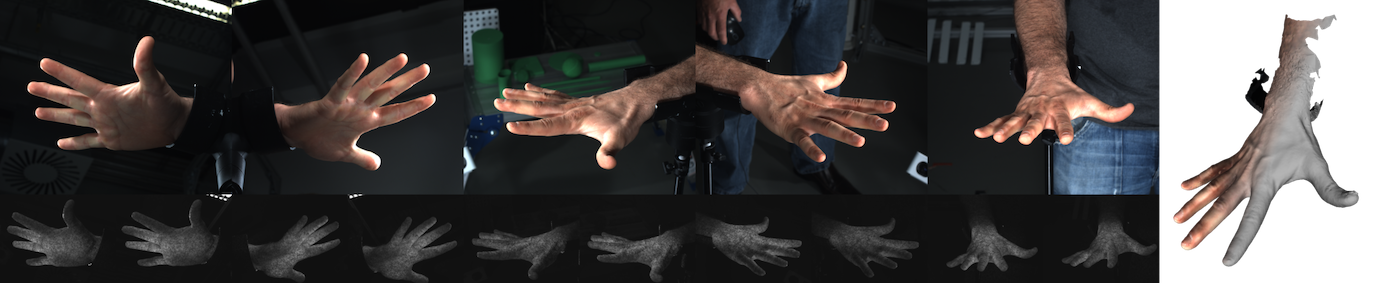}
  \caption{Data collection with the 3dMDhand System. The system
    captures a hand with the wrist resting on a hand stand, and performing
    different articulated motions. 
	The active stereo system is composed by
    5 scanning units, each collecting one color image (top row) with flash-based
    illumination and two grayscale images (bottom row) using speckle
    illumination. The result is a high quality, textured scan of the hand
    (right most image).}
  \label{fig:data_collection}
\end{figure*}

Given the difficulty of capturing hands with bodies, we advocate for a two-stage approach in the creation of a dexterous full body model. 
First, we collect a large number of scans of hands in isolation. 
These scans are obtained with a scanner configured specifically to capture hands with a fixed wrist position. 
This allows us to capture the nuances of hand deformation.
We then train a hand model using an iterative process of aligning a template to the scans using the model and learning a model from the registered scans.
In the second stage, we integrate this hand model with the full body to obtain a single, dexterous and fully articulated body
model. 

\mano is based on SMPL~\cite{Loper:SIGASIA:2015} and is compatible with the full body SMPL model. 
In order to facilitate the integration of the hand models with the full body model, we take the hand vertices from the full body model as a
template. 
The general formulation of a SMPL model $M$, taken from the original paper \cite{Loper:SIGASIA:2015} for completeness, is as follows:
\begin{align}
  M(\shape, \pose) &= W(T_P(\shape, \pose), J(\shape), \pose, \bweights) \\
  T_P(\shape, \pose) &= \template + B_S(\shape) + B_P(\pose)
  \label{eq:initial_model}
\end{align}
where a skinning function $W$ (in our case LBS) is applied to an articulated rigged body mesh with shape $T_P$, joint locations $J$ defining a kinematic tree, pose $\pose$, shape $\shape$, and blend weights $\bweights$. 

Unlike standard LBS models, with a SMPL model the mesh $T_P$ that is posed and skinned, is a function of the pose and shape of the hand.
Shape blendshape function $B_S$ allows the base shape, in this case of the hand, to vary with identity.
The pose blend shape function, $B_P$, captures deformations of the mesh as a function of the bending of the joints. 
Traditional LBS models are overly smooth and suffer from ``collapse'' at the joints.
For \mano, we learn corrective blend shapes that correct these artifacts, resulting in more natural-looking finger bending.

Specifically the pose and shape blend shapes are defined as the linear combination of a set of deformations, i.e. vertex offsets:
\begin{eqnarray}
	B_P(\pose; \mathcal{P}) & = & \sum \limits_{n=1}^{9K} \left(R_n(\pose) - R_n(\pose^*) \right) \textbf{P}_n,		\label{eq:blendShapes_POSE}
    \\
	B_S(\shape; \mathcal{S}) & = & \sum \limits_{n=1}^{|\shape|} \beta_n \textbf{S}_n.								\label{eq:blendShapes_SHAPE}
\end{eqnarray}
Here $\textbf{P}_n\in\mathcal{P}$ are the pose blend shapes and $K$ is the number of parts in the hand model.
These are not directly controlled by the hand part rotations but rather by the elements of the part rotation matrices as in SMPL.
$R_n(\pose)$ indexes into the $n^{th}$ element of a vector of concatenated rotation matrix elements and $\pose^*$ is the zero pose.

\input{tex_FIG/FIG_shape_pca___SIGGASIA}

The shape blend shapes are computed from a set of registered hand shapes, normalized to the zero pose,  using principal component analysis (PCA) as illustrated in Fig.~\ref{fig:pcaPCsShape}.
Consequently here, the $\beta_n$ are linear coefficients and the vectors $\textbf{S}_n\in\mathcal{S}$ are principal components in a low-dimensional shape basis that we learn below.

The joint locations, $J(\shape)$, also depend on shape parameters. 
These are learned as a sparse linear regression matrix $\mathcal{J}$ from mesh vertices as in SMPL.
We refer the reader to~\cite{Loper:SIGASIA:2015} for more details.

Before we can learn the model parameters $(\mathcal{S},\mathcal{P},\mathcal{W},\mathcal{J},\template)$,  we first need hand scans of many people in many poses and then we need to register our template mesh to these to put them in correspondence.
We describe these steps below.

\subsection{Hand data}\label{sec:HAND_model}

Our hand data was captured with a 3dMDhand System \cite{3dMDhand} (Fig.~\ref{fig:data_collection}, 3dMD LLC, Atlanta). 
The system is composed of five scanning units, each containing one color camera with resolution $2448 \times 2048$, two grayscale cameras with resolution $1624 \times 1236$ and two speckle projectors that provide the illumination for the grayscale images. 
The scans have a resolution of approximately $50,000$ vertices, include a texture map, and provide an accuracy of 0.2 millimeters root mean square (RMS) error according to the manufacturer.

\input{tex_FIG/FIG_handpose_protocol}

Using this system, we collected data from both left and right hands of a total of $31$ subjects, providing us a total number of $2018$ scans.
We mirror the left-hand scans to appear as right hands. 
Mirroring enables us to train a single consistent hand model.  
We later mirror back the learned right hand model to create a left hand model.

Each subject was captured performing three types of poses:
a set of joint exploration poses, the $31$ poses from the grasp taxonomy of Feix et al.~\shortcite{Feix:HumanMachine:2016}, and a few mixed
poses. 
One can see a complete set of poses in Fig.~\ref{fig:hand_poses}. 
Each subject performed a subset of the protocol depending on their availability.

For each scan, we manually remove the geometry corresponding to the arm and handstand. 
This manual cropping takes approximately $30$ seconds per scan. 
We then automatically segment out the objects in the poses involving object grasps using color information, since they are painted green. 
More specifically, a vertex is classified as object vertex simply if $G>max(R, 20)$ in 8-bit RGB space, where $G$ and $R$ are the values of the green and red channel correspondingly. 
This leaves the scans of hands grasping objects with significant holes.

\subsection{Registration}\label{sec:handModel_RegistrationBootstraping}
The next step to building a hand model is to register, or align, a template to all the hand scans, bringing them into correspondence.
The process of registering noisy hand scan data is challenging. 
Mesh registration is a challenging problem in general, but hands are especially difficult due to a high degree of self-similar structures (fingers) and a high level of self and object occlusion. 
While some previous work uses landmarks to simplify registration, we do not do so because it is impractical when registering a large
number of scans. 
Instead, we bootstrap the creation of the model by manually curating the registrations.
Specifically, we start with a crude hand model, use it to register all our scans, manually curate good ones, learn an improved model, and repeat.
This gradually improves the alignments and the model.

We treat the registration process as an optimization problem, where we minimize the distance between the scan and a registered mesh, $\algn$, with respect to the registration vertex locations, while keeping the registration \emph{likely} according to the model.
We do not provide full details here because the approach is similar to previous work \cite{Hirshberg:ECCV:2012,Loper:SIGASIA:2015}, with the addition of the manual curation.
Specifically we minimize
\begin{align}
  E(\shape, \pose, \algn; \scanverts) &= \lambda_g E_g + \lambda_c E_c + \lambda_\theta E_\theta + \lambda_\beta E_\beta 
  \label{eq:initial_registration}\\
  E_g(\algn; \scanverts) &= \sum_{\scanvert \in \scanverts} \rho(\min_{\valgnsurf \in \algnsurf} \| \scanvert - \valgnsurf\|) 
  \label{eq:initial_registration_Eg}\\
  E_c(\algn, \shape, \pose) &= \sum_i \| D_i(\algn) - D_i(M(\shape, \pose)) \|^2 
  \label{eq:initial_registration_Ec}\\
  E_\theta(\pose) &= \| \pose \|^2 
  \label{eq:initial_registration_Etheta}\\
  E_\beta(\shape) &= \| \shape \|^2 .
  \label{eq:initial_registration_Ebeta}
\end{align}

The energy is composed by four main components: data (or geometry) term $E_g$, coupling term $E_c$, pose prior $E_\theta$ and shape prior $E_\beta$.
The data term $E_g$ represents the point-to-plane distance between the vertices $\scanvert$ on the scan $\scanverts$ and the surface of the registration $\algnsurf$ (where calligraphic denotes a continuous surface), robustified with a Geman-McClure error function, $\rho$, \cite{GemanMcClure1987}. 

The coupling term $E_c$ encourages the registered mesh, $\algn$, to be similar to the model $M$, whose parameters, $\shape, \pose$, we also optimize.
We define similarity here in terms of the differences between the edges of the model and the registered mesh. 
The edges are given by the function $D$, which is simply a linear mapping.
By directly optimizing the vertices in the registration we can go beyond the limitations of the model and obtain more faithful registrations, while the coupling term keeps the registration conservatively close to the model. 

The shape prior term, $E_\beta$, penalizes the Mahalanobis distance between the optimized shape parameters and the distributions of hand shapes in the CAESAR dataset~\cite{caesar}. 
Our shape space is orthogonal since it is the result of performing PCA of unposed registrations.
We scale its basis vectors according to the square root of the explained variance $\sqrt{cov}$, which requires the coefficients $\shape$ to be scaled by its inverse, effectively standardizing them.
As a result (see \cite{mahalanobisScaledPCA}), the Mahalanobis distance is conveniently computed just as the norm of the shape parameters $\shape$. 

For the pose prior $E_\theta$, we define specific priors for each pose (since they are known given the protocol). 
We initialize this with a prior that penalizes deviations from the neutral pose and then refine with Gaussian priors for each pose.

The objective in Eq.~\ref{eq:initial_registration} is highly non-convex. 
To optimize it, we use dogleg, a quasi-Newton least-squares optimizer~\cite{nocedal2006numerical}. 
The gradients are computed with automatic differentiation with  OpenDR~\cite{Loper:ECCV:2014}.

We iterate the whole process of registration and model building twice, with a final visual inspection in between to define the training database of well-registered meshes. 
After inspection,  $1554$ registrations of $2018$ scans were deemed successful and are included in the released dataset. 
This inspection takes only approximately $1$-$2$ seconds per scan, as it is a simple binary labeling task. 
For registration, in the first iteration we use the initial pose prior, while in the second we use the pose specific Gaussian prior, as described above. 
Registration takes approximately $30$ seconds per frame on a $3.7$ GHz Quad-Core Intel Xeon E5 computer using 4 threads, where only the closest-point search is multithreaded. 

Figure~\ref{fig:crazyPoses}  shows a variety of hand scans from one subject and the corresponding aligned meshes.
Figure~\ref{fig:neutralPoseManySubjects} shows a variety of hand shapes from different subjects in the same, relatively flat, pose.
These figures give a sense of the level of detail in the aligned meshes.

\input{tex_FIG/FIG_example_scans_registrations}

\subsection{Hand Model}
Given this set of curated registrations, the goal is to learn the parameters of a SMPL-style hand model so that it fits the registrations.
We start with the same strategy as used for the body in \cite{Loper:SIGASIA:2015}.
The model parameters, namely $(\mathcal{S},\mathcal{P},\mathcal{W},\mathcal{J},\template)$, are optimized iteratively keeping the rest constant. 
As in SMPL, personalized templates $\boldsymbol{\hat{T}}_i$ are used for optimizing the pose related components $(\mathcal{P},\mathcal{W})$, and its linear decomposition performed with PCA populates the shape space $\mathcal{S}$ and template $\template$.
However, we modify a number of components in SMPL to take into account differences between the hands and the body.

\mano contains $15$ joints plus the global orientation.  
Unlike most joints in the body, many hand articulations are anatomically restricted to one degree of freedom, while our model considers them as ball joints for simplicity. 
Hand pose is therefore effectively over-parameterized.
Since most of the parameters in a SMPL model belong to the pose-dependent blend shapes $B_P$ (which grows linearly with number of joints), the regularization of $B_P$ for a highly articulated object like the hand is important to avoid overfitting.
One of the effects of this overfitting is the non-locality of pose-dependent deformations obtained when a model is trained under the procedure described in~\cite{Loper:SIGASIA:2015}.

Consequently, we reformulate the model to have a stronger regularization that promotes pose-dependent blend shapes $B_P$ that are local to the joints that influence them.
Since pose-dependent blend shapes map pose rotation elements to vertex displacements, a natural way to achieve this is by penalizing the dependency of vertex displacements on joints whose rotation centers are far away. 

More concretely, we replace the constant cost $\lambda_P$ associated with all pose blend shape elements in SMPL (Eq.~$14$~\cite{Loper:SIGASIA:2015}) by a cost that depends on the distance between the input joint and the output vertex
\begin{align}
  \Lambda_P(i, j) &= \mathcal{J}_j D_{\textrm{geo}}(t_i, \boldsymbol{\bar{T}})
  \label{eq:new_regularization_posedep}
\end{align}
where $D_{\textrm{geo}} \in \mathbb{R}^N$ is the geodesic distance between a particular vertex $t_i$ and the rest of the vertices in the template mesh $\boldsymbol{\bar{T}}$, and $\mathcal{J}_j \in \mathbb{R}^N$ denotes the joint regressor matrix for joint $j$.
$\Lambda_P$ establishes a cost for each pair of input joint and output vertex. 
Since each input joint spans $9$ scalar inputs corresponding to its rotation matrix, and each output vertex corresponds to $3$ scalars, each element in $\Lambda_P$ is expanded into a $3\times 9$ block before computing the product with $\mathcal{P}$. 
The geodesic distance to a weighted average of points is not defined; therefore, we replace it by the weighted average of the geodesic distances.
This new regularization scheme results in more local and regularized pose-dependent deformations, as illustrated in the {\bf supplemental video}. 

\begin{figure}[t]
    \includegraphics[width=1.0\linewidth]{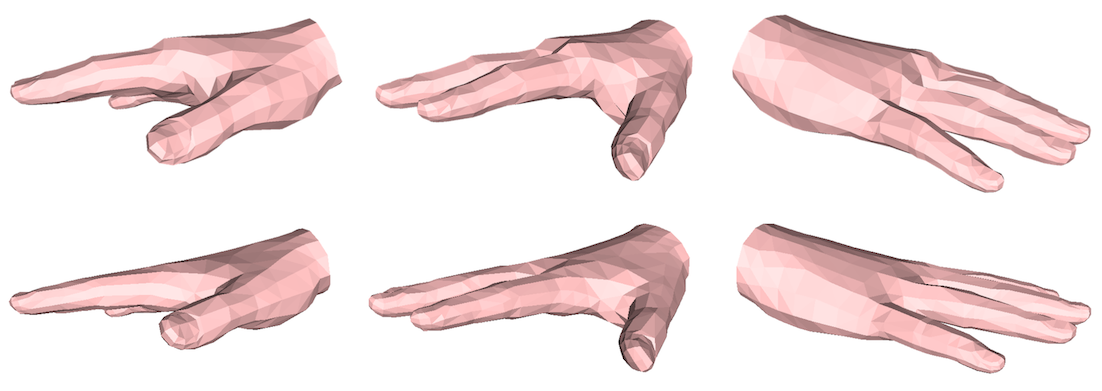}
    \caption{
        Three different views (columns) of the average pose-subject templates $\hat{\boldsymbol{\mathrm{T}}_i}$ for a model optimized with constant (first row) and vertex-joint-dependent (second row) regularization. 
        Pay attention to the bumps in the knuckles in the first row, which are attenuated considerably in the second.
    }
    \label{fig:old_vs_new_vtmpl}
\end{figure}

Second, the zero pose of our model represents a flat hand (as in the CAESAR dataset \cite{caesar}, for compatibility with SMPL), which is far from the mean pose in our dataset (see Fig.~\ref{fig:pcaPCsPose}).
The deformations entailed by the difference between the zero and mean poses result in unnatural templates due to the optimization order of the model.
More specifically, the initial optimization of the personalized templates, $\boldsymbol{\hat{T}}_i$, is performed using a model without pose blend shapes, $B_P$, which are optimized afterwards. 
The templates then absorb the pose-dependent corrections between the zero and mean pose, resulting in unnaturally extruded knuckles (see top row in Fig.~\ref{fig:old_vs_new_vtmpl}).
For this reason the template optimization should not take into account registrations with extreme poses until the pose dependent deformations work reasonably. 
We express this by weighting the registrations used in the optimization of $\boldsymbol{\hat{T}}_i$ according to how much we believe they can be potentially affected by the pose-dependent blend shapes, namely
\begin{align}
    w_i &= \left[ \sum_{n=1}^{9K} \|R_n(\vec{\theta}) - R_n(\vec{\theta}^*)\|^2_F \right]^{-2}
    \label{eq:weight_template}
\end{align}
where $R_n$ represents the $n^{th}$ element of the concatenation of rotation matrices corresponding to the pose axis angles $\vec{\theta}$ or the rest pose $\vec{\theta}^*$ and $\| \cdot \|_F$ denotes the Frobenius norm. 
We choose to use the Frobenius norm of the difference of rotation axes since it has a clear Euclidean interpretation, while it is also rotation invariant and has no problems with periodicity, as opposed to other representations like the difference of angles. 
The use of double squaring is a simple heuristic to heavily penalize deviations from the rest pose. 
The weights of Equation \eqref{eq:weight_template} are only applied in the first iteration of $\boldsymbol{\hat{T}}_i$ optimization, when the pose-dependent blend shapes are zero, and essentially penalize deviations from the rest pose that activate strong pose blend shapes. 

\input{tex_FIG/FIG_poseblendshapes_onoff___SIGGASIA}

Third, we exploit the symmetry of left and right hands by creating a single right hand model from right and (mirrored) left data.  
We then mirror the resulting model to obtain the left hand model.
This allows us to virtually double the amount of training data, which helps limit overfitting.
Details of the mirroring are included in the {\bf supplementary material}

Finally, another difference with respect to SMPL is that, after the initial set of registrations, we do not use the CAESAR data for our shape space. 
We instead compute our shape space using the neutral poses from our pose subjects. 
Figure~\ref{fig:neutralPoseManySubjects} shows some examples.
The reason for this is that CAESAR registrations have some systematic bias due to the combination of the pose, occlusions, and technology used to create the dataset.

\input{tex_FIG/FIG_pose_pca___SIGGASIA}

With these changes in the training procedure, we now train \mano following~\cite{Loper:SIGASIA:2015}. 
This step takes approximately $42$ hours on a single $3.7$ GHz Quad-Core Intel Xeon E5 computer using 8 threads. 
Learned components of our model, like the pose-dependent corrective blend shapes are critical to correct the errors of linear blend skinning and to capture natural bending of the fingers; see Fig.~\ref{fig:poseBlendShapesOnOff}.

\paragraph{Hand pose embedding.}
In order to make the model practical for the purpose of scan registration, we will try to expose a set of parameters that efficiently explain the most common hand poses.
Since research in neuroscience has shown that most hand pose variance lies on low-dimensional manifolds~\cite{Santello:Neuroscience:1998}, we will parameterize each hand posture by a set of coefficients that map a low-dimensional manifold. 
The only requirement for this manifold is to be differentiable in order to apply the chain rule within our optimization framework. 
Striving for simplicity, we choose a linear mapping obtained with PCA on the poses in our dataset, in the axis angle representation (after mapping all axis angles to the range $(-\pi, \pi]$).
The principal components (PCs) of our pose space can be seen in~Fig.~\ref{fig:pcaPCsPose}. 
For high-quality data, high accuracy can be achieved using many PCs. 
For noisy data though, one needs fewer PCs in a trade-off between accuracy (many PCs) and robustness (few PCs). 
For our data, we observed that $6$ components per hand suffice to model most common hand poses, capturing approximately $81\%$ of the variance in the training poses (see Experiments). 

\subsection{\smplH: Model integration}\label{sec:SMPLH_modelIntegration}

Unlike previous hand models, our goal is to obtain an integrated dexterous full body model that can be registered to full body scans. 
We therefore need to integrate the right and left hand models with a full SMPL body model. 
The basic integration is straightforward: the components of the model related to the finger articulation (blend weights, pose-dependent deformations and joint-regressors) are taken from the trained \mano models, while the components related to the rest of the body are taken from the full body SMPL model, including the wrist. 
In order to capture the correlations between body and hand shape, we use the shape space from the full body model \cite{Loper:SIGASIA:2015, caesar} to determine the shape of the hands.

Similarly to the SMPL model, this body+hands model is fully compatible with animation packages and game engines, and its simplicity makes it fast and simple to use.
The full articulation of the hands comes at a cost though: the dimensionality of the pose space has increased more than twofold.
This is rather counterproductive, since most of the pose degrees of freedom are devoted to a small area of the body in which typically data is either noisy or missing.
Consequently, rather than modeling the full dimensionality of the hand pose space, we use a 6-dimensional linear embedding described above for each hand.
Then, the kinematic structure of SMPL has $66$ degrees of freedom without the hands and \smplH then has $78$.

%% file: tex_FIG/FIG_shape_pca___SIGGASIA.tex
\newcommand{\ImgSquizPcaShapeWHma}{-12.0mm}
\newcommand{\ImgSquizPcaShapeWHta}{-10.0mm}
\newcommand{\ImgSquizPcaShapeWHtb}{-08.0mm}
\newcommand{\ImgSquizPcaShapeWHHa}{-16.0mm}
\newcommand{\ImgSquizPcaShapeWHHb}{-15.0mm}
\newcommand{\ImgSquizPcaShapeWVva}{-04.0mm}
\newcommand{\ImgSquizPcaShapeWVvb}{-03.0mm}
\newcommand{\ImgSquizPcaShapeWVVa}{-09.0mm}
\newcommand{\ImgSquizPcaShapeWSSa}{0.103}
\newcommand{\ImgSquizPcaShapeWSSb}{0.10}

\newcommand{\ImgSquizPcaShapeHini}{-04.0mm}

\newcommand{\ImgSquizShapeEngineeringViewpointsVa}{-5mm}
\newcommand{\ImgSquizShapeEngineeringViewpointsVb}{-8mm}

\newcommand{\stdNoShapePLU}{3}
\newcommand{\stdNoShapeMIN}{\stdNoShapePLU}

\begin{figure}
	\footnotesize
	\hspace*{\ImgSquizPcaShapeHini}
	\begin{tabular}{c c c c c c c c c c c c}
		\multirow{-4}{*}{	\includegraphics[trim=090mm 000mm 105mm 000mm, clip=true,  height=\ImgSquizPcaShapeWSSa \textwidth]{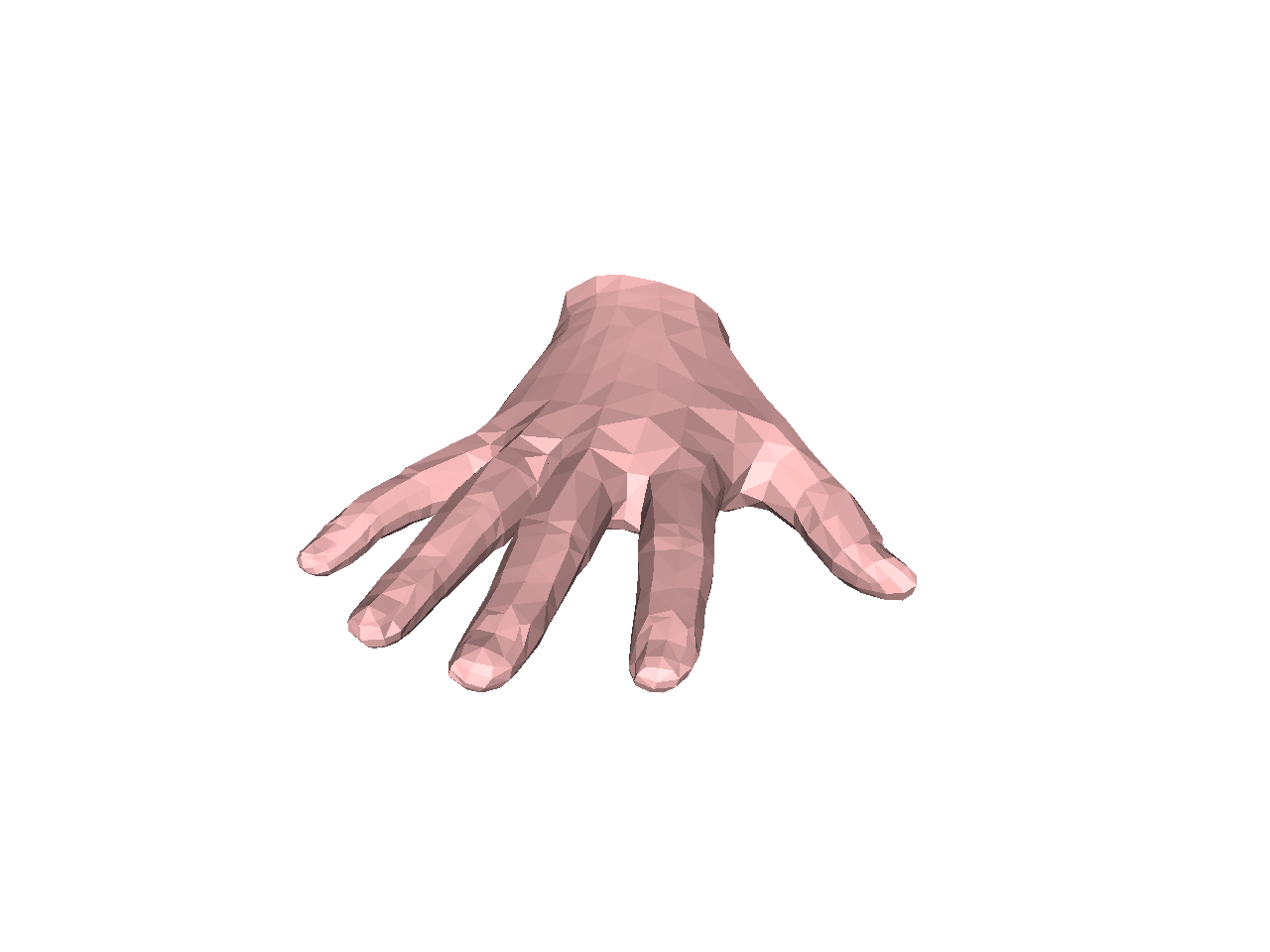}	}	&		\hspace*{\ImgSquizPcaShapeWHma}
							\includegraphics[trim=000mm 000mm 000mm 000mm, clip=false, height=\ImgSquizPcaShapeWSSa \textwidth]{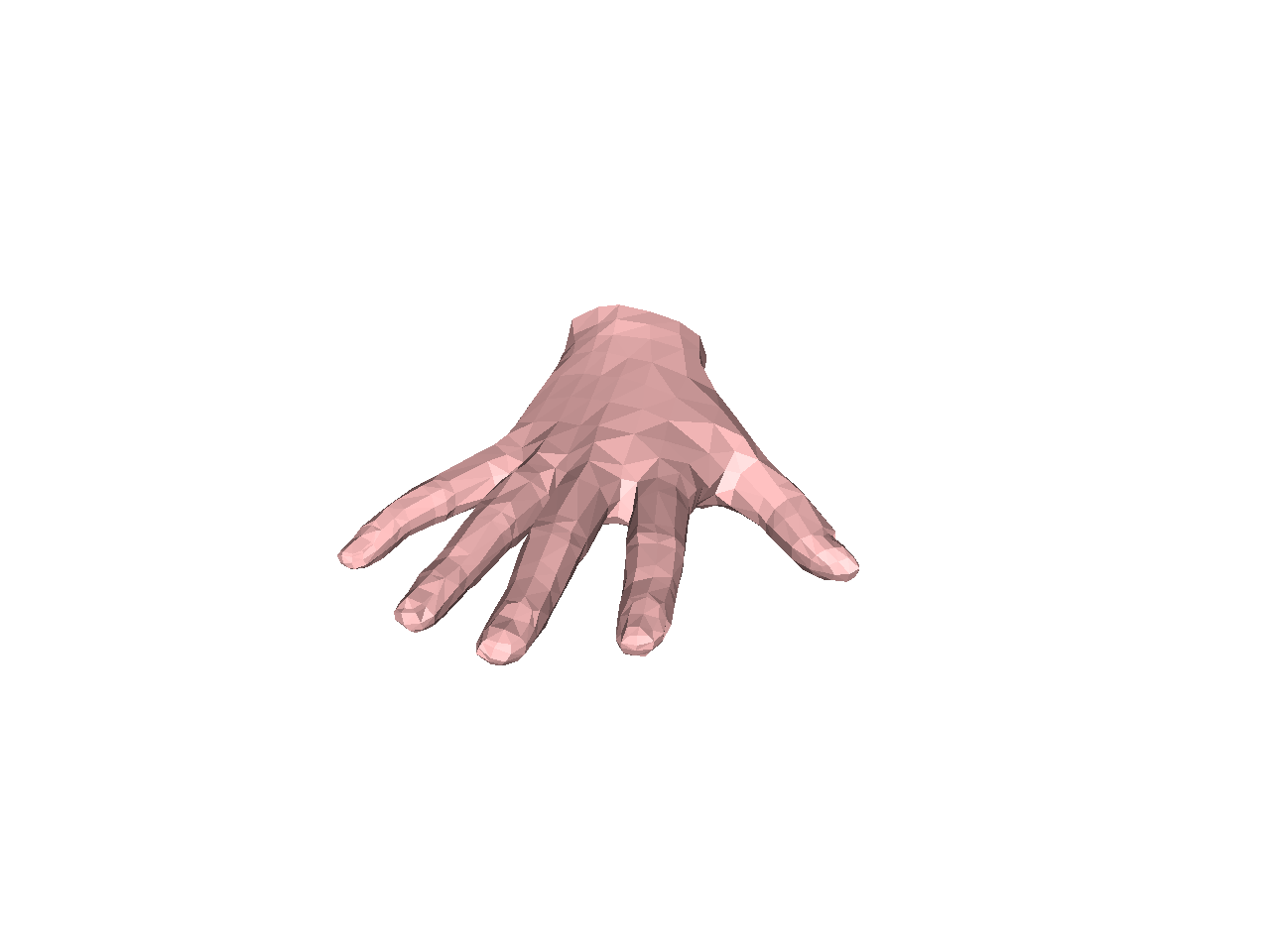}		&		\hspace*{\ImgSquizPcaShapeWHHb}
							\includegraphics[trim=000mm 000mm 000mm 000mm, clip=false, height=\ImgSquizPcaShapeWSSa \textwidth]{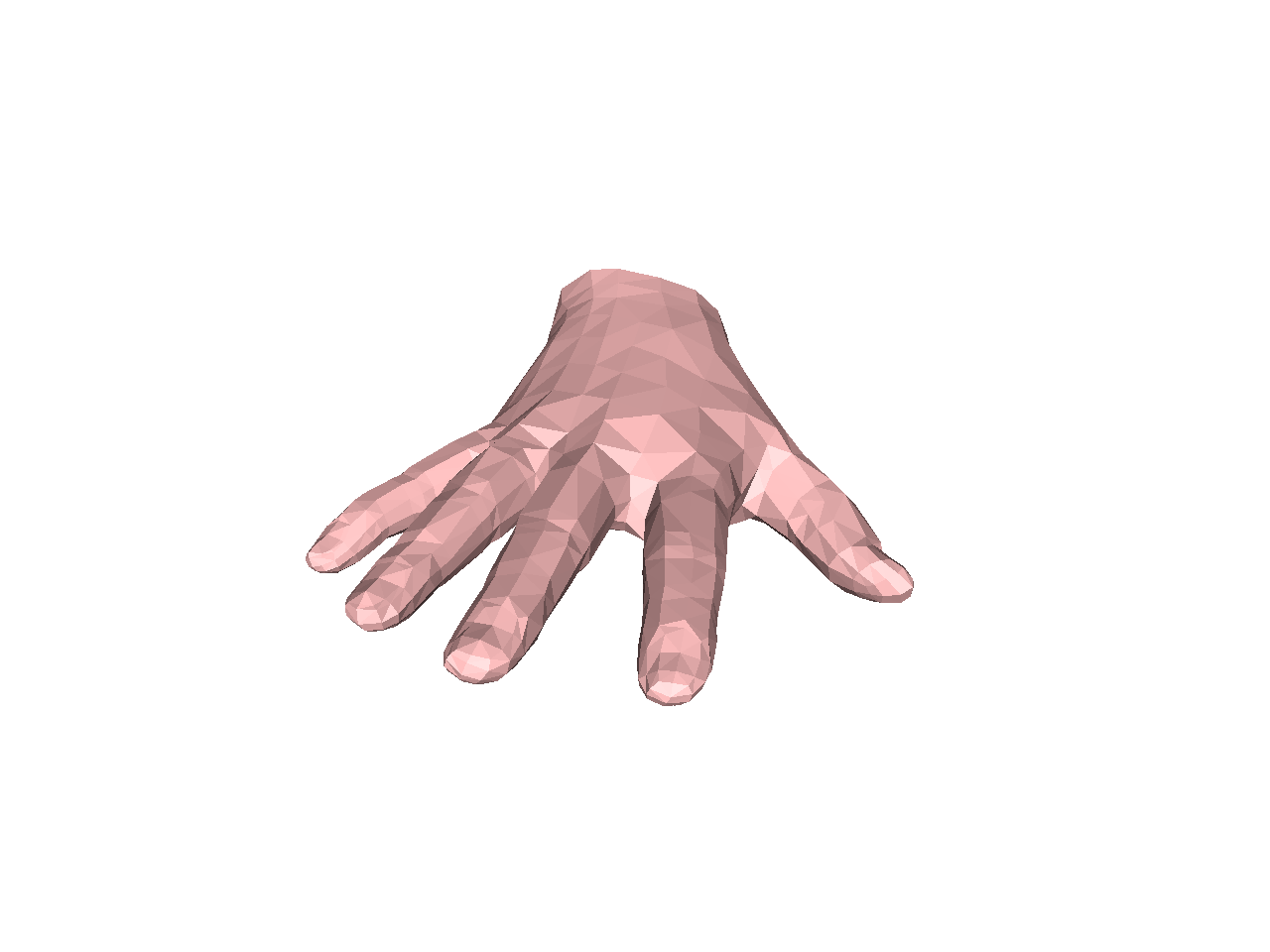}		&		\hspace*{\ImgSquizPcaShapeWHHa}
							\includegraphics[trim=000mm 000mm 000mm 000mm, clip=false, height=\ImgSquizPcaShapeWSSa \textwidth]{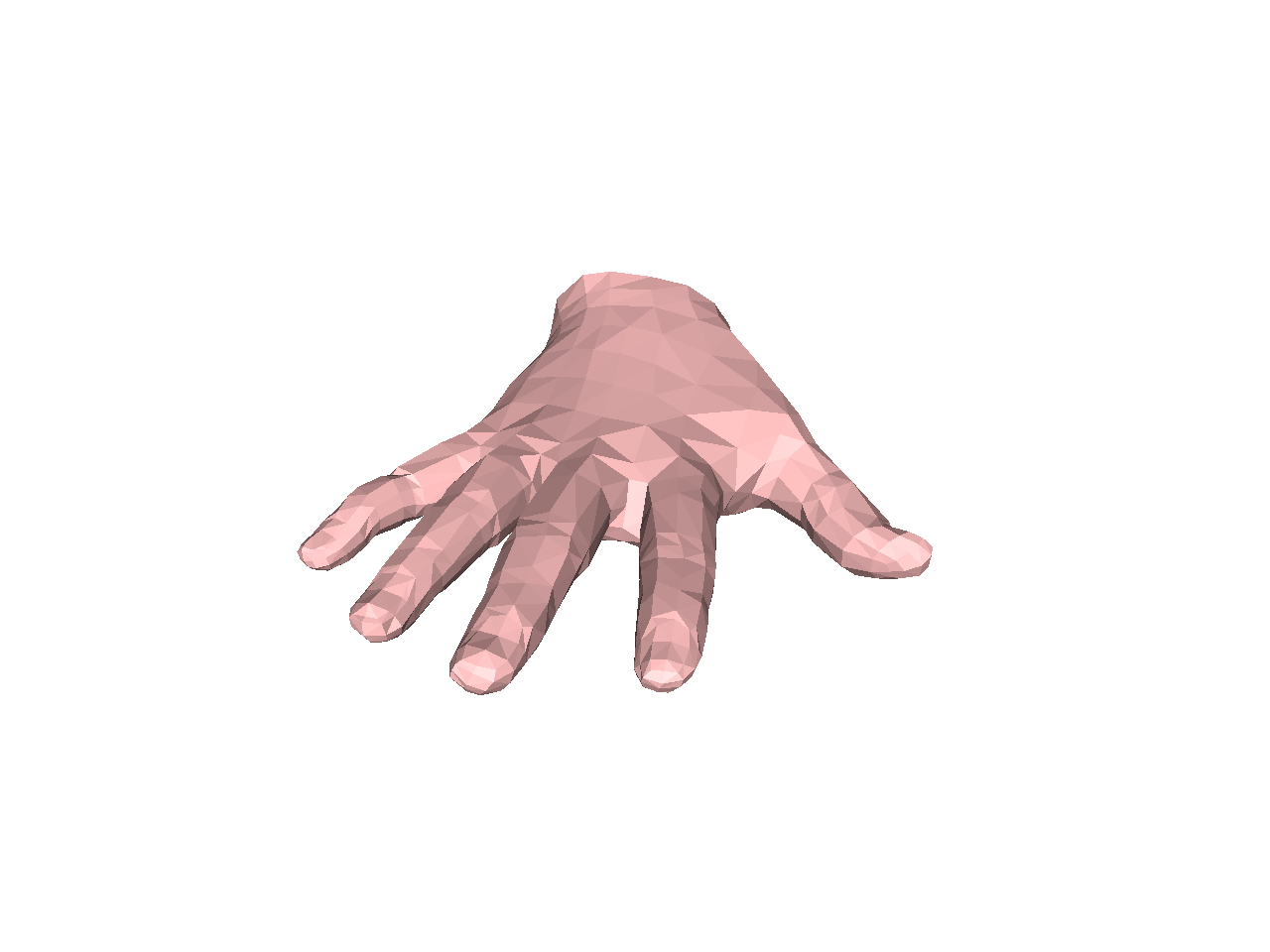}		&		\hspace*{\ImgSquizPcaShapeWHHa}
							\includegraphics[trim=000mm 000mm 000mm 000mm, clip=false, height=\ImgSquizPcaShapeWSSa \textwidth]{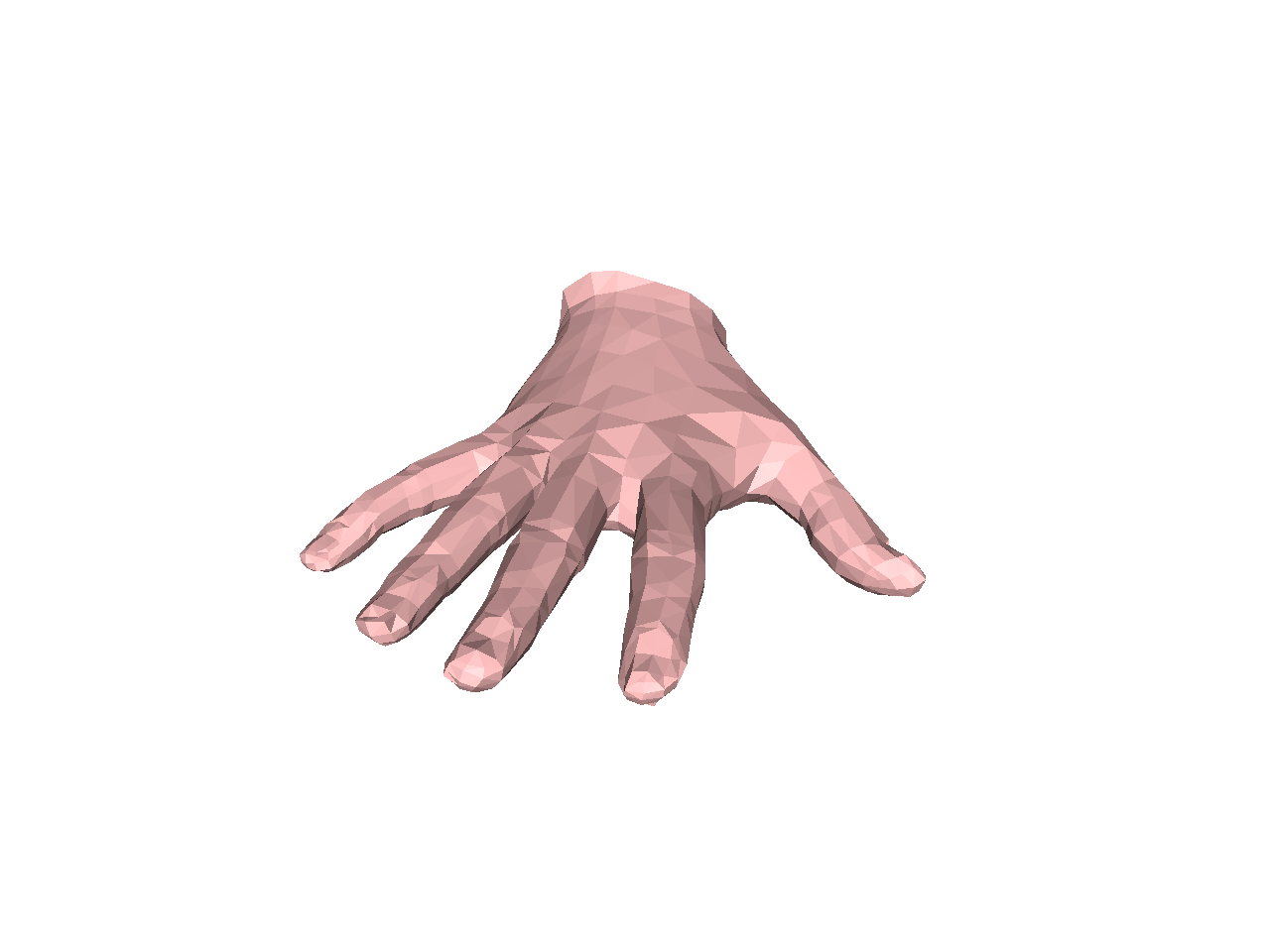}		&		\hspace*{\ImgSquizPcaShapeWHHa}
							\includegraphics[trim=000mm 000mm 000mm 000mm, clip=false, height=\ImgSquizPcaShapeWSSa \textwidth]{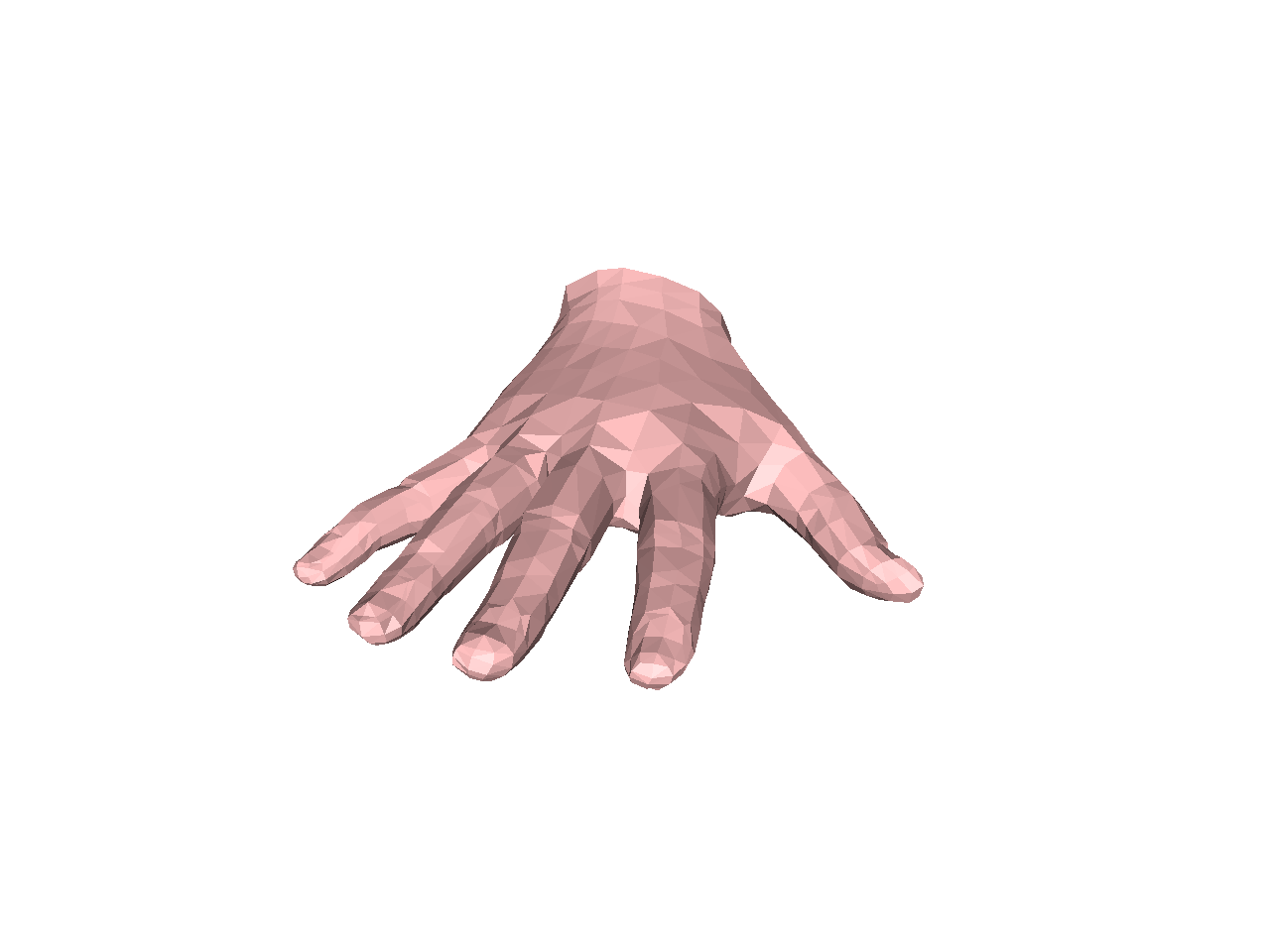}		&		\hspace*{\ImgSquizPcaShapeWHta}
							\multirow{-6}{*}{\hspace{6mm}\parbox{09mm}{mean\\+\stdNoShapePLU ~std\\}}																														\\				[\ImgSquizPcaShapeWVVa]
							{}																																															&		\hspace*{\ImgSquizPcaShapeWHma}
							\includegraphics[trim=000mm 000mm 000mm 000mm, clip=false, height=\ImgSquizPcaShapeWSSa \textwidth]{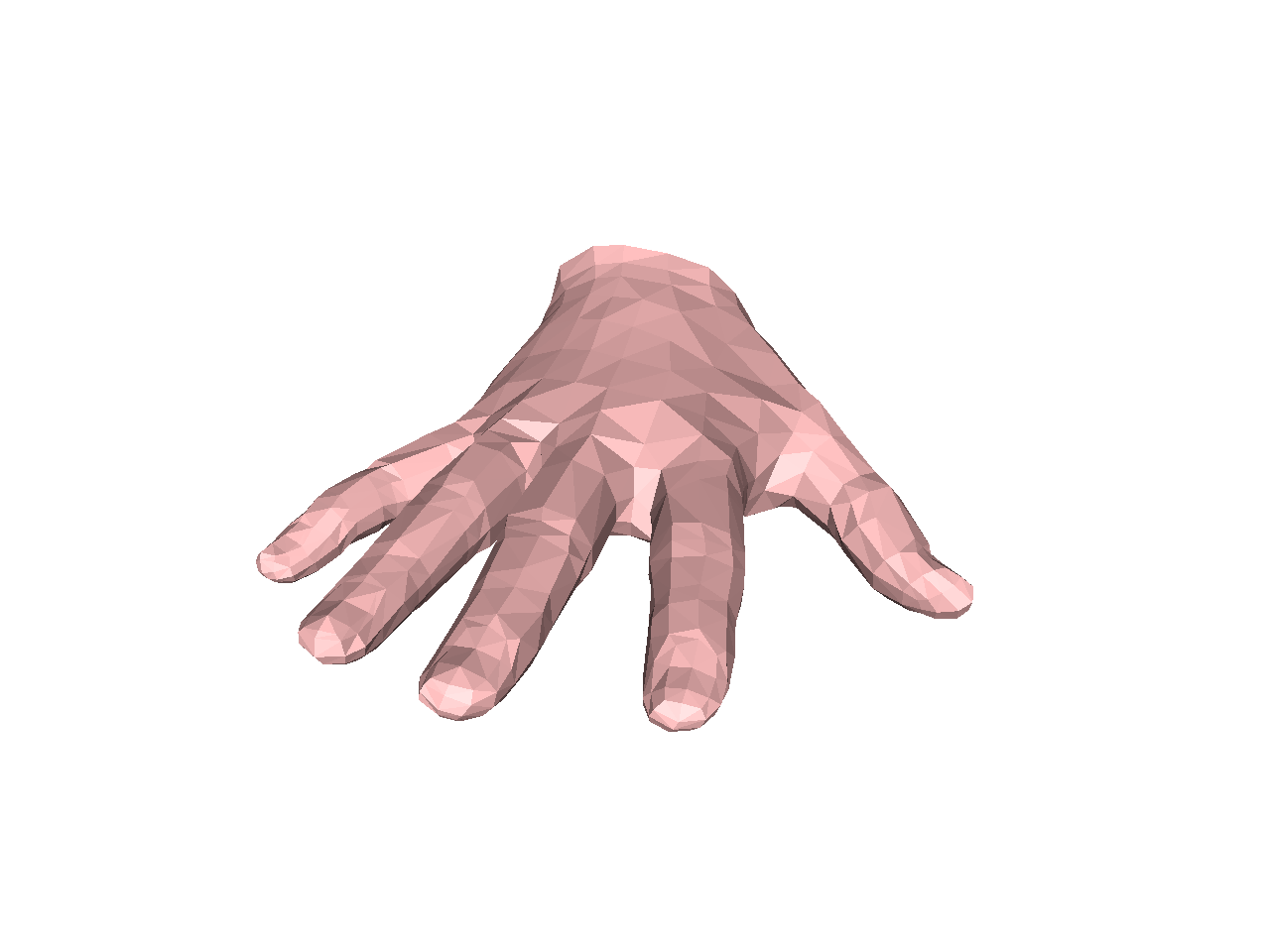}		&		\hspace*{\ImgSquizPcaShapeWHHa}
							\includegraphics[trim=000mm 000mm 000mm 000mm, clip=false, height=\ImgSquizPcaShapeWSSa \textwidth]{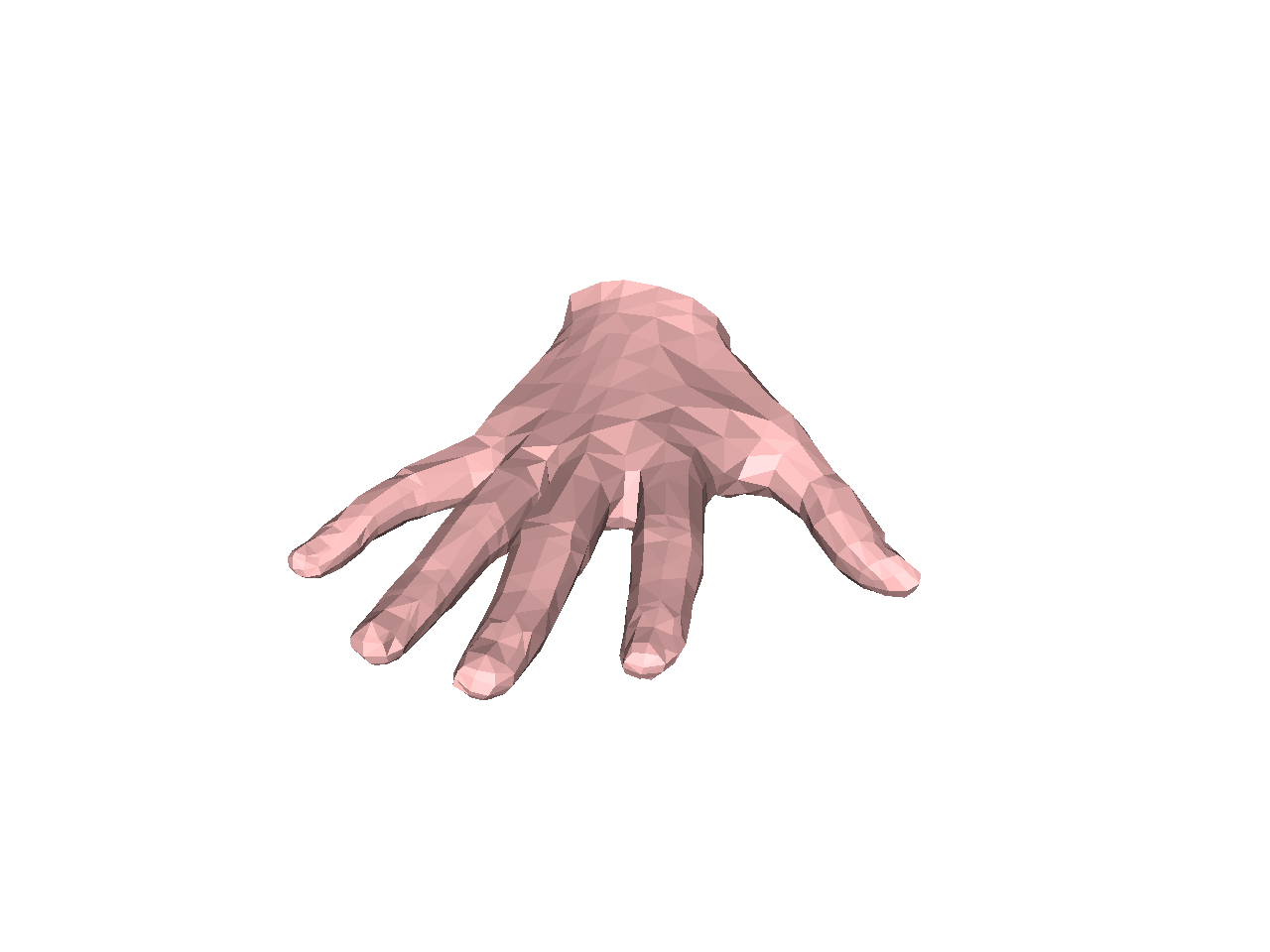}		&		\hspace*{\ImgSquizPcaShapeWHHa}
							\includegraphics[trim=000mm 000mm 000mm 000mm, clip=false, height=\ImgSquizPcaShapeWSSa \textwidth]{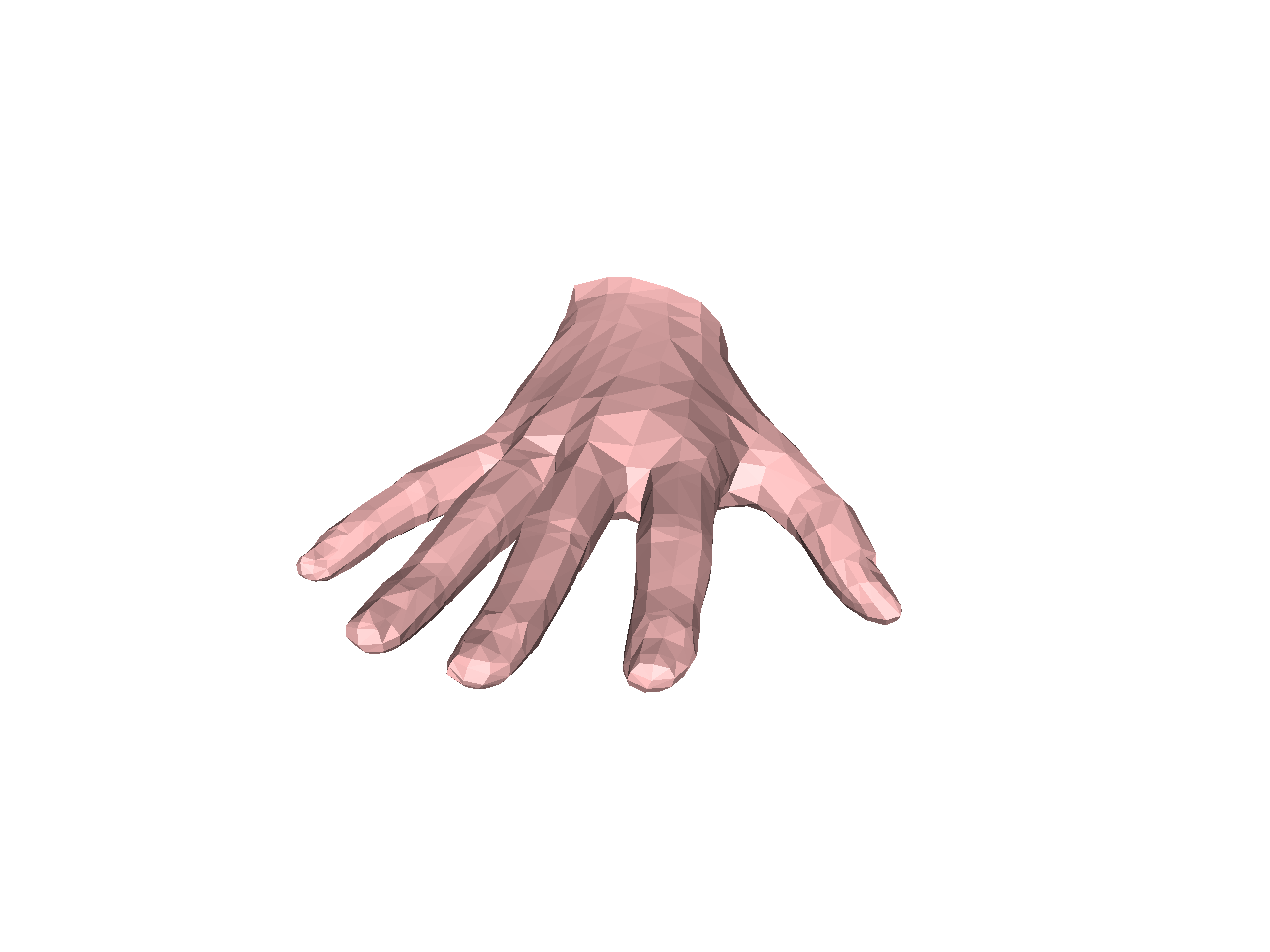}		&		\hspace*{\ImgSquizPcaShapeWHHa}
							\includegraphics[trim=000mm 000mm 000mm 000mm, clip=false, height=\ImgSquizPcaShapeWSSa \textwidth]{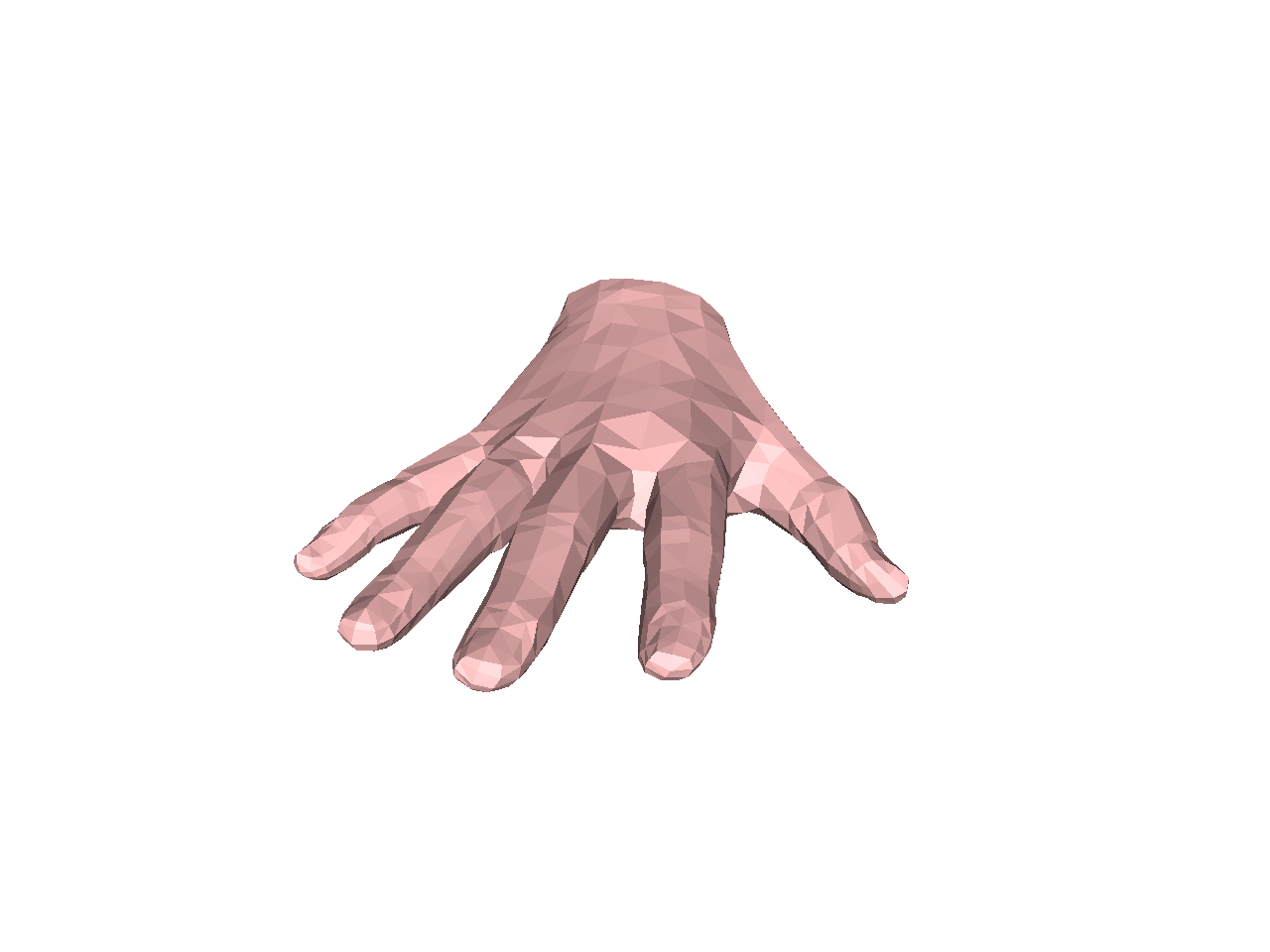}		&		\hspace*{\ImgSquizPcaShapeWHHa}
							\includegraphics[trim=000mm 000mm 000mm 000mm, clip=false, height=\ImgSquizPcaShapeWSSa \textwidth]{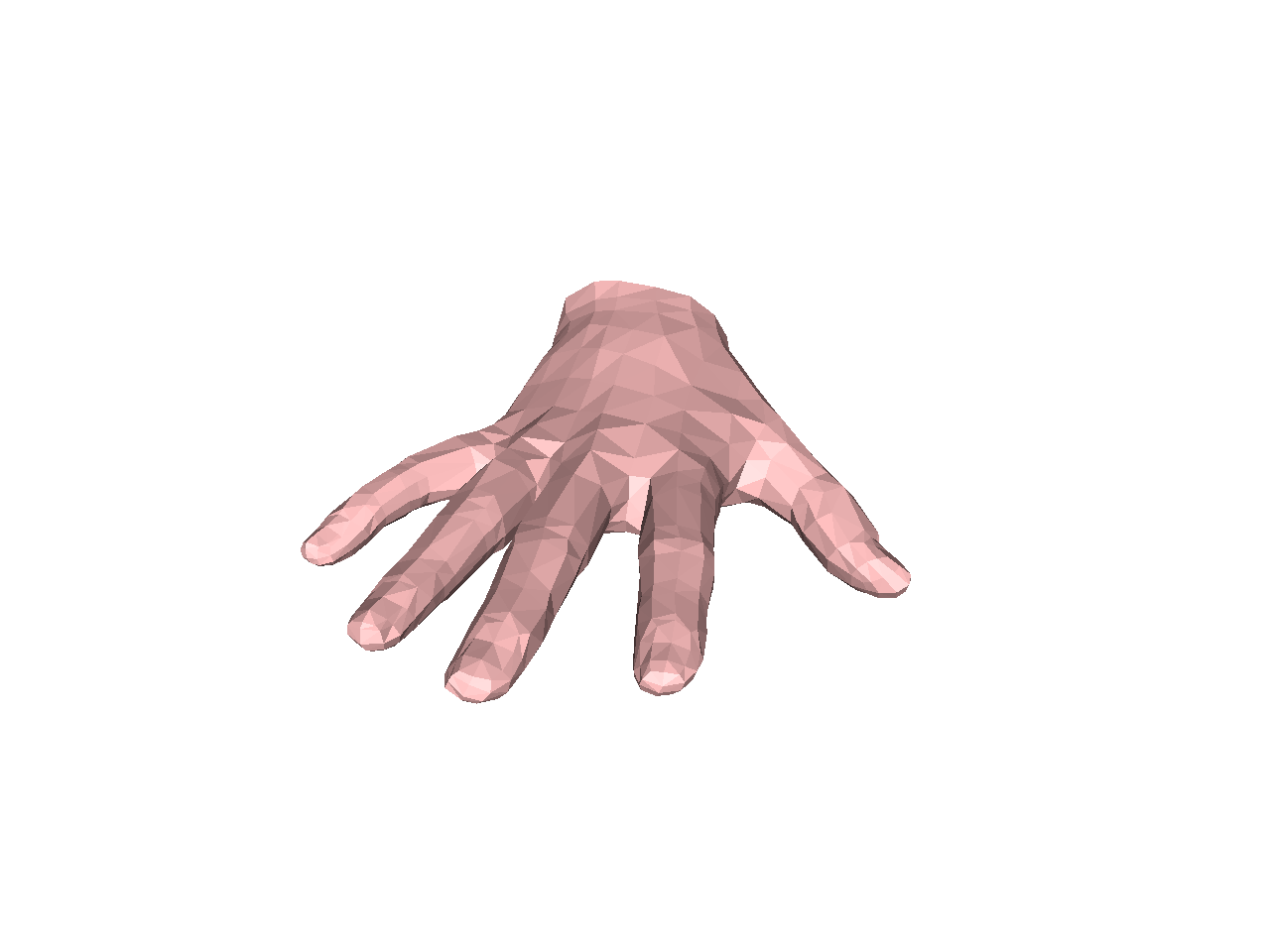}		&		\hspace*{\ImgSquizPcaShapeWHta}
							\multirow{-6}{*}{\hspace{6mm}\parbox{09mm}{mean\\-\stdNoShapeMIN ~std\\}}																														\\				[\ImgSquizPcaShapeWVva]
		\multirow{-4}{*}{
							\includegraphics[trim=090mm 000mm 105mm 000mm, clip=true,  height=\ImgSquizPcaShapeWSSb \textwidth]{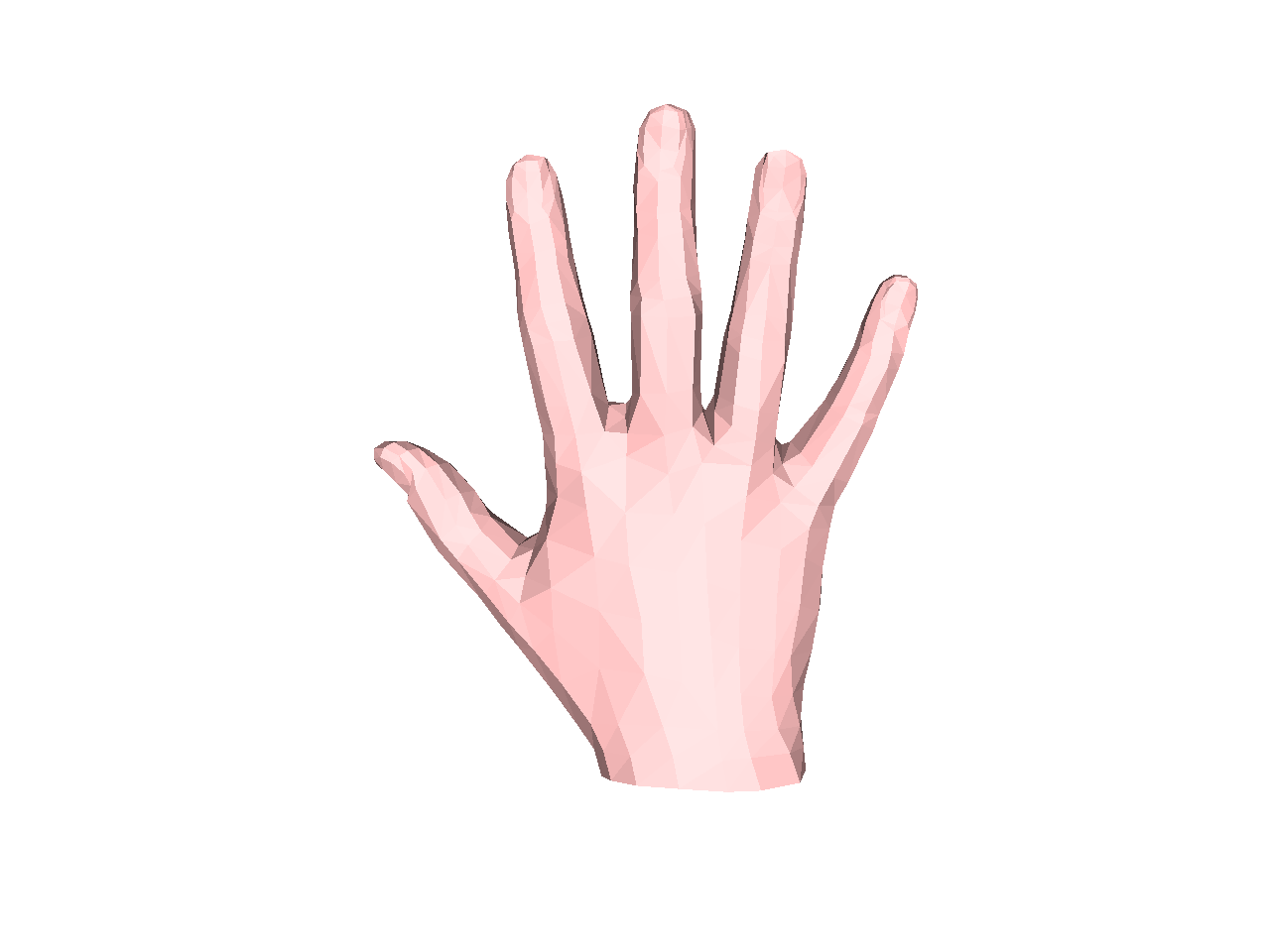}  }																													&		\hspace*{\ImgSquizPcaShapeWHma}
							\includegraphics[trim=000mm 000mm 000mm 000mm, clip=false, height=\ImgSquizPcaShapeWSSb \textwidth]{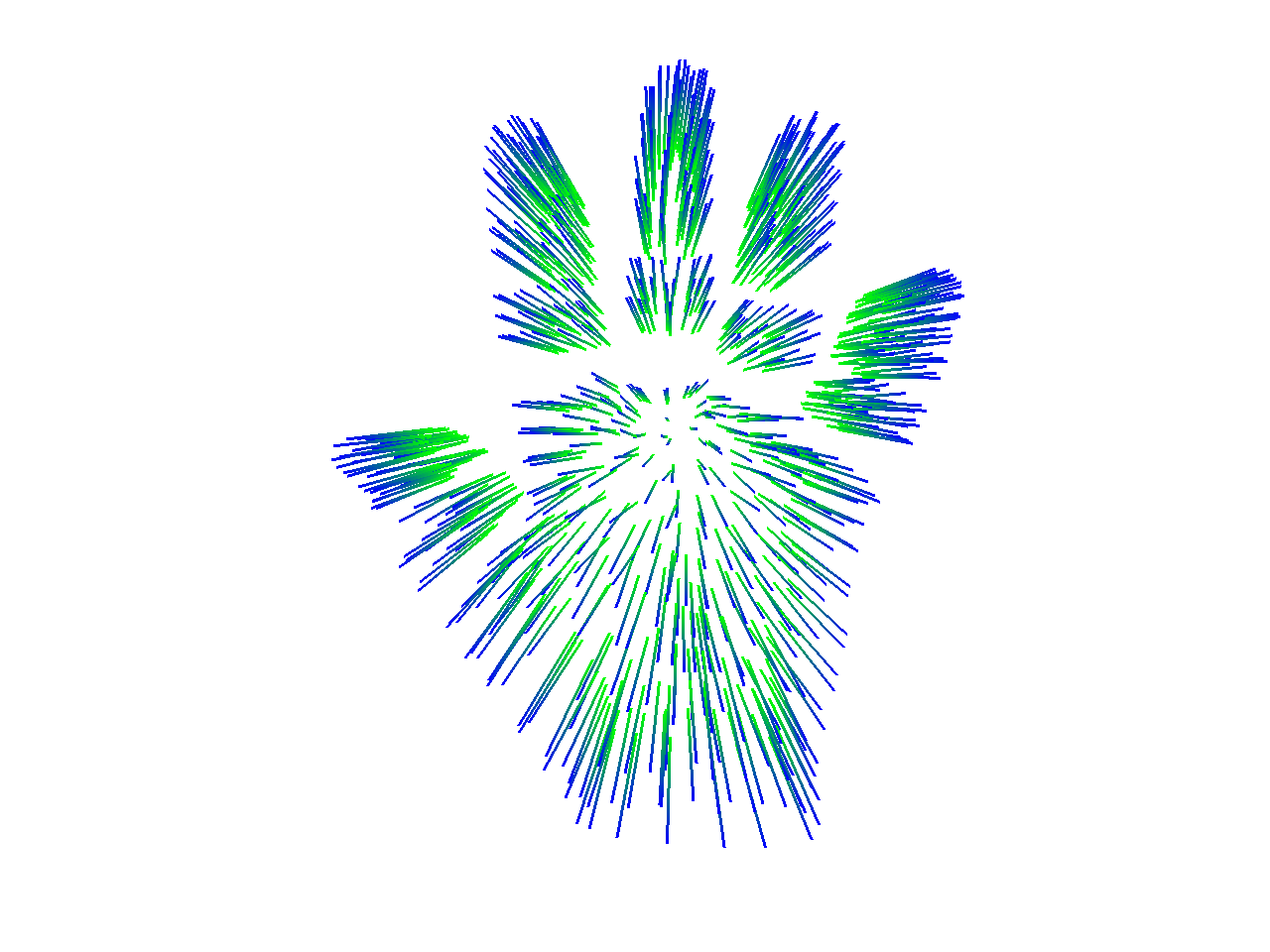}	&		\hspace*{\ImgSquizPcaShapeWHHb}
							\includegraphics[trim=000mm 000mm 000mm 000mm, clip=false, height=\ImgSquizPcaShapeWSSb \textwidth]{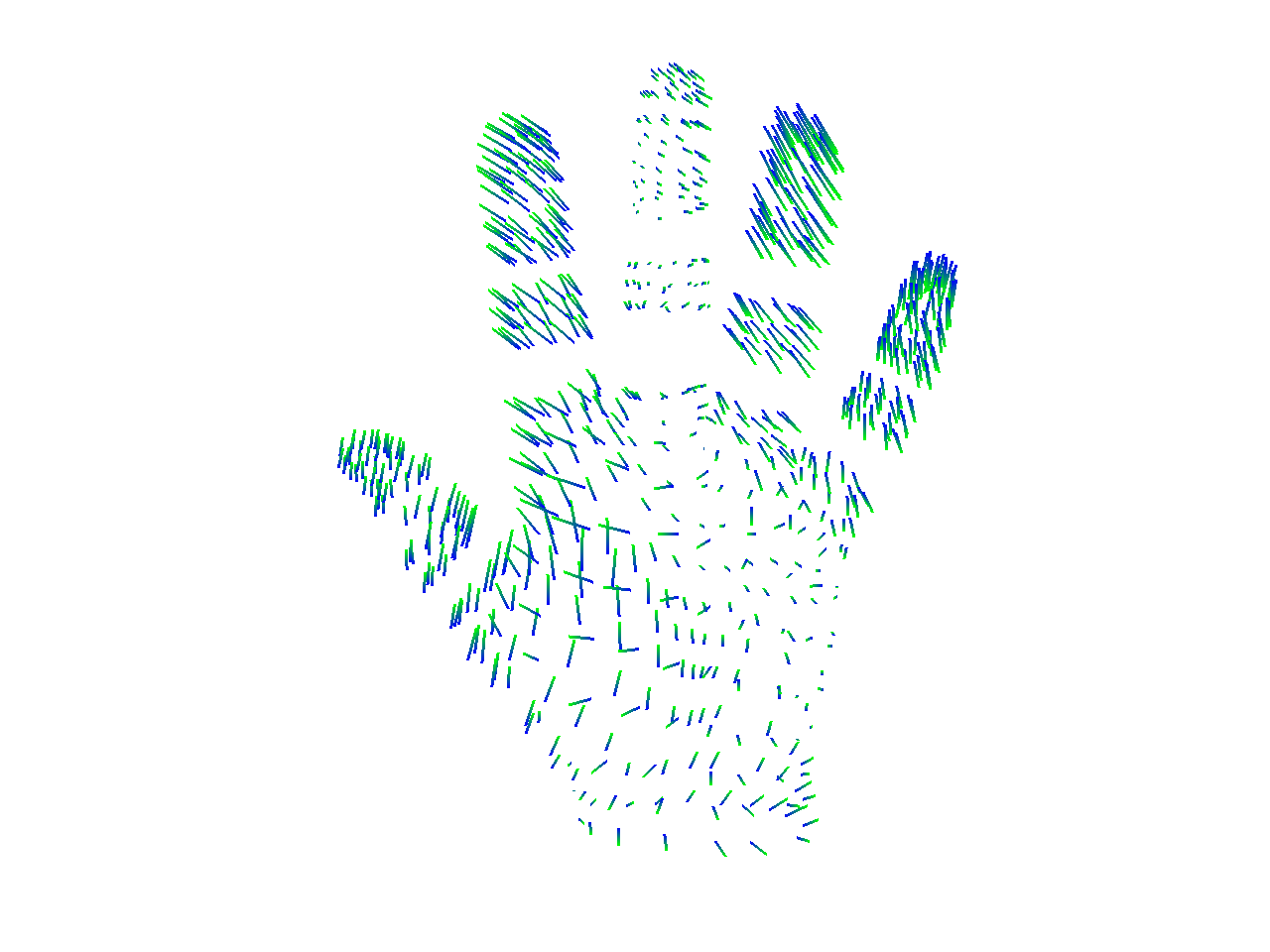}	&		\hspace*{\ImgSquizPcaShapeWHHa}
							\includegraphics[trim=000mm 000mm 000mm 000mm, clip=false, height=\ImgSquizPcaShapeWSSb \textwidth]{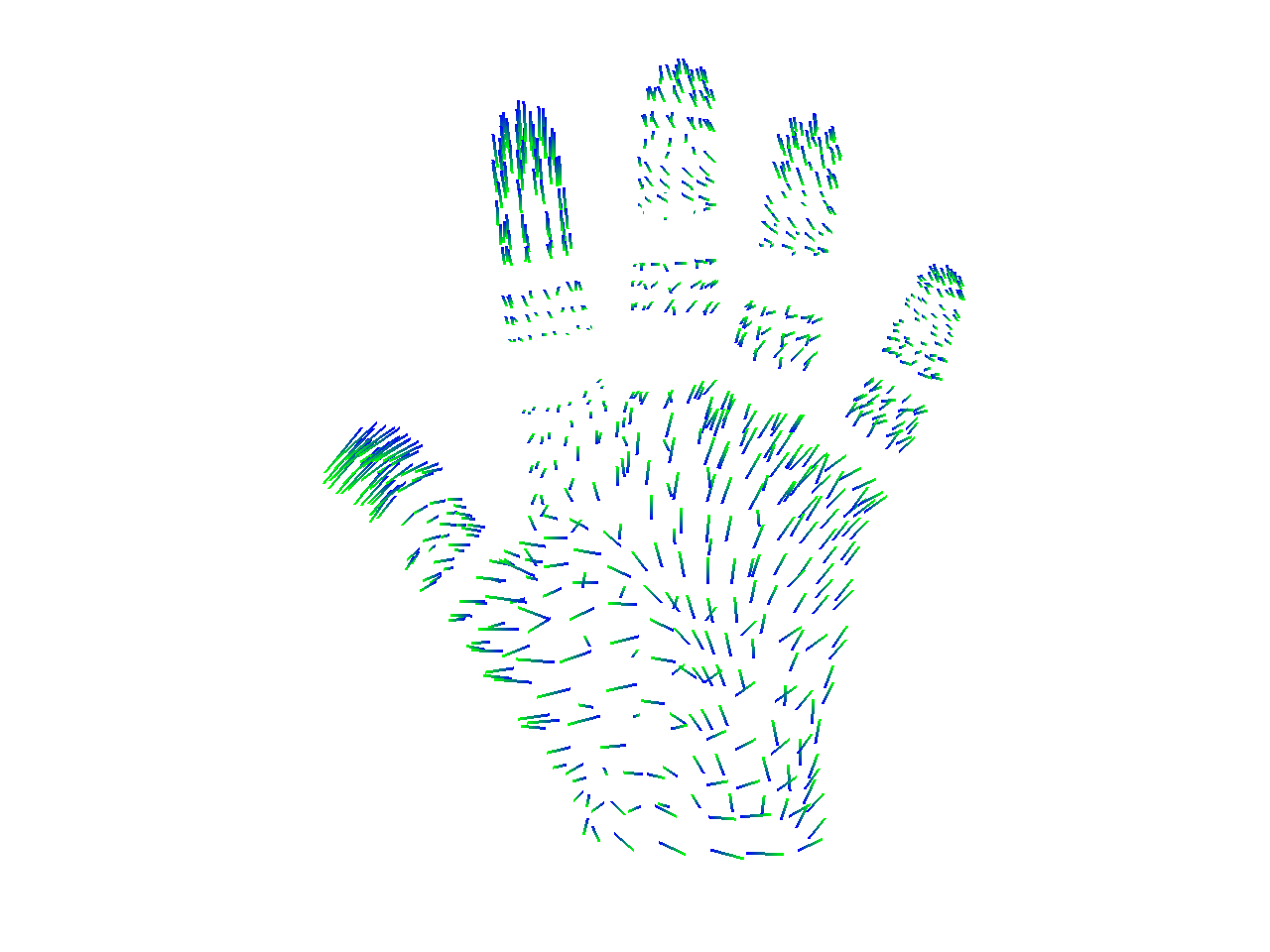}	&		\hspace*{\ImgSquizPcaShapeWHHa}
							\includegraphics[trim=000mm 000mm 000mm 000mm, clip=false, height=\ImgSquizPcaShapeWSSb \textwidth]{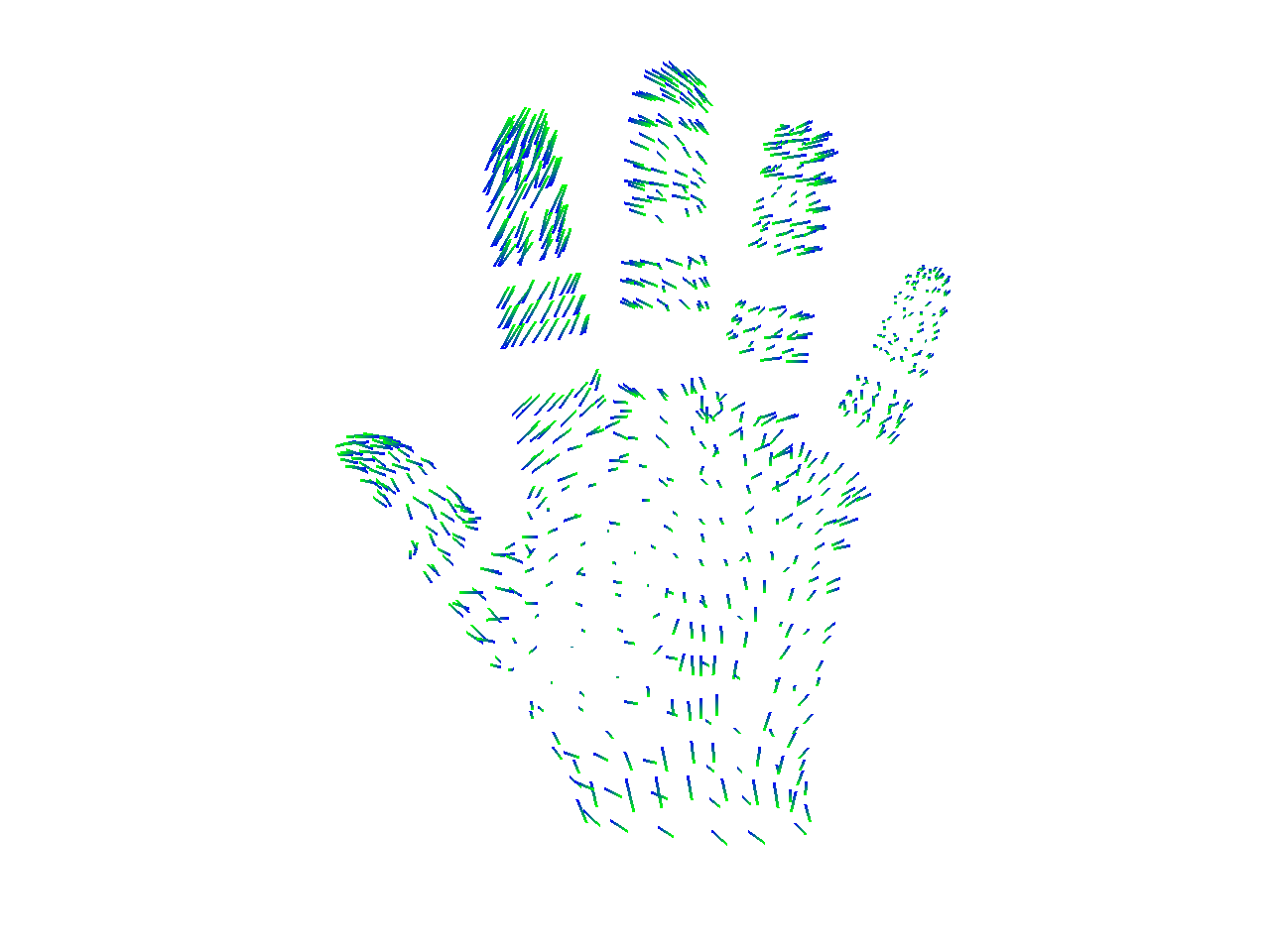}	&		\hspace*{\ImgSquizPcaShapeWHHa}
							\includegraphics[trim=000mm 000mm 000mm 000mm, clip=false, height=\ImgSquizPcaShapeWSSb \textwidth]{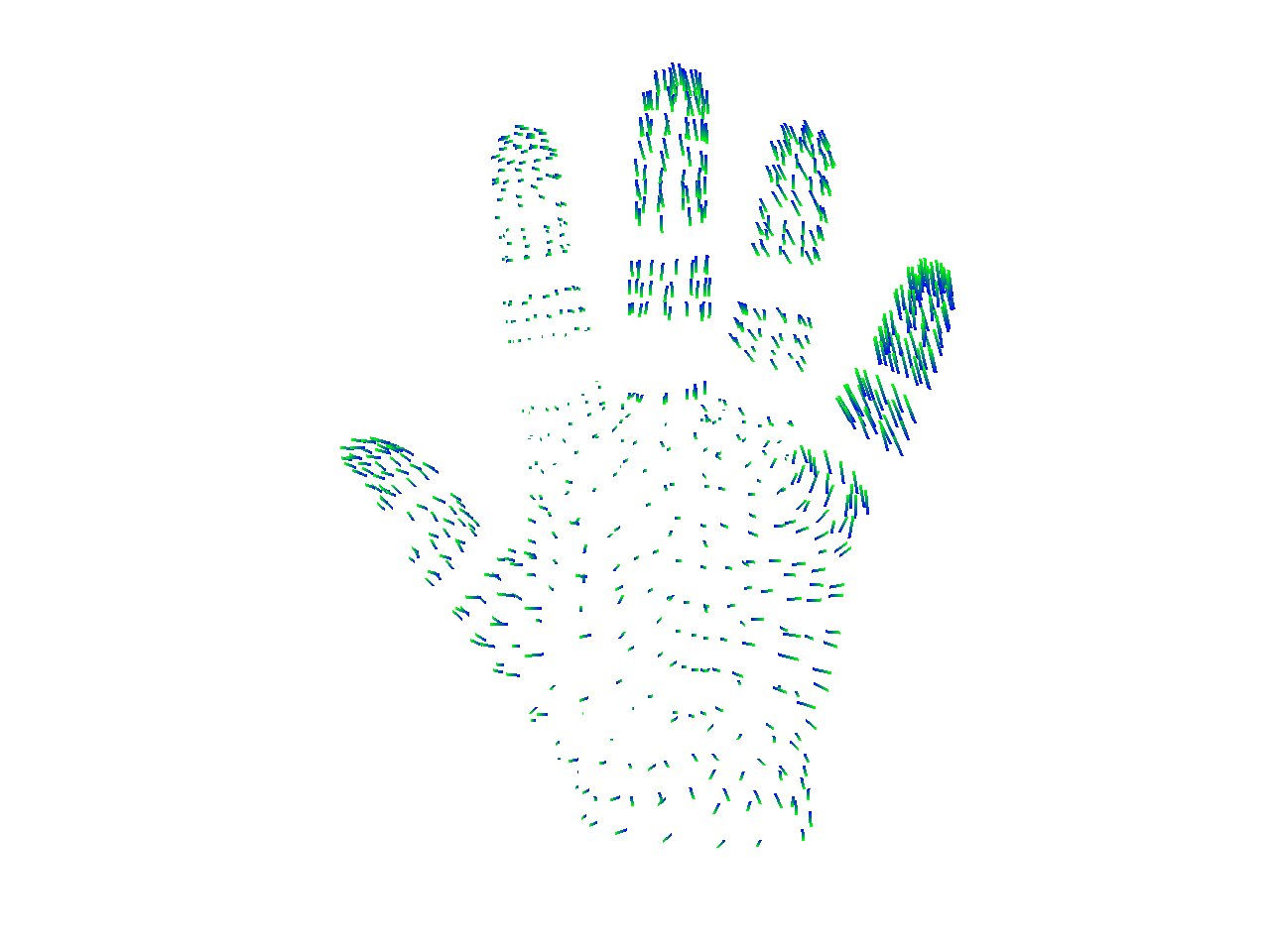}	&		\hspace*{\ImgSquizPcaShapeWHtb}
							\multirow{-5}{*}{\hspace{3mm}\parbox{09mm}{mean\\$\mycpm$ \stdNoShapePLU ~std\\ \\ \\}}																																																					\\				[\ImgSquizPcaShapeWVva]
							{}																																																																										&		\hspace*{\ImgSquizPcaShapeWHma}
							\includegraphics[trim=000mm 000mm 000mm 000mm, clip=false, height=\ImgSquizPcaShapeWSSb \textwidth]{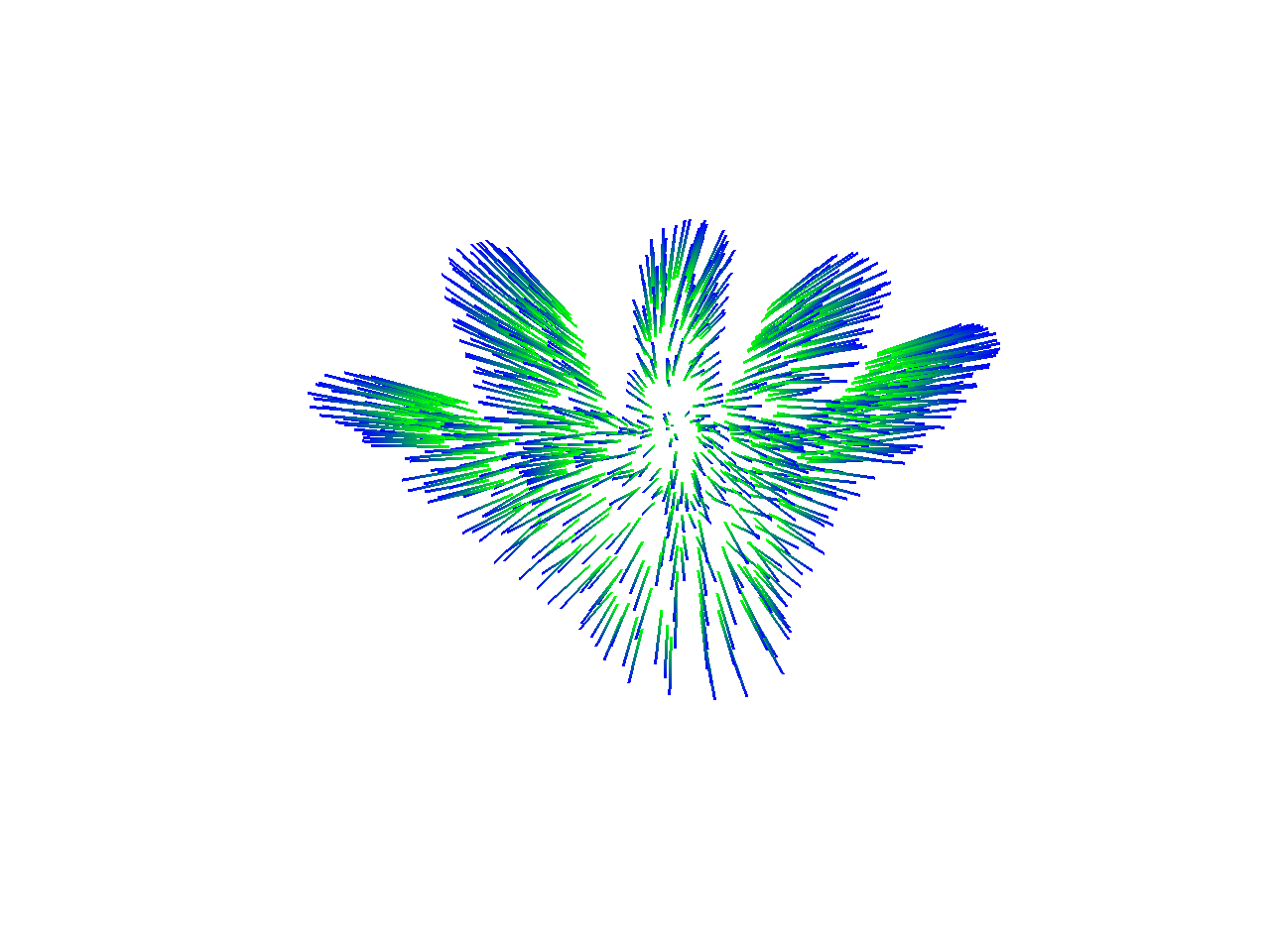}	&		\hspace*{\ImgSquizPcaShapeWHHa}
							\includegraphics[trim=000mm 000mm 000mm 000mm, clip=false, height=\ImgSquizPcaShapeWSSb \textwidth]{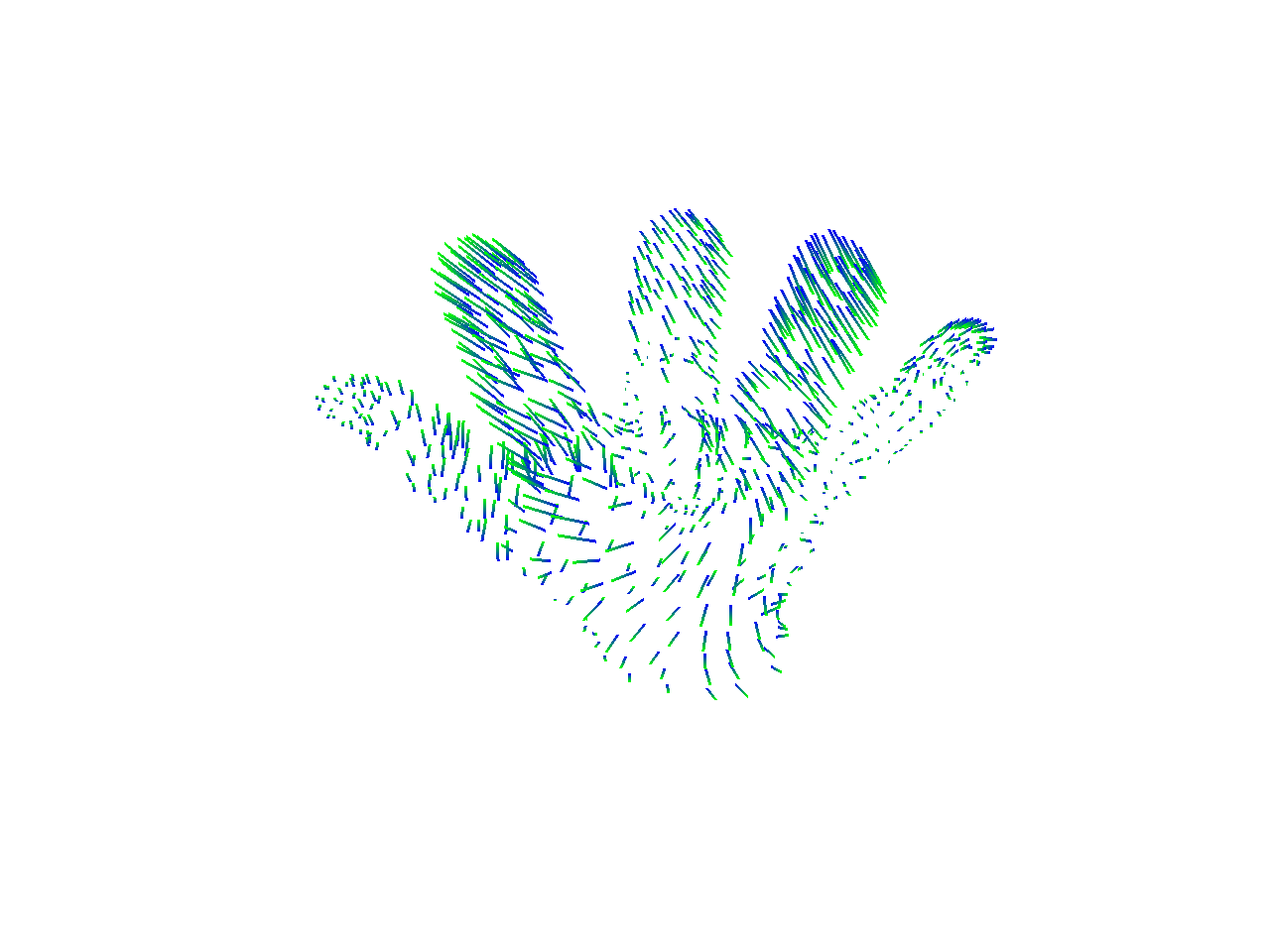}	&		\hspace*{\ImgSquizPcaShapeWHHa}
							\includegraphics[trim=000mm 000mm 000mm 000mm, clip=false, height=\ImgSquizPcaShapeWSSb \textwidth]{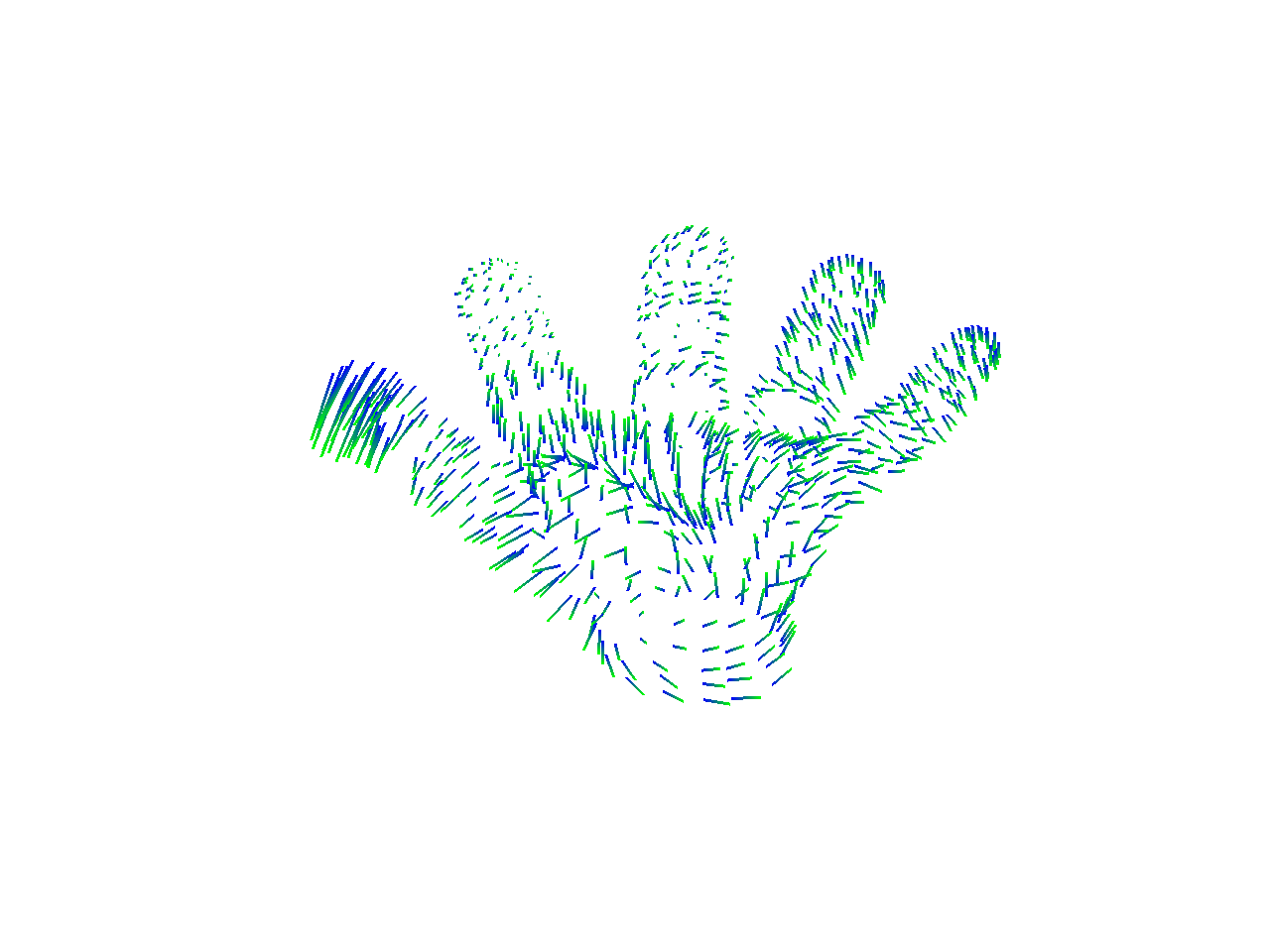}	&		\hspace*{\ImgSquizPcaShapeWHHa}
							\includegraphics[trim=000mm 000mm 000mm 000mm, clip=false, height=\ImgSquizPcaShapeWSSb \textwidth]{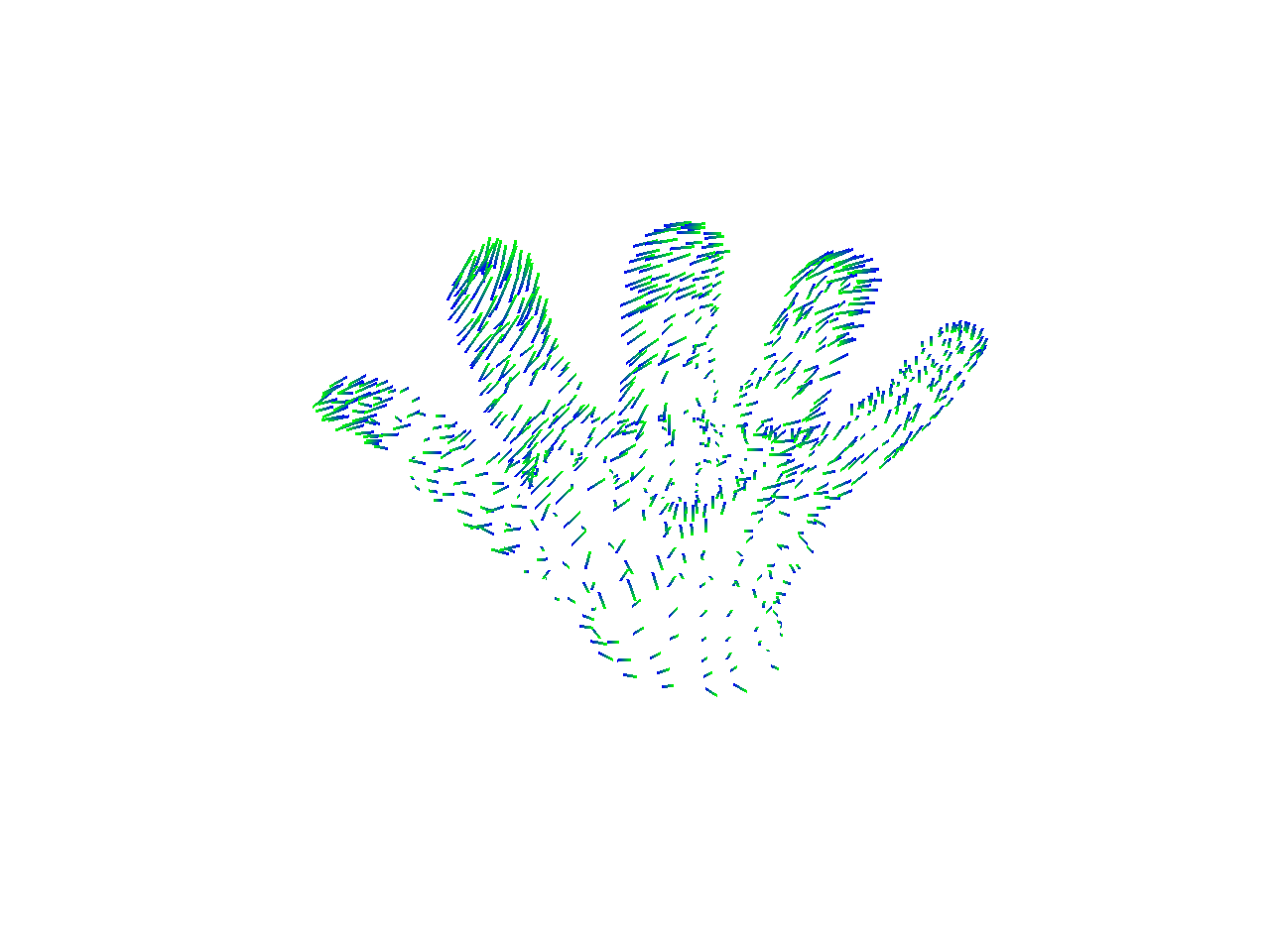}	&		\hspace*{\ImgSquizPcaShapeWHHa}
							\includegraphics[trim=000mm 000mm 000mm 000mm, clip=false, height=\ImgSquizPcaShapeWSSb \textwidth]{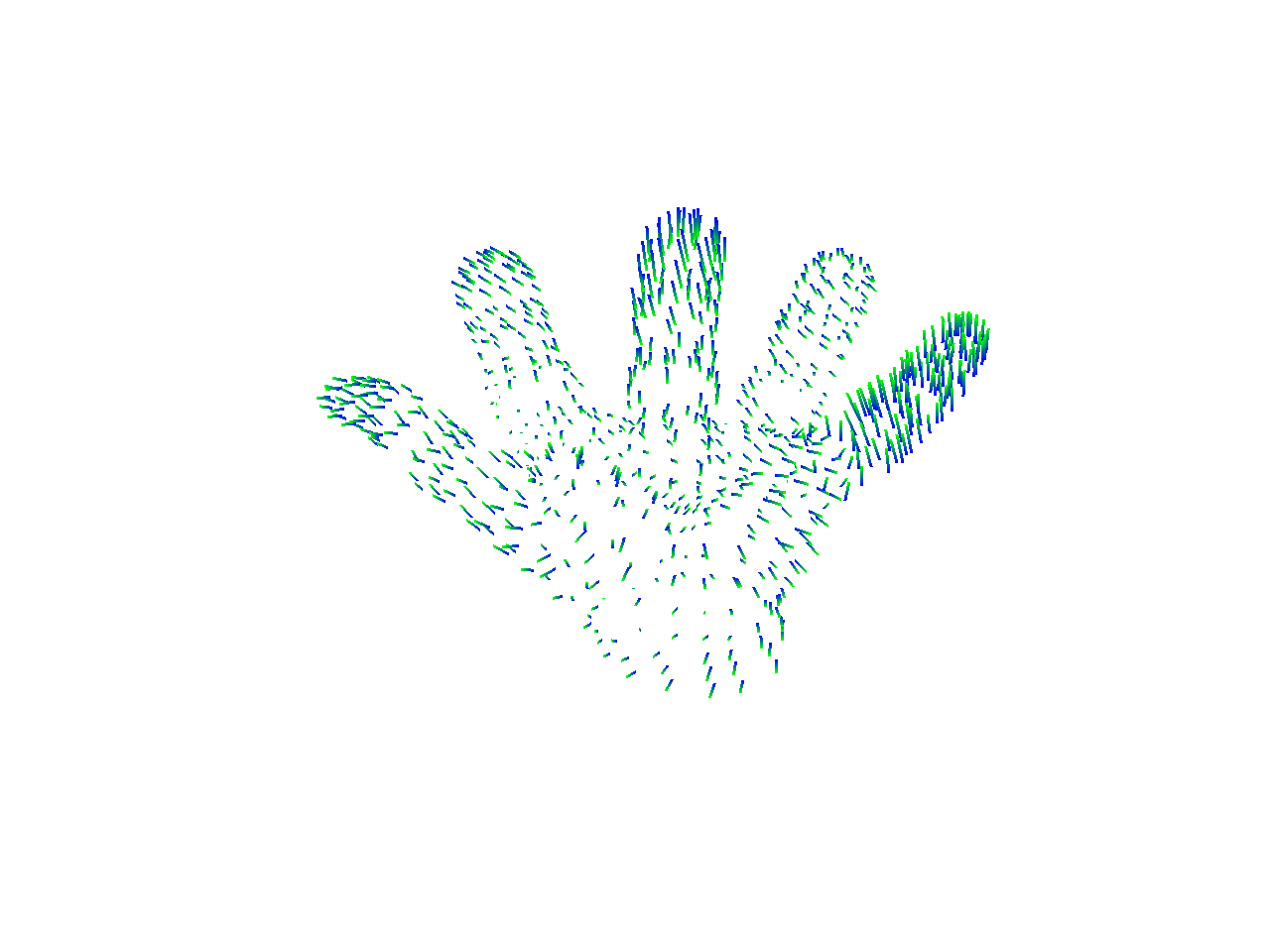}	&		\hspace*{\ImgSquizPcaShapeWHtb}
							\multirow{-5}{*}{\parbox{07mm}{mean\\$\mycpm$ \stdNoShapePLU ~std\\ \\}}																																																									\\				[\ImgSquizPcaShapeWVvb]
							\hspace{6mm}\parbox{10mm}{Mean Shape}																																																																	&		\hspace*{\ImgSquizPcaShapeWHma}
							\hspace{1mm}{PC 1}																																																																						&		\hspace*{\ImgSquizPcaShapeWHHa}
							\hspace{1mm}{PC 2}																																																																						&		\hspace*{\ImgSquizPcaShapeWHHa}
							\hspace{1mm}{PC 3}																																																																						&		\hspace*{\ImgSquizPcaShapeWHHa}
							\hspace{1mm}{PC 4}																																																																						&		\hspace*{\ImgSquizPcaShapeWHHa}
							\hspace{1mm}{PC 5}																																																																						&		\hspace*{\ImgSquizPcaShapeWHta}
							{}
	\end{tabular}
	\caption{
				PCA shape space. 
				Each column depicts the effect of one of the first five \emph{principal components} (PCs) of the learned hand shape space.
				The effect of each PC is shown by adding $\mycpm \stdNoShapePLU$ standard deviations (std) to the mean shape (left-most image), as indicated.
				See the Supplemental Video.
	}
	\label{fig:pcaPCsShape}
\end{figure}

%% file: tex_FIG/FIG_handpose_protocol.tex
\begin{figure*}[t]
	\includegraphics[width=1.0\linewidth]{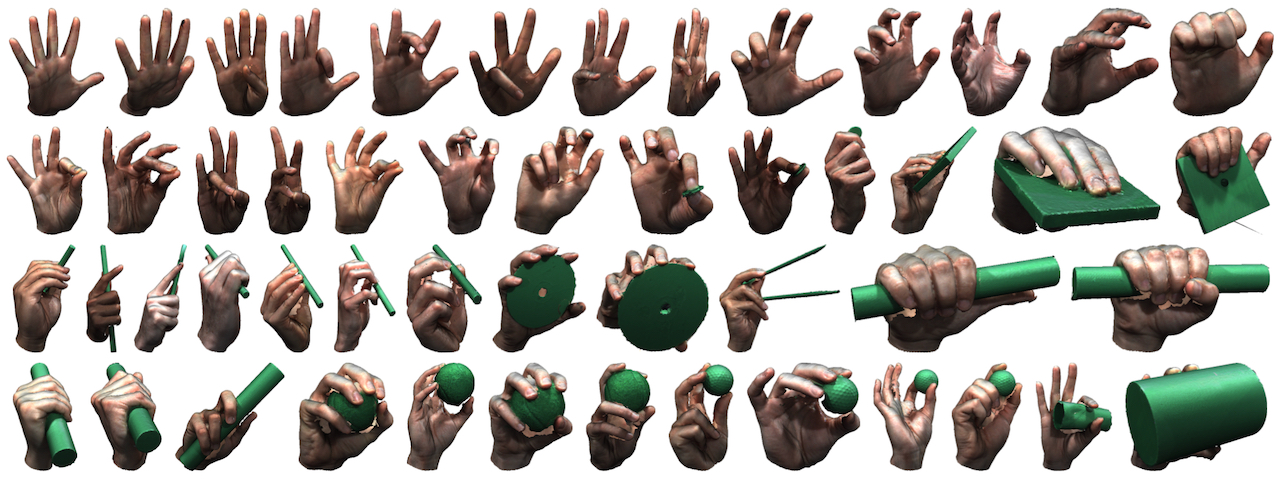}
	\caption{
		Hand capture protocol.  
		Each of the 31 subjects performed most of the 51 hand poses shown here. 
		Images here are shown from multiple subjects.
	}
	\label{fig:hand_poses}
\end{figure*}

%% file: tex_FIG/FIG_example_scans_registrations.tex
\begin{figure*}[t]
  \includegraphics[width=1.0\linewidth]{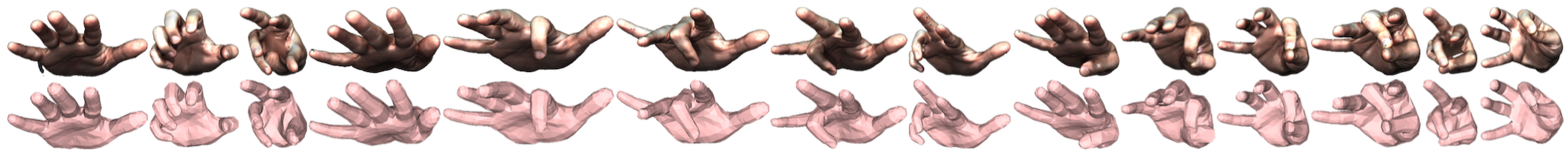}
   \caption{
   				Examples of the captured poses for a single subject.
   				The \emph{top} row shows some of the captured \emph{scans}, while
   				the \emph{bottom} row shows the corresponding mesh \emph{alignments} with \emph{pink} color.
   }
\label{fig:crazyPoses}
\end{figure*}

\begin{figure*}[t]
   \includegraphics[width=1.0\linewidth]{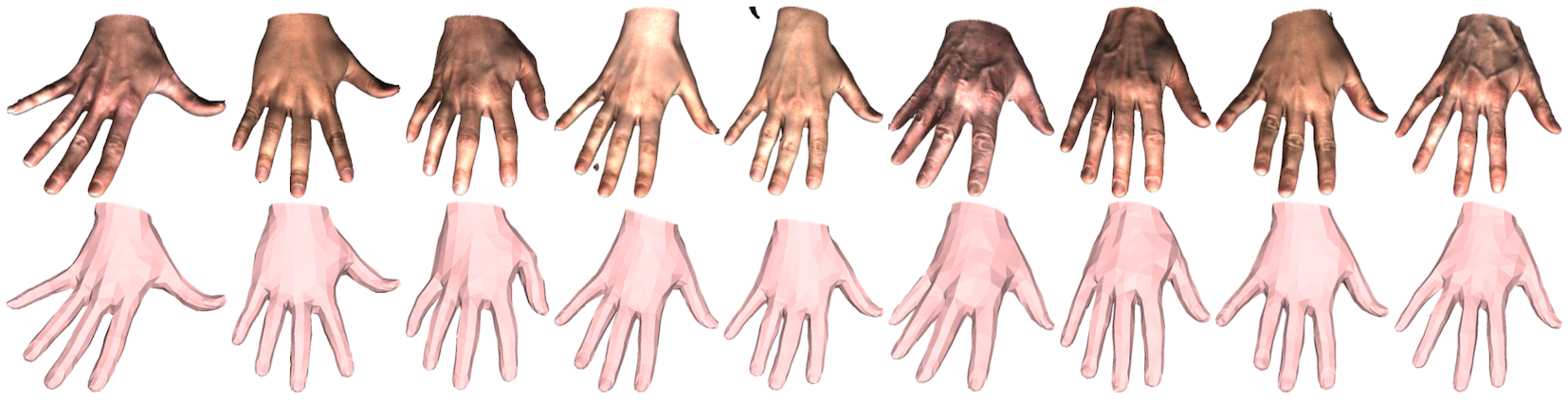}
   \caption{
   			Examples of the captured \emph{``flat-hand''} pose for various subjects.
   			The \emph{top} row shows some of the captured \emph{scans}, while the \emph{bottom} row shows the corresponding mesh \emph{alignments} with \emph{pink} color.
   }
\label{fig:neutralPoseManySubjects}
\end{figure*}

%% file: tex_FIG/FIG_poseblendshapes_onoff___SIGGASIA.tex
\newcommand{\hhh}{23mm}
\newcommand{\hsp}{\hspace*{-4.7mm}}

\begin{figure*}[t]
	\includegraphics[width=1.0\linewidth]{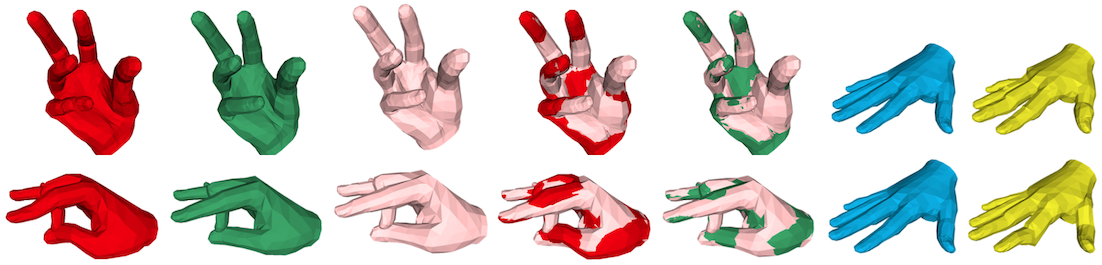}
    \caption{
		{\bf The effect of pose blendshapes}.
		From left to right, 
		model without pose-dependent blendshapes in red, 
		model with pose-dependent blendshapes in green, 
		registration in pink, 
		two 	overlays, 
		personalized template in cyan, 
		personalized template plus pose-dependent blendshapes in yellow. 
		Pose blend shapes prepare the hand template for the subsequent articulation, to correct for artifacts of the latter. 
		Therefore the deformations of the template in yellow correct the artifacts in red	(best viewed on screen) and result in a good model after posing, shown with green color. 
		Each row shows two different poses; 
		the first one shows a large correction on the bending of the pinky finger, while 
		the second one shows the correction of the articulation in the middle finger.
    }
    \label{fig:poseBlendShapesOnOff}
\end{figure*}

%% file: tex_FIG/FIG_pose_pca___SIGGASIA.tex
\newcommand{\stdNoPosePLU}{2}
\newcommand{\stdNoPoseMIN}{\stdNoPosePLU}

\newcommand{\ImgSquizPcaPoseWHma}{-12.0mm}
\newcommand{\ImgSquizPcaPoseWHta}{-10.0mm}
\newcommand{\ImgSquizPcaPoseWHHa}{-16.0mm}
\newcommand{\ImgSquizPcaPoseWVva}{-04.0mm}
\newcommand{\ImgSquizPcaPoseWVVa}{-09.0mm}
\newcommand{\ImgSquizPcaPoseWSS}{0.110}

\begin{figure*}
	\footnotesize
	\begin{tabular}{c c c c c c c c c c c c}
		\multirow{-4}{*}{	\includegraphics[trim=090mm 000mm 105mm 000mm, clip=true,  height=\ImgSquizPcaPoseWSS \textwidth]{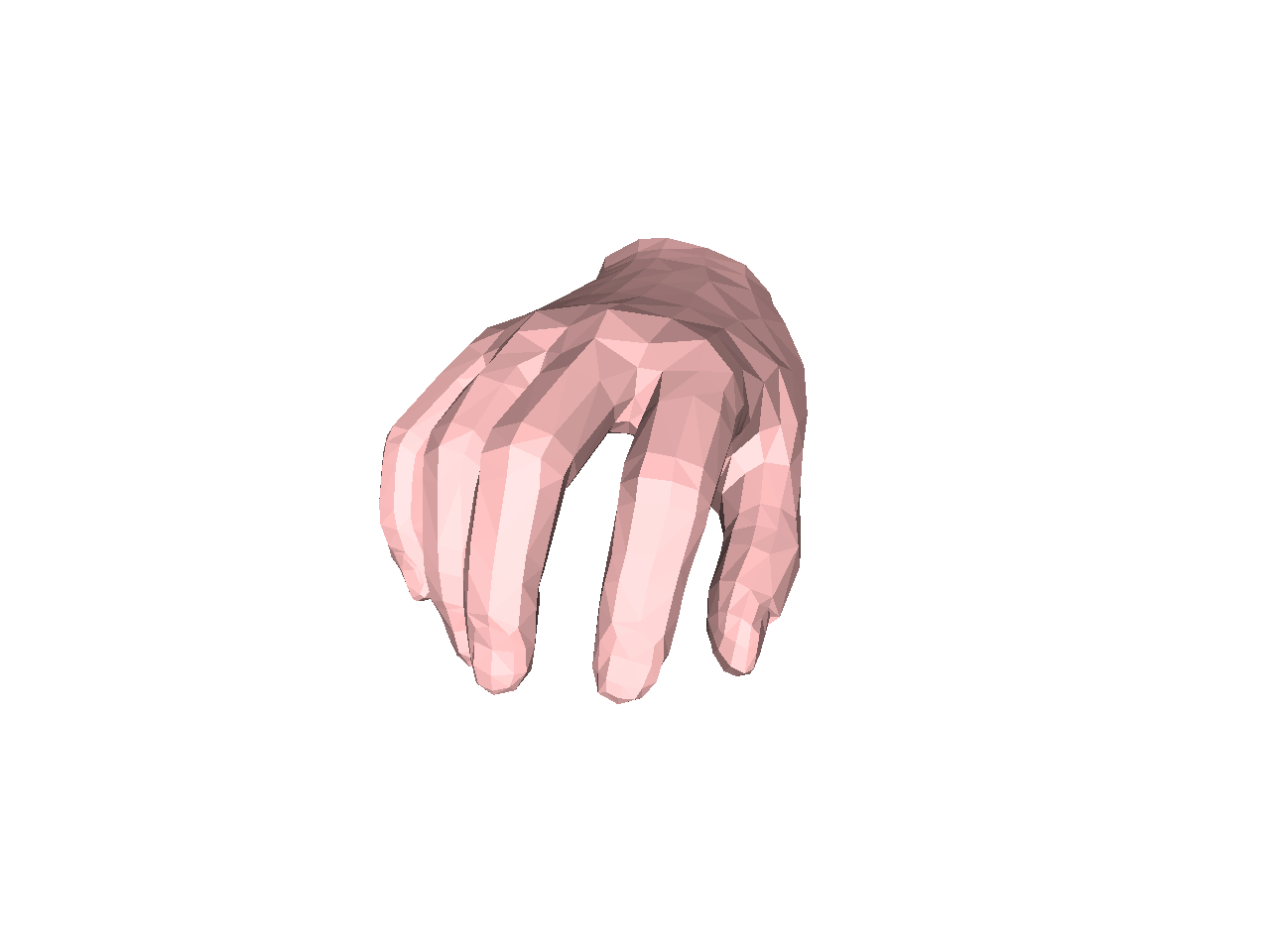}	}	&		\hspace*{\ImgSquizPcaPoseWHma}
							\includegraphics[trim=000mm 000mm 000mm 000mm, clip=false, height=\ImgSquizPcaPoseWSS \textwidth]{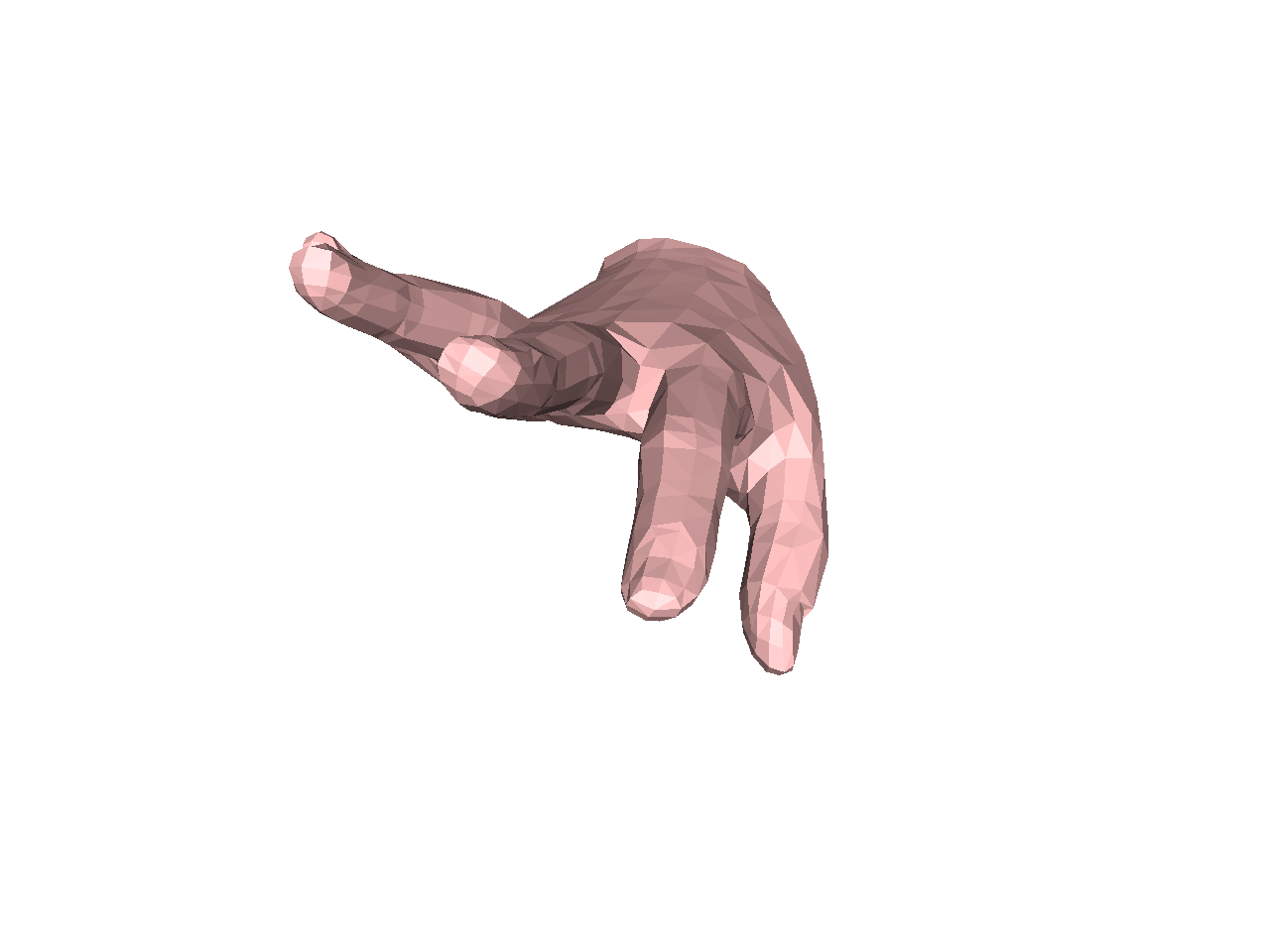}		&		\hspace*{\ImgSquizPcaPoseWHHa}
							\includegraphics[trim=000mm 000mm 000mm 000mm, clip=false, height=\ImgSquizPcaPoseWSS \textwidth]{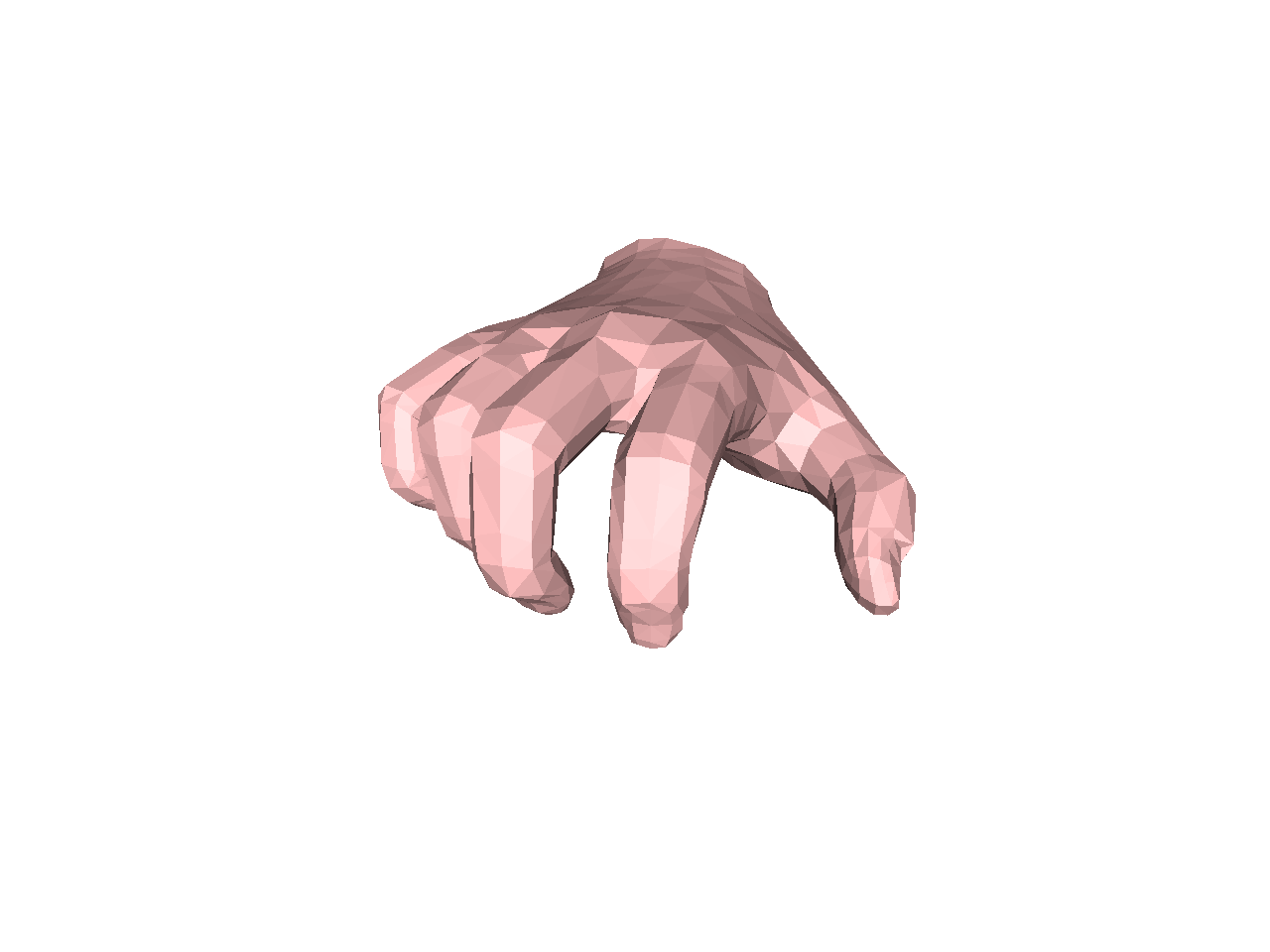}		&		\hspace*{\ImgSquizPcaPoseWHHa}
							\includegraphics[trim=000mm 000mm 000mm 000mm, clip=false, height=\ImgSquizPcaPoseWSS \textwidth]{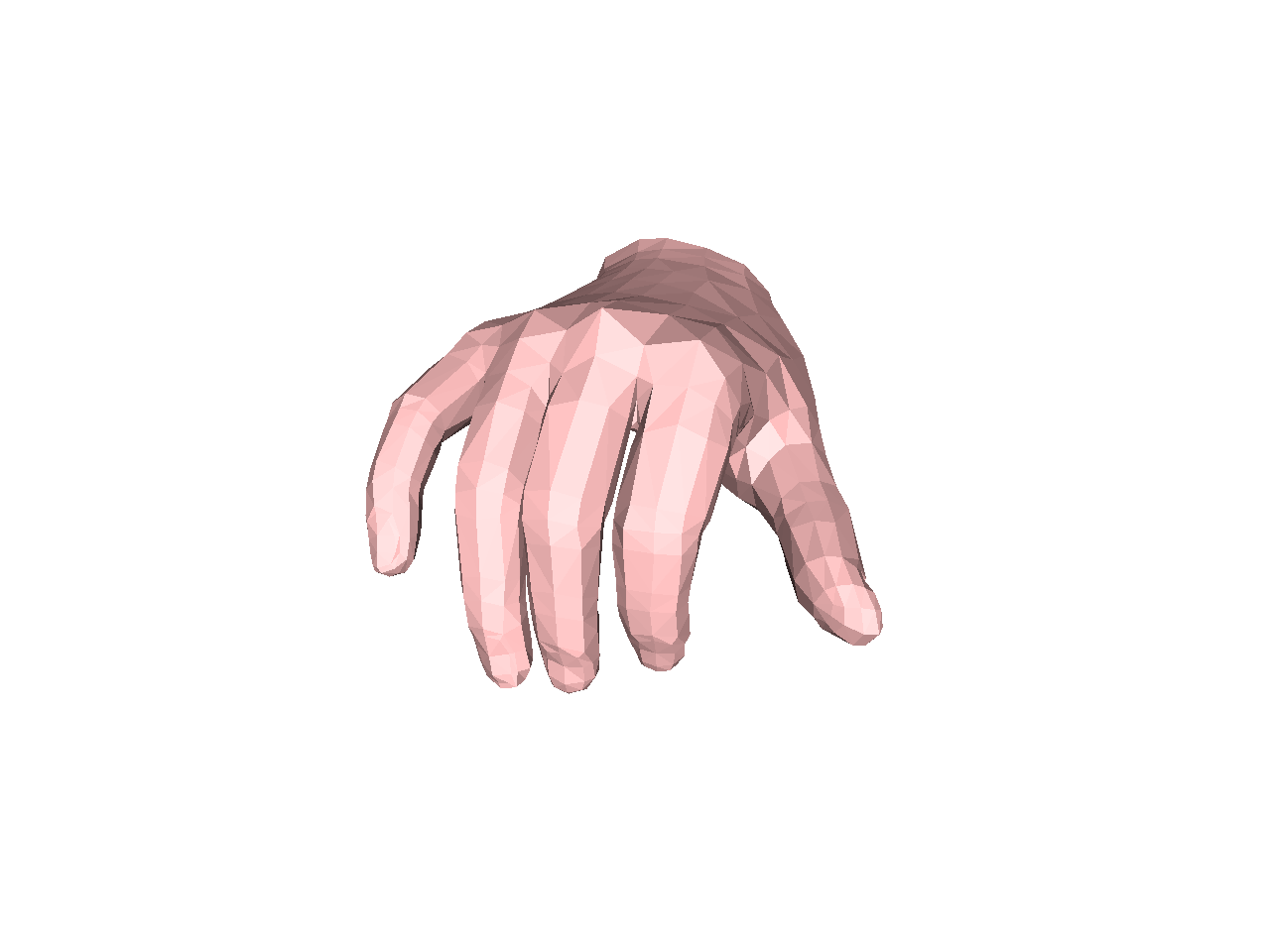}		&		\hspace*{\ImgSquizPcaPoseWHHa}
							\includegraphics[trim=000mm 000mm 000mm 000mm, clip=false, height=\ImgSquizPcaPoseWSS \textwidth]{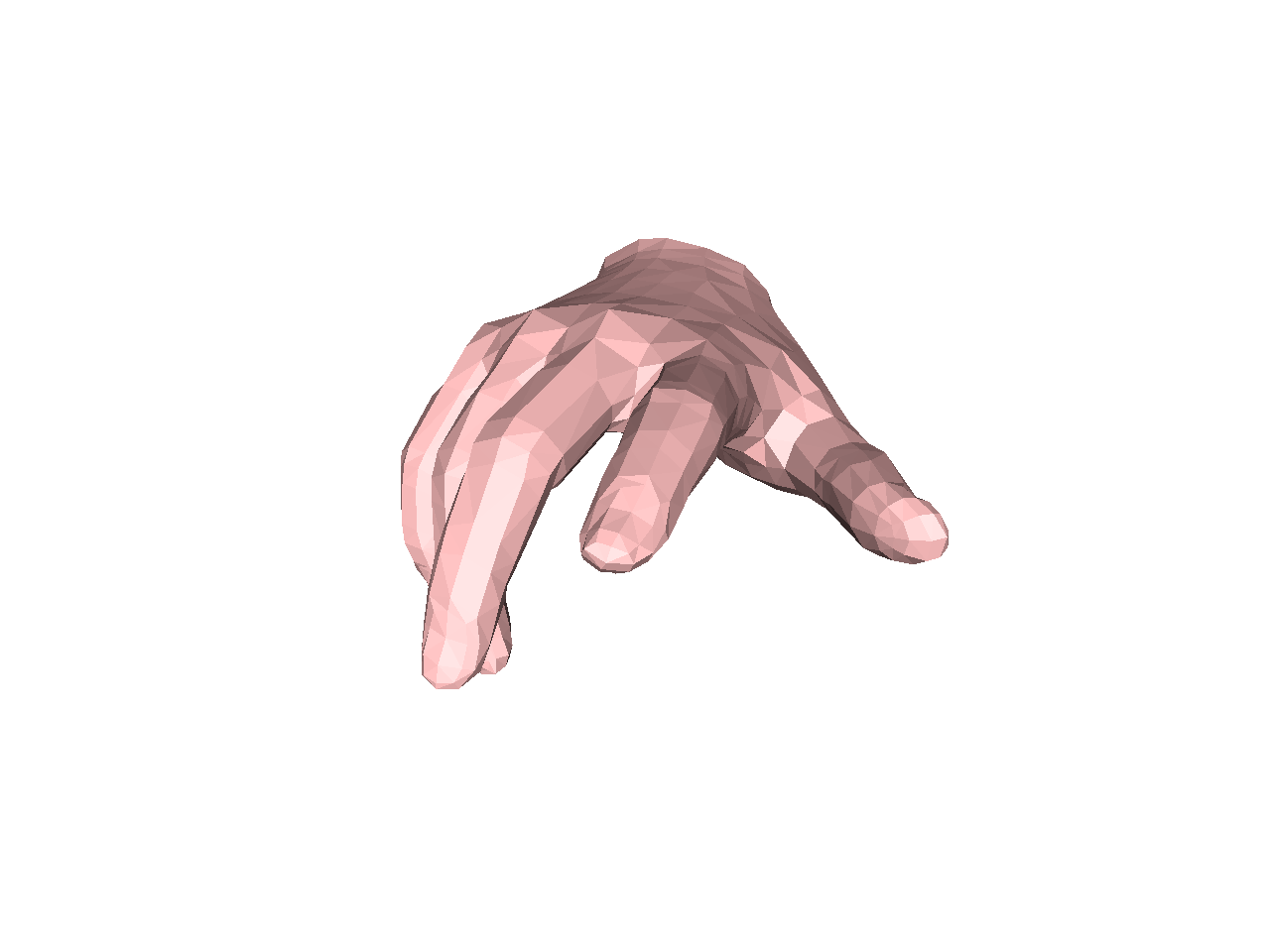}		&		\hspace*{\ImgSquizPcaPoseWHHa}
							\includegraphics[trim=000mm 000mm 000mm 000mm, clip=false, height=\ImgSquizPcaPoseWSS \textwidth]{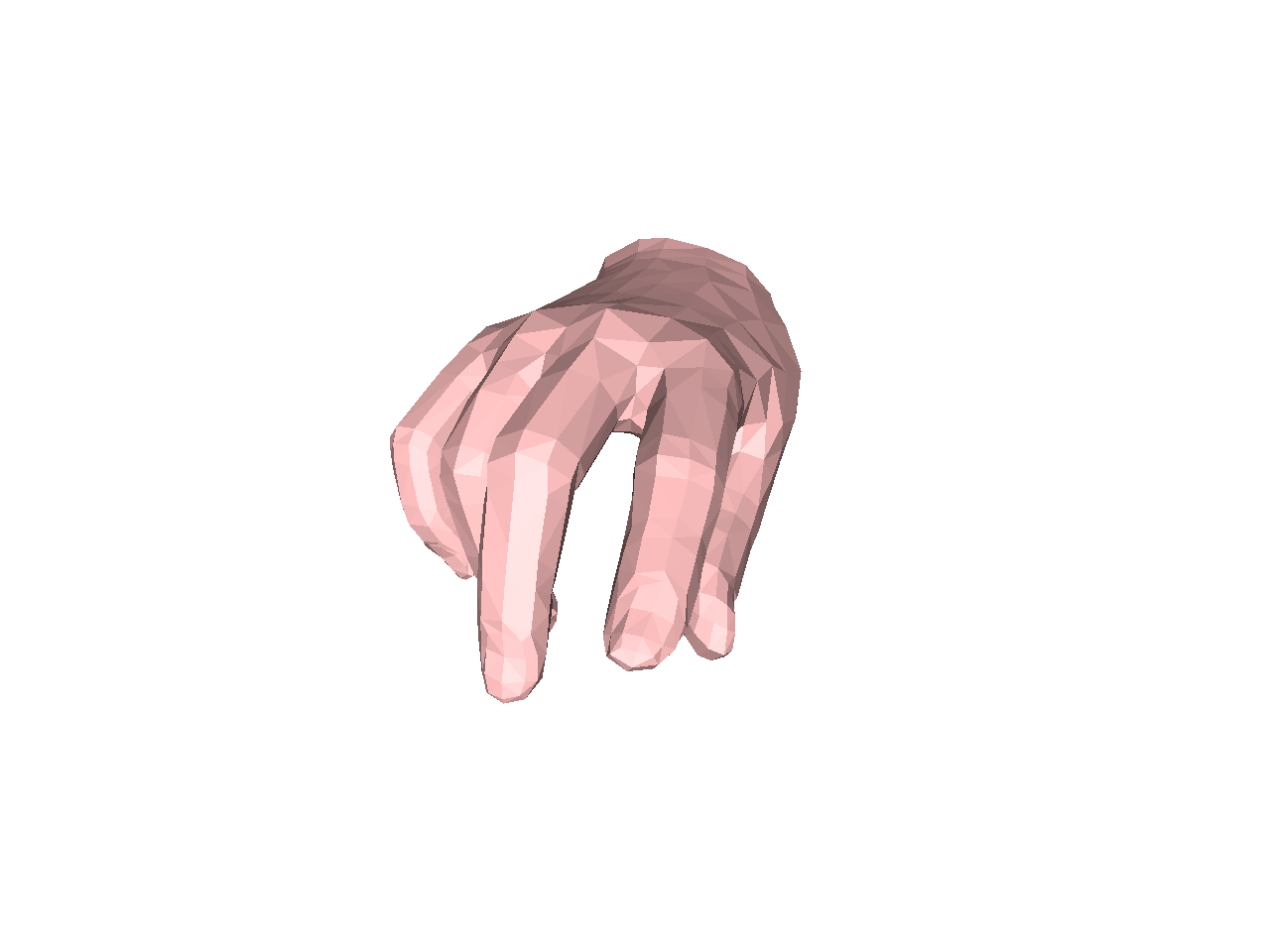}		&		\hspace*{\ImgSquizPcaPoseWHHa}
							\includegraphics[trim=000mm 000mm 000mm 000mm, clip=false, height=\ImgSquizPcaPoseWSS \textwidth]{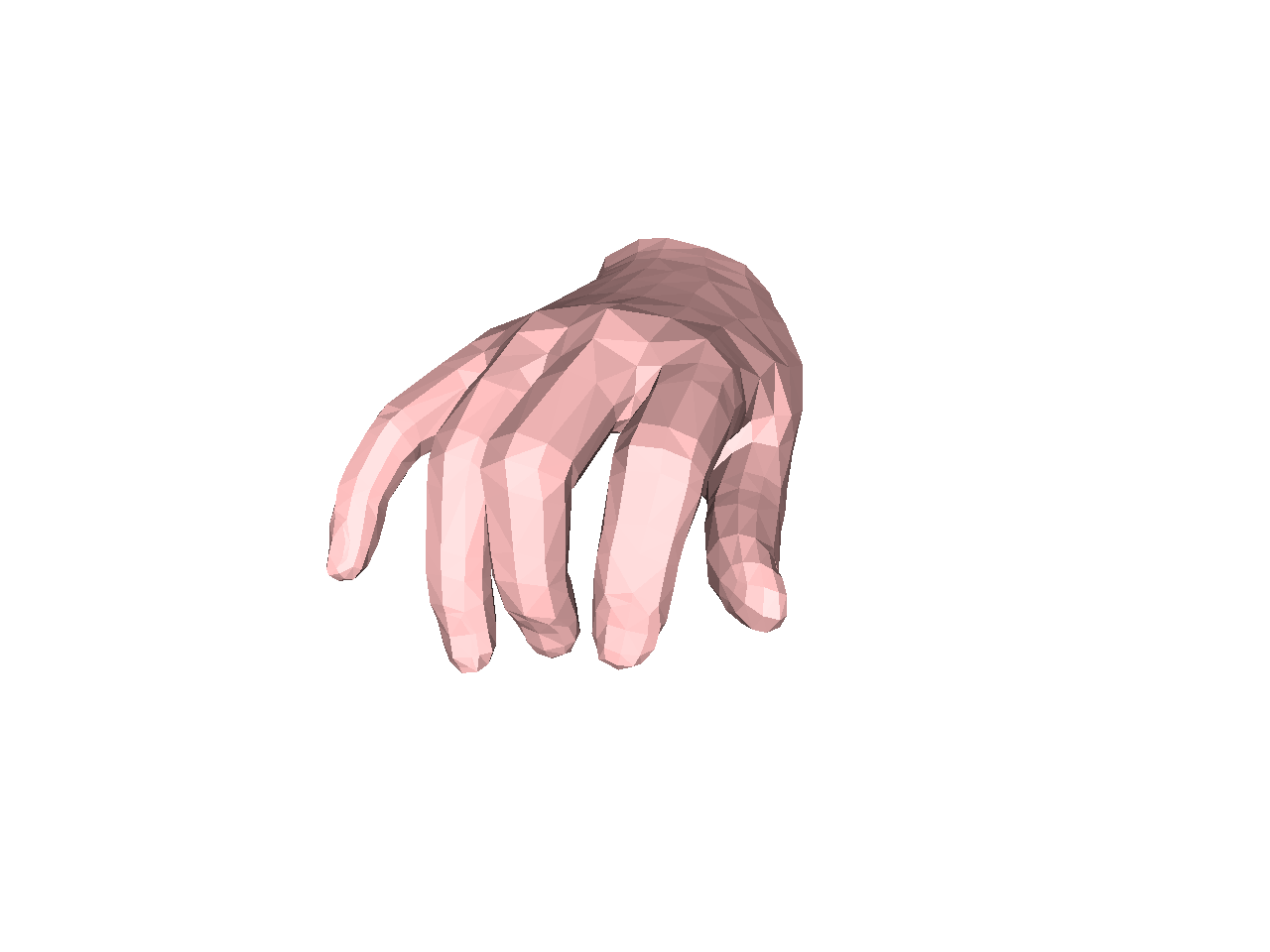}		&		\hspace*{\ImgSquizPcaPoseWHHa}
							\includegraphics[trim=000mm 000mm 000mm 000mm, clip=false, height=\ImgSquizPcaPoseWSS \textwidth]{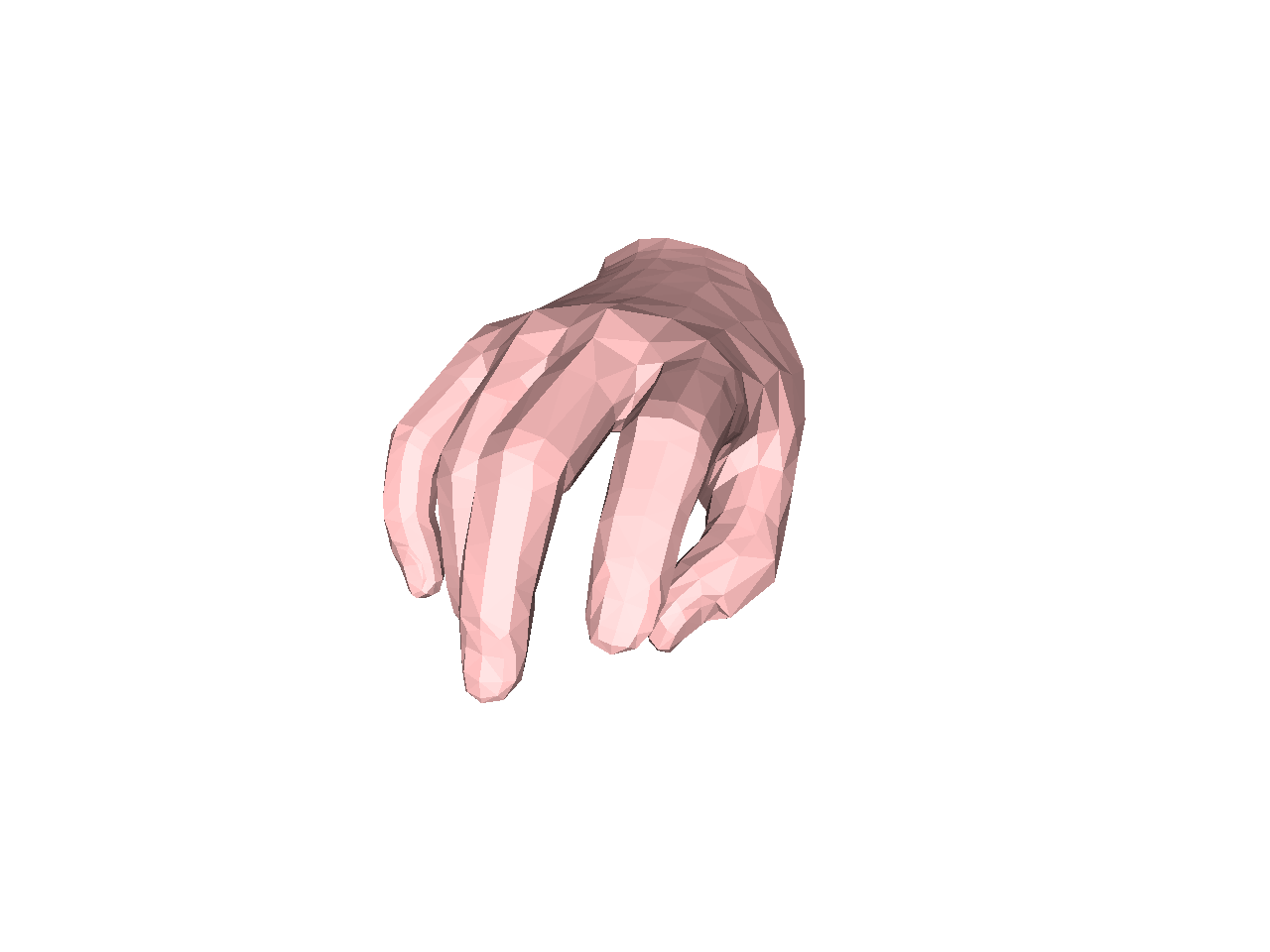}		&		\hspace*{\ImgSquizPcaPoseWHHa}
							\includegraphics[trim=000mm 000mm 000mm 000mm, clip=false, height=\ImgSquizPcaPoseWSS \textwidth]{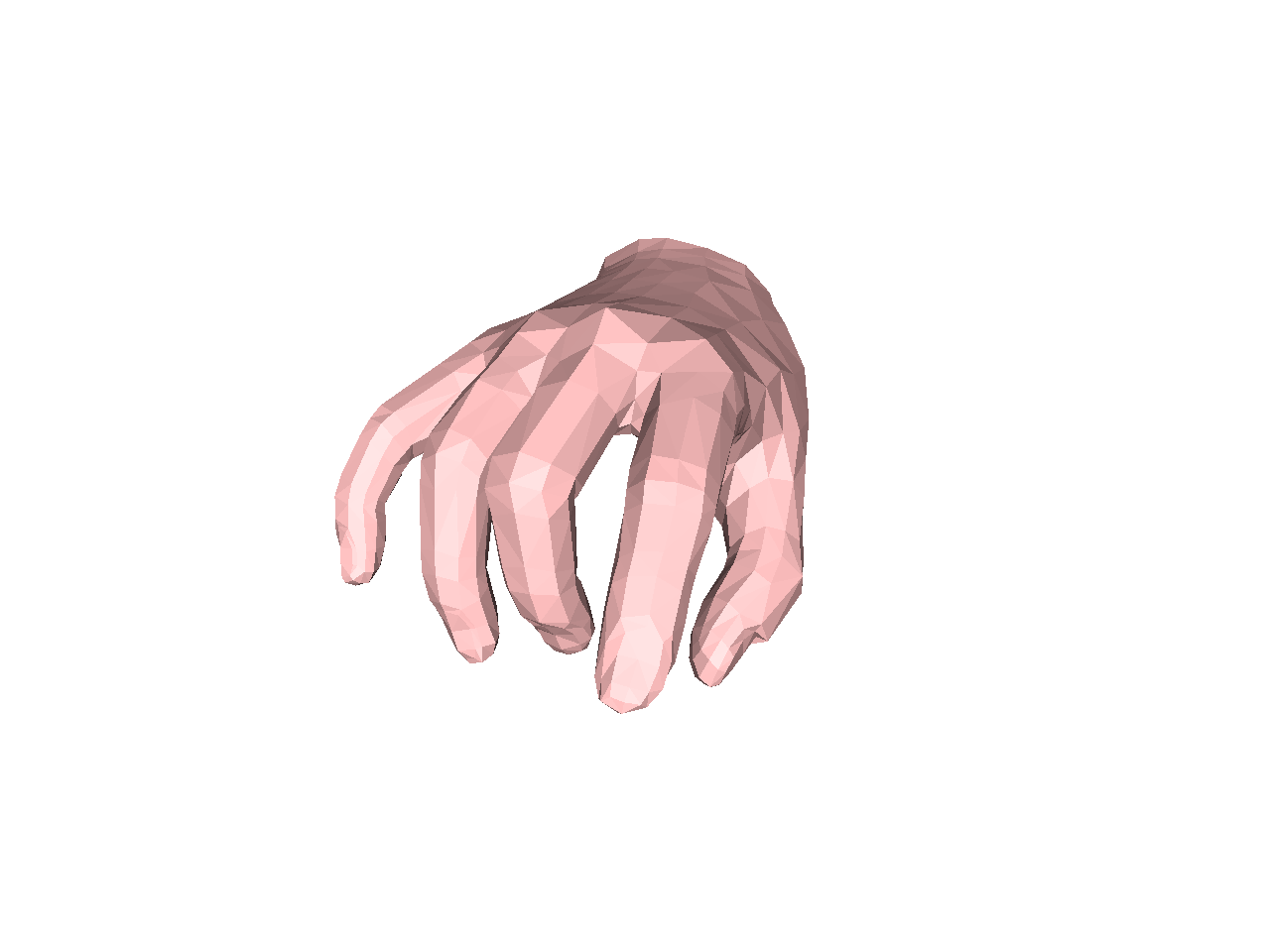}		&		\hspace*{\ImgSquizPcaPoseWHHa}
							\includegraphics[trim=000mm 000mm 000mm 000mm, clip=false, height=\ImgSquizPcaPoseWSS \textwidth]{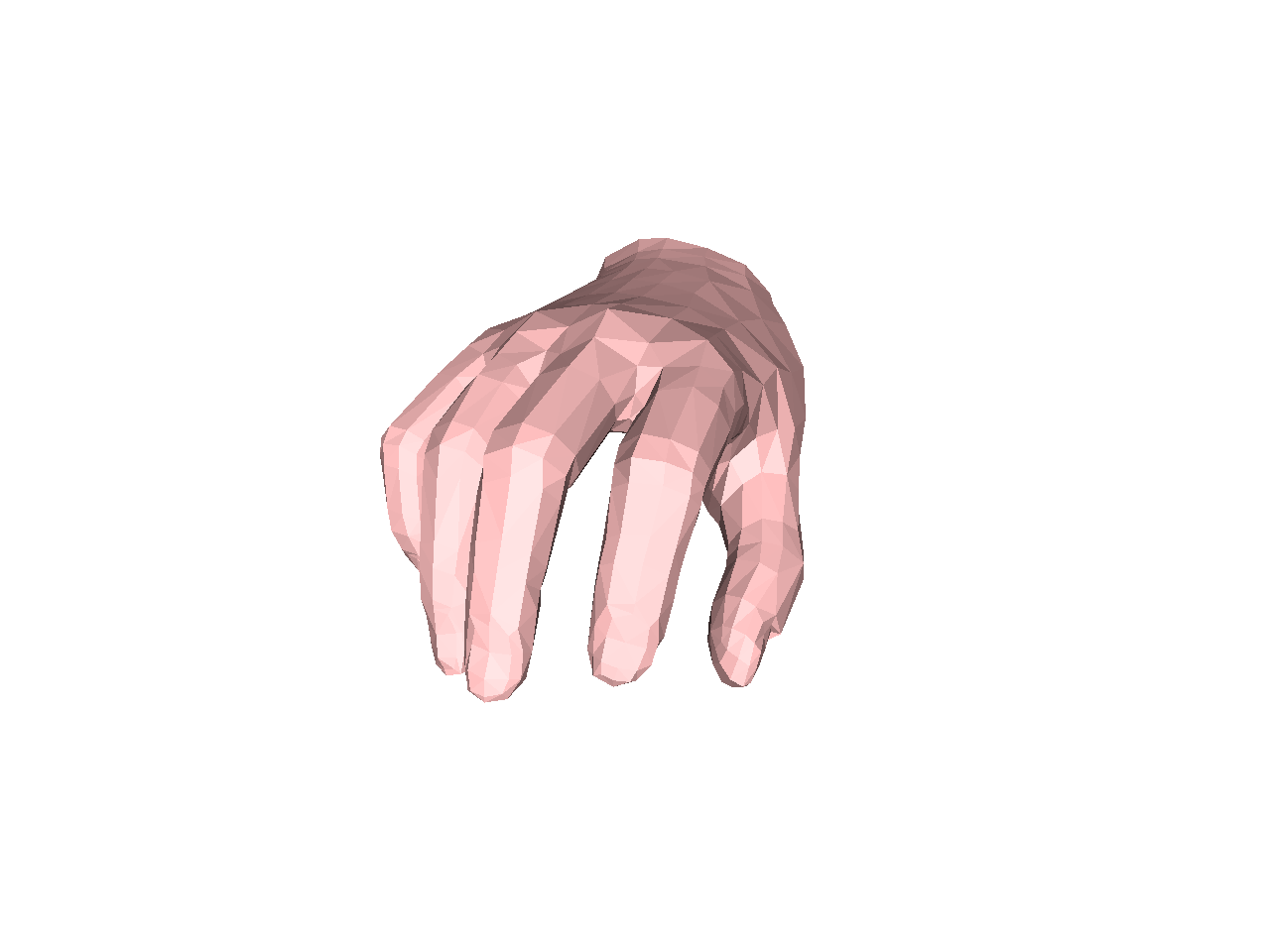}		&		\hspace*{\ImgSquizPcaPoseWHHa}
							\includegraphics[trim=000mm 000mm 000mm 000mm, clip=false, height=\ImgSquizPcaPoseWSS \textwidth]{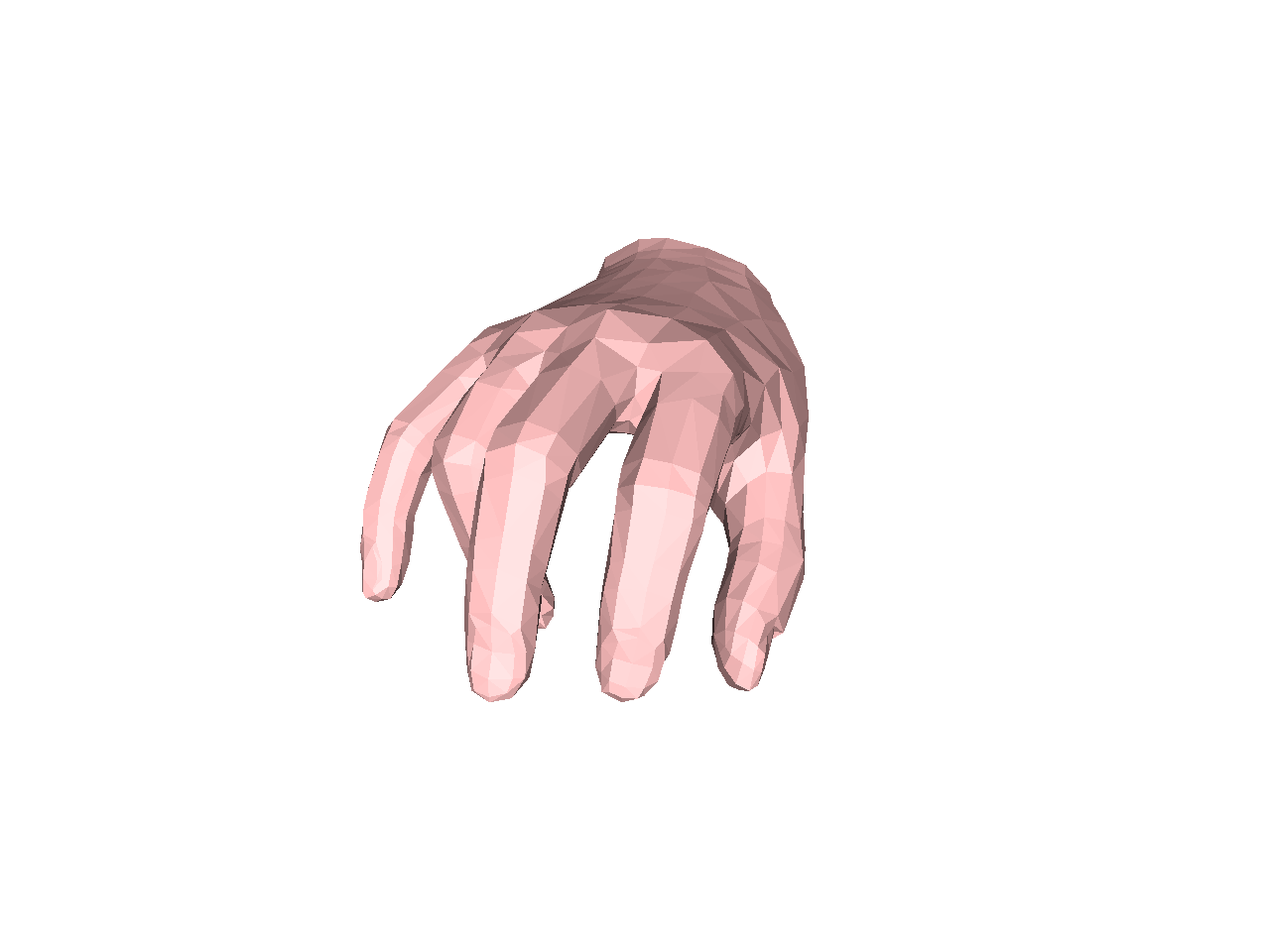}		&		\hspace*{\ImgSquizPcaPoseWHta}
							\multirow{-6}{*}{\parbox{09mm}{mean +\stdNoPosePLU ~std\\ \\}}																															\\				[\ImgSquizPcaPoseWVVa]
							{}																																														&		\hspace*{\ImgSquizPcaPoseWHma}
							\includegraphics[trim=000mm 000mm 000mm 000mm, clip=false, height=\ImgSquizPcaPoseWSS \textwidth]{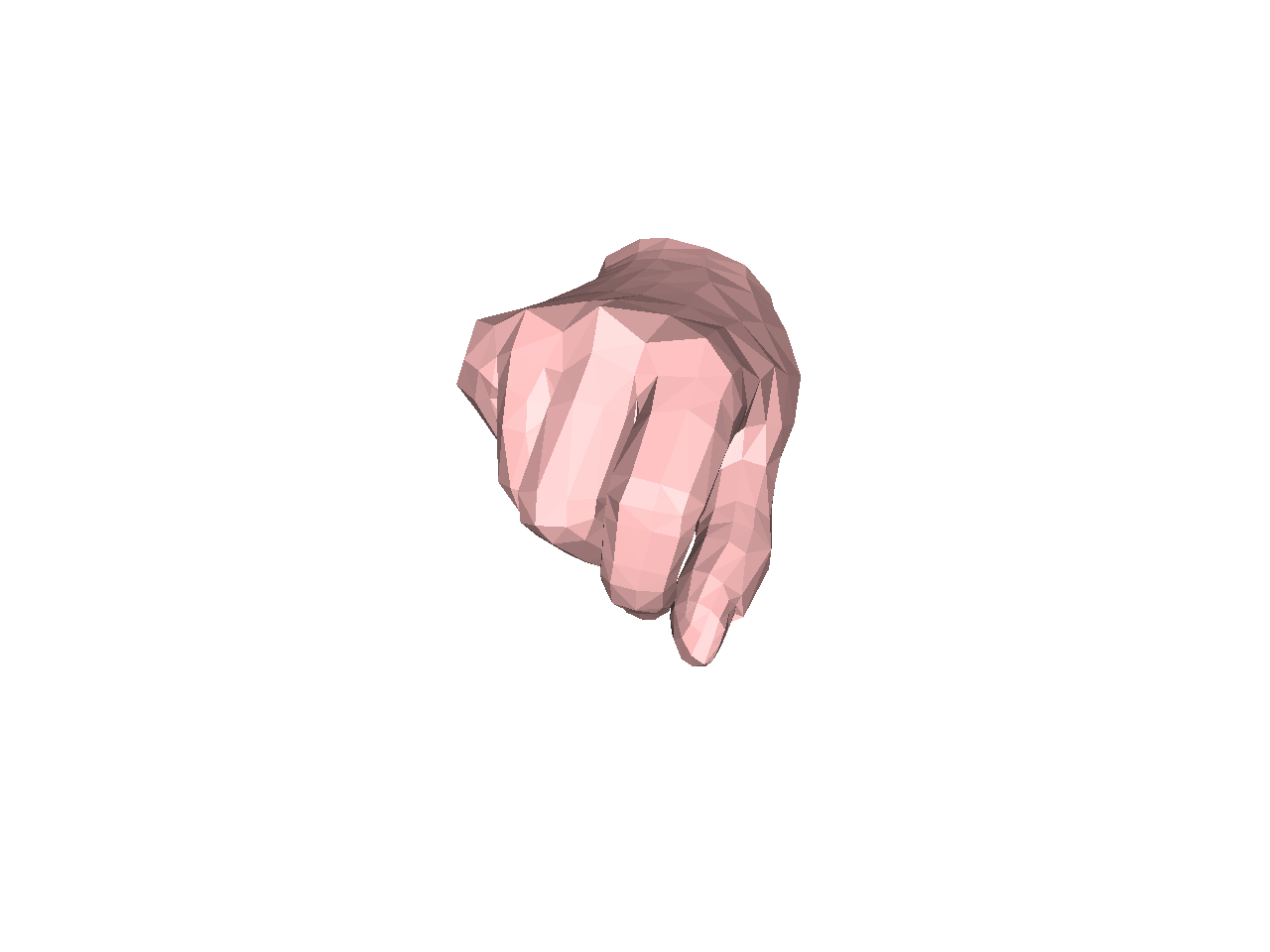}		&		\hspace*{\ImgSquizPcaPoseWHHa}
							\includegraphics[trim=000mm 000mm 000mm 000mm, clip=false, height=\ImgSquizPcaPoseWSS \textwidth]{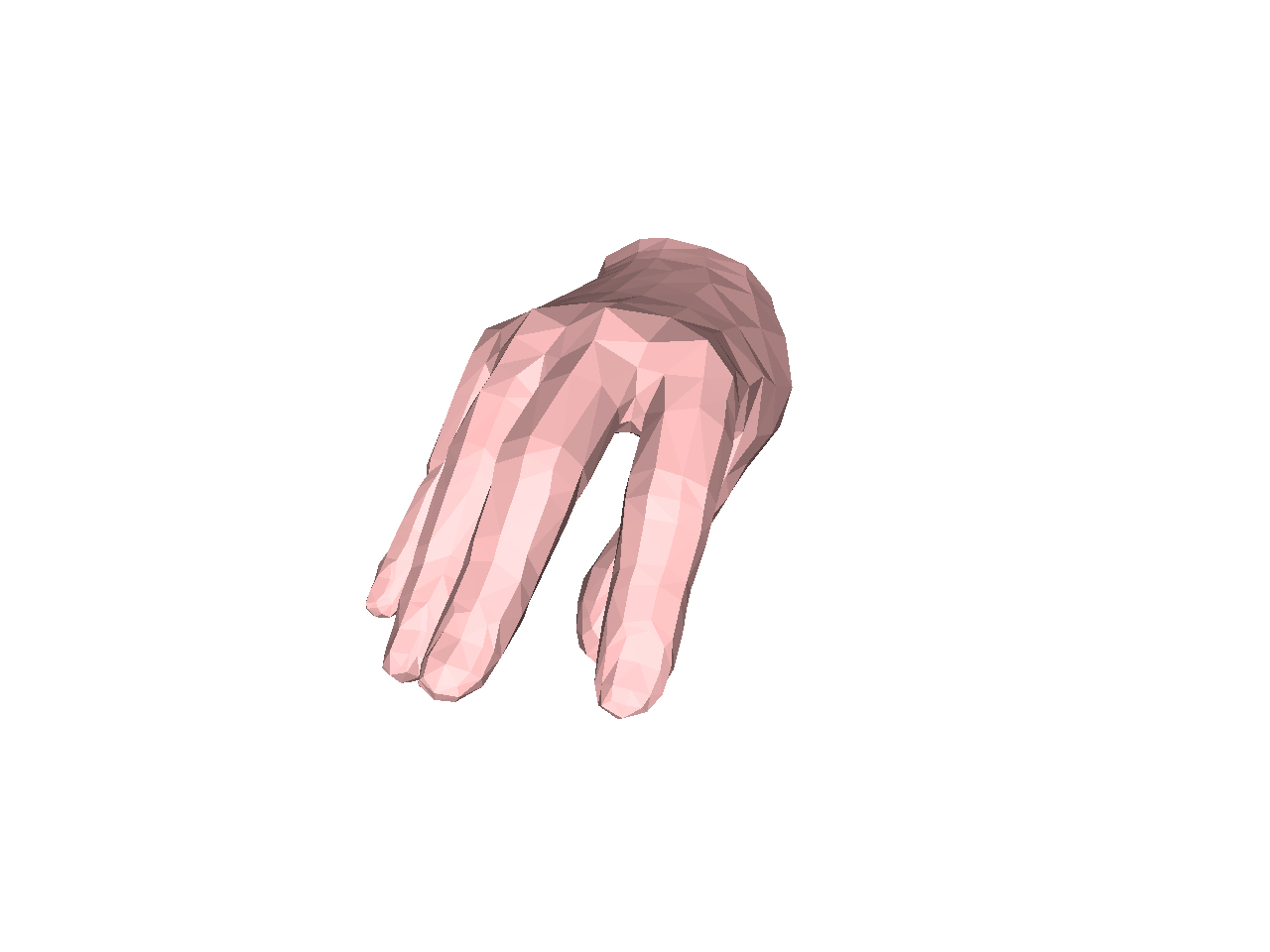}		&		\hspace*{\ImgSquizPcaPoseWHHa}
							\includegraphics[trim=000mm 000mm 000mm 000mm, clip=false, height=\ImgSquizPcaPoseWSS \textwidth]{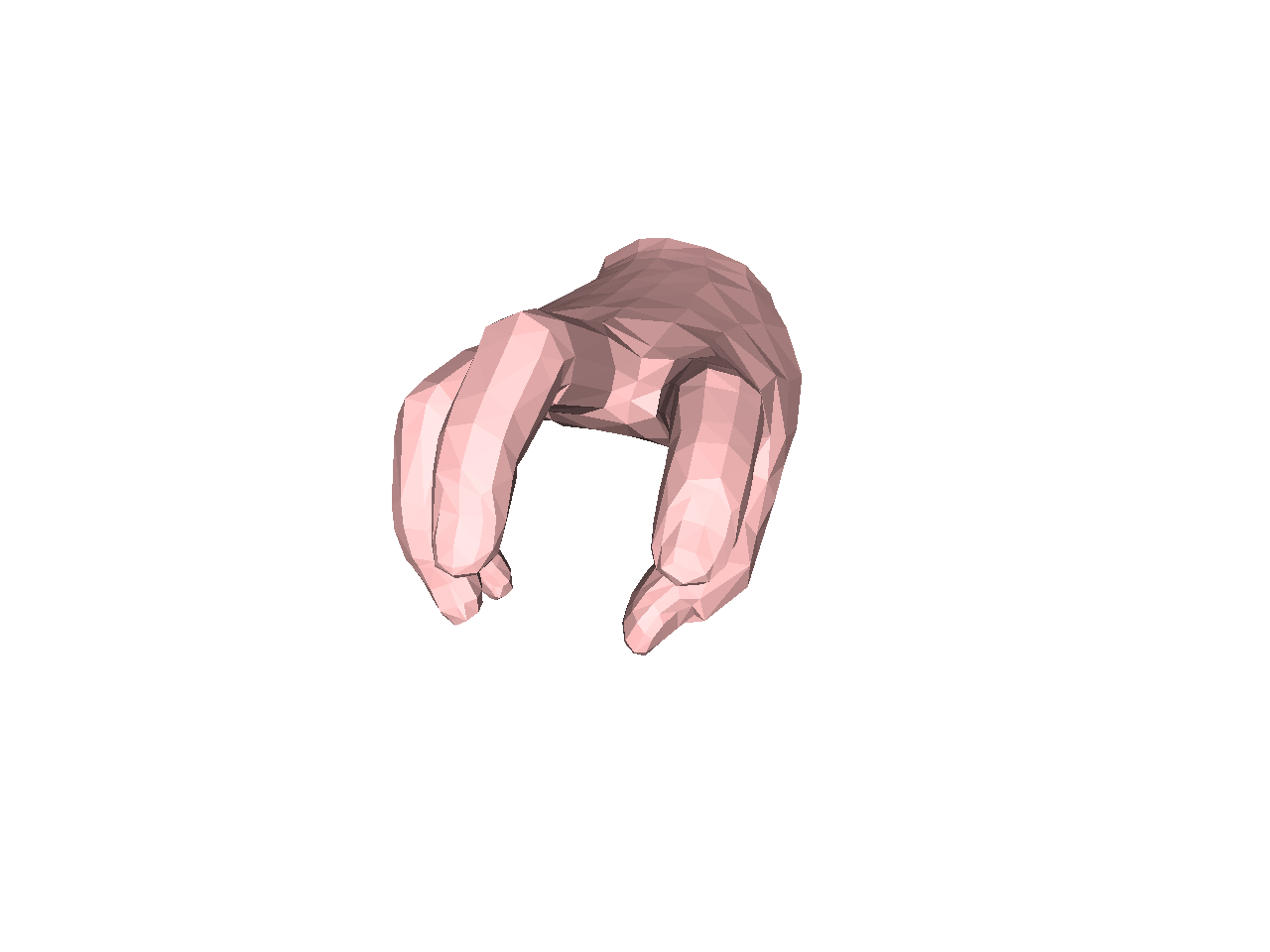}		&		\hspace*{\ImgSquizPcaPoseWHHa}
							\includegraphics[trim=000mm 000mm 000mm 000mm, clip=false, height=\ImgSquizPcaPoseWSS \textwidth]{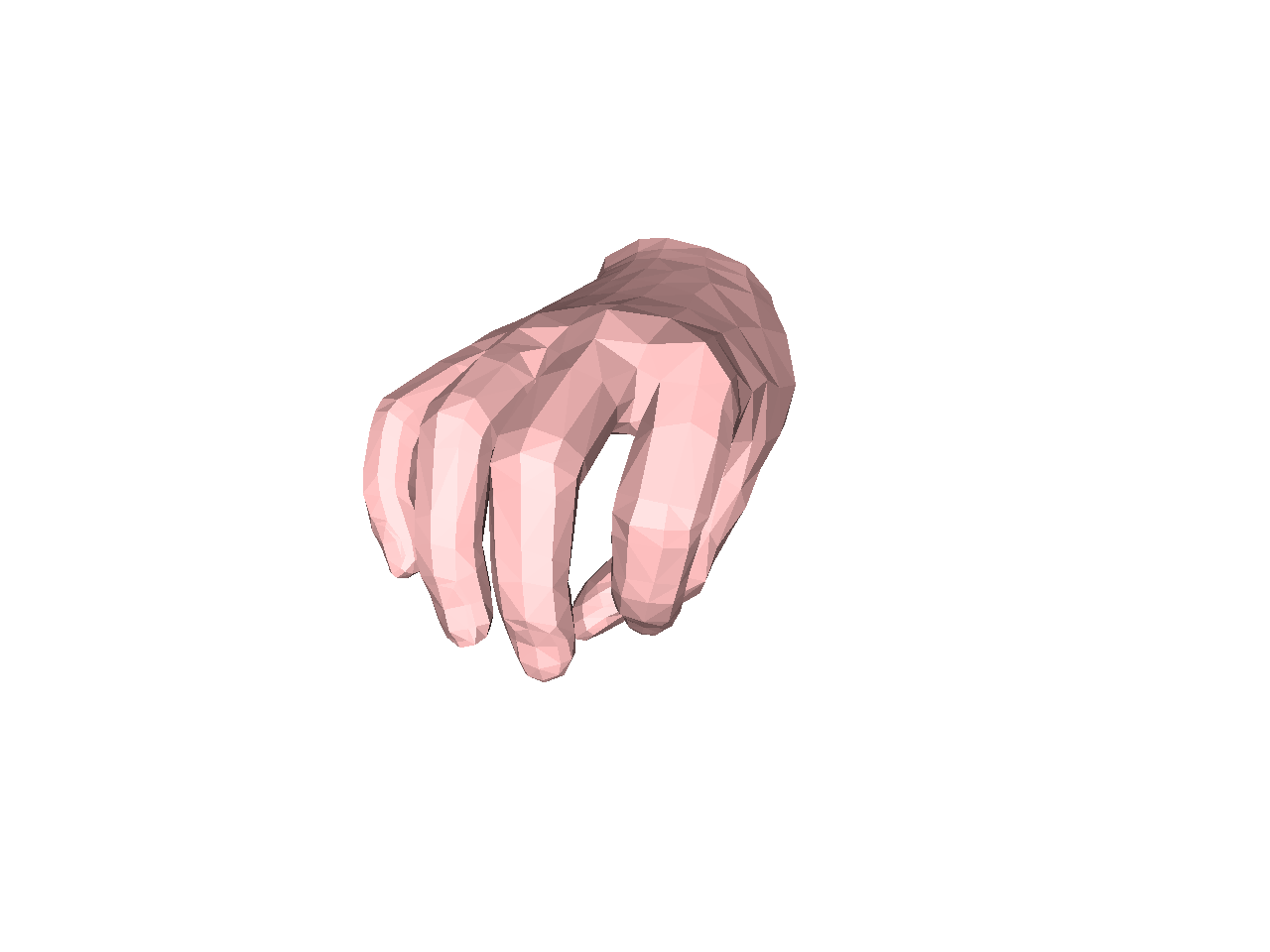}		&		\hspace*{\ImgSquizPcaPoseWHHa}
							\includegraphics[trim=000mm 000mm 000mm 000mm, clip=false, height=\ImgSquizPcaPoseWSS \textwidth]{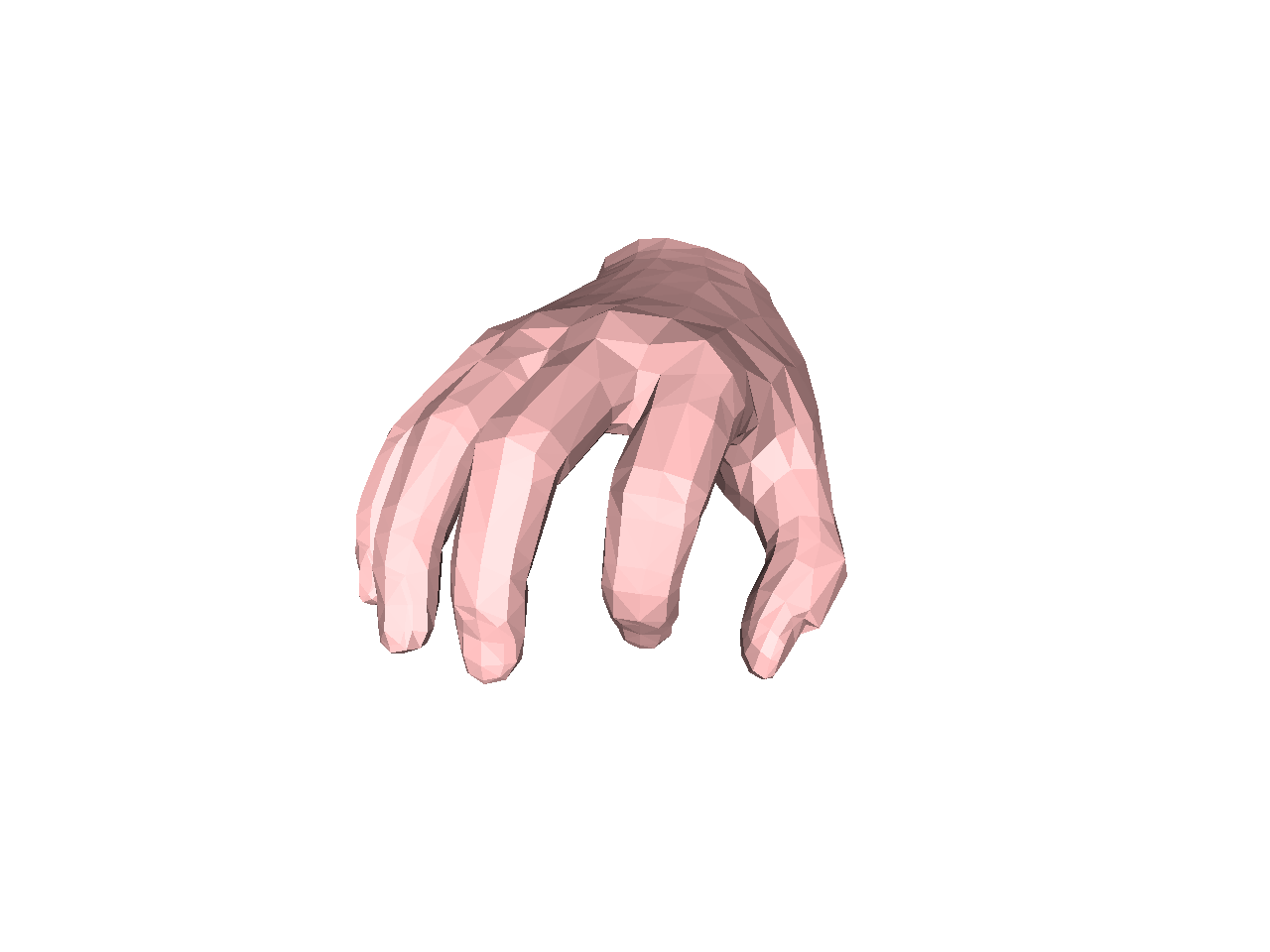}		&		\hspace*{\ImgSquizPcaPoseWHHa}
							\includegraphics[trim=000mm 000mm 000mm 000mm, clip=false, height=\ImgSquizPcaPoseWSS \textwidth]{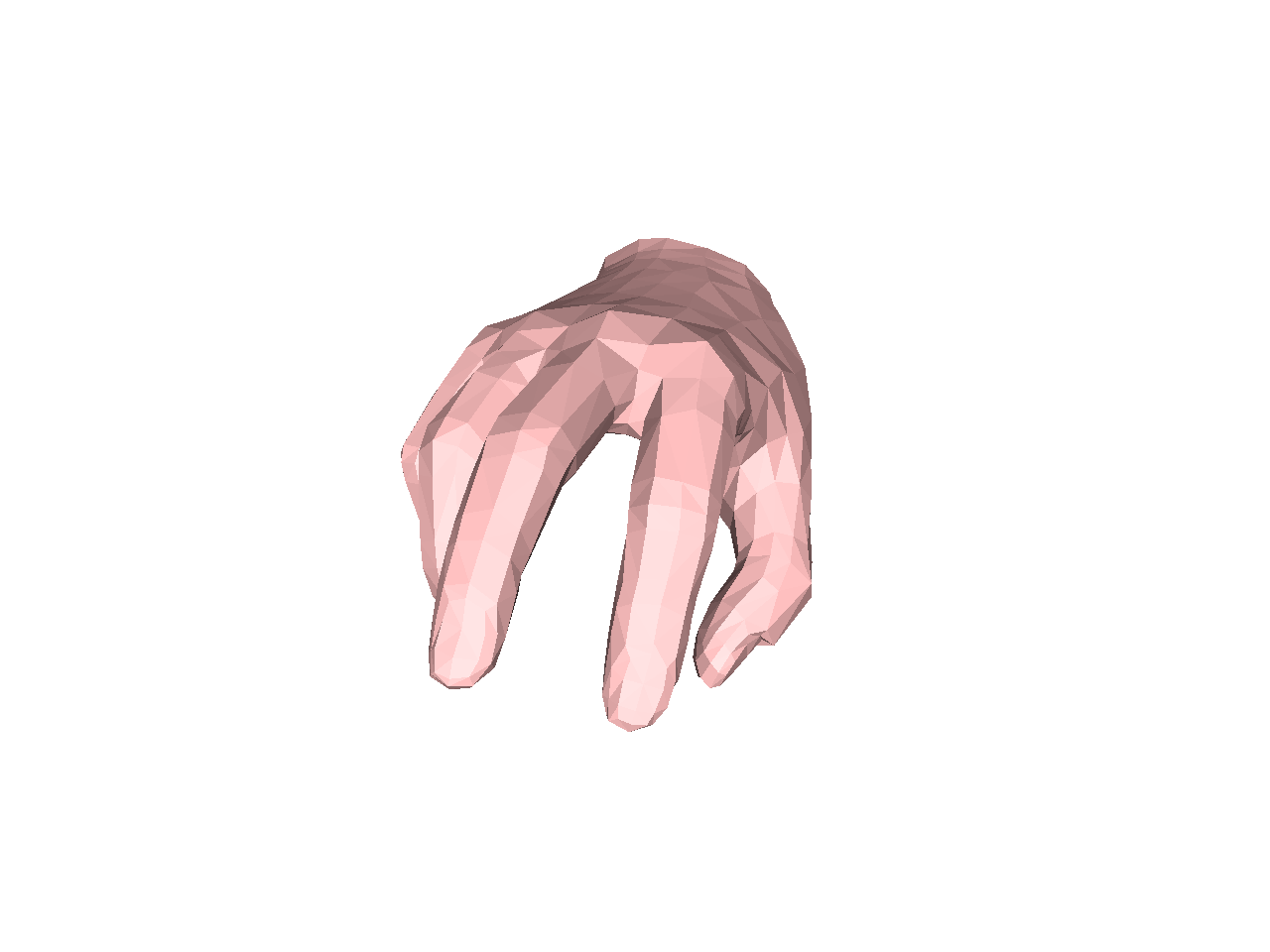}		&		\hspace*{\ImgSquizPcaPoseWHHa}
							\includegraphics[trim=000mm 000mm 000mm 000mm, clip=false, height=\ImgSquizPcaPoseWSS \textwidth]{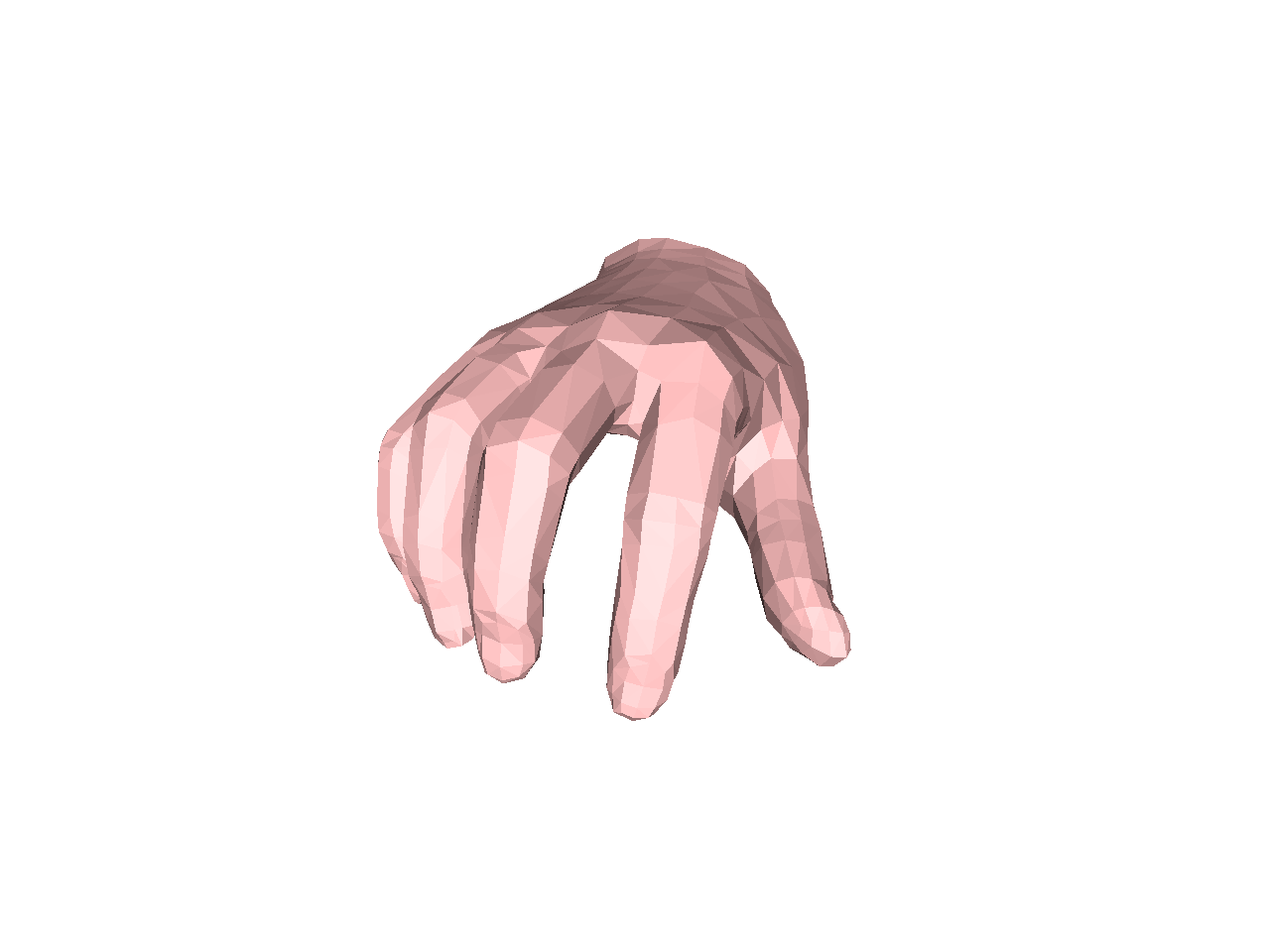}		&		\hspace*{\ImgSquizPcaPoseWHHa}
							\includegraphics[trim=000mm 000mm 000mm 000mm, clip=false, height=\ImgSquizPcaPoseWSS \textwidth]{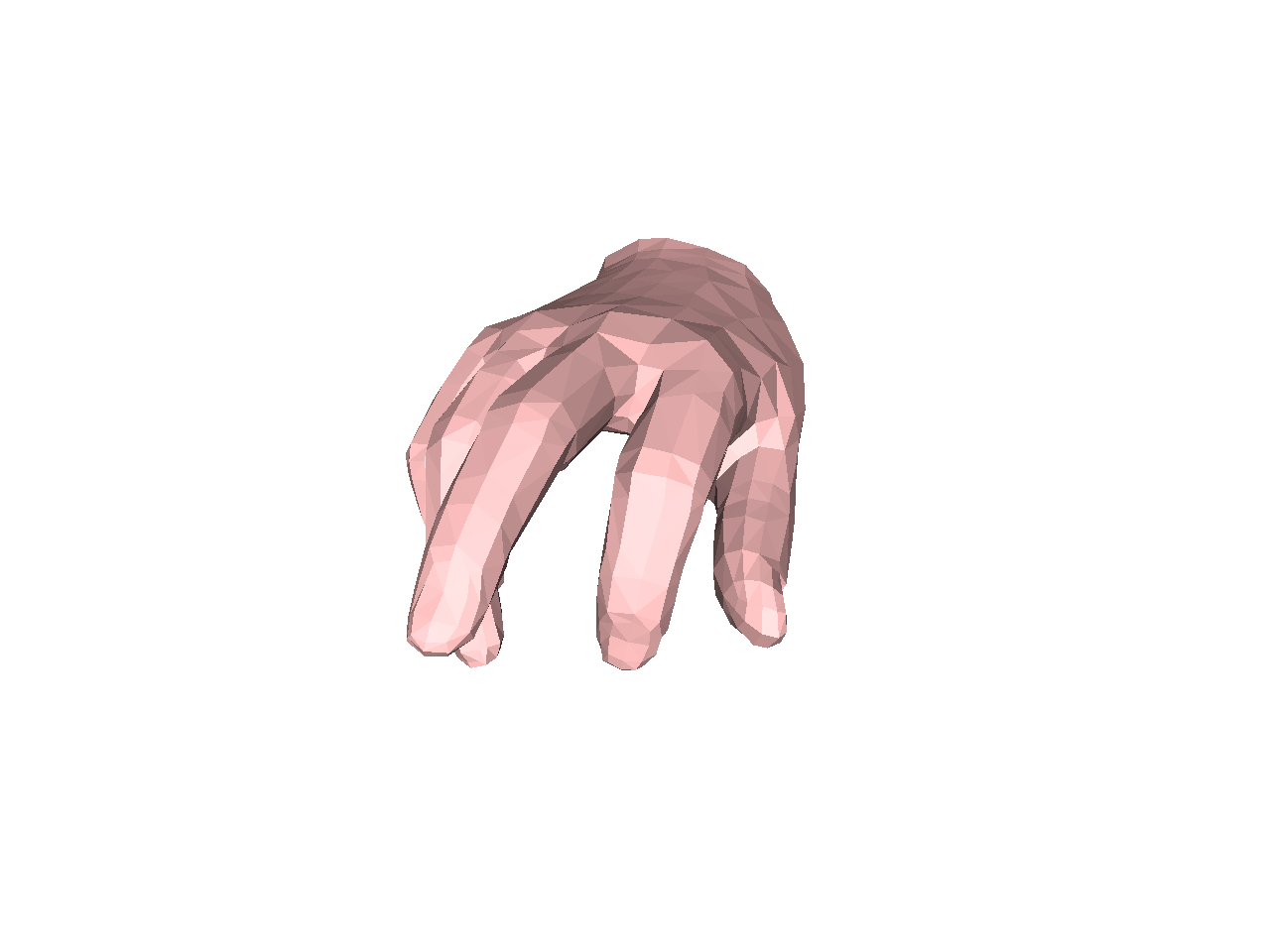}		&		\hspace*{\ImgSquizPcaPoseWHHa}
							\includegraphics[trim=000mm 000mm 000mm 000mm, clip=false, height=\ImgSquizPcaPoseWSS \textwidth]{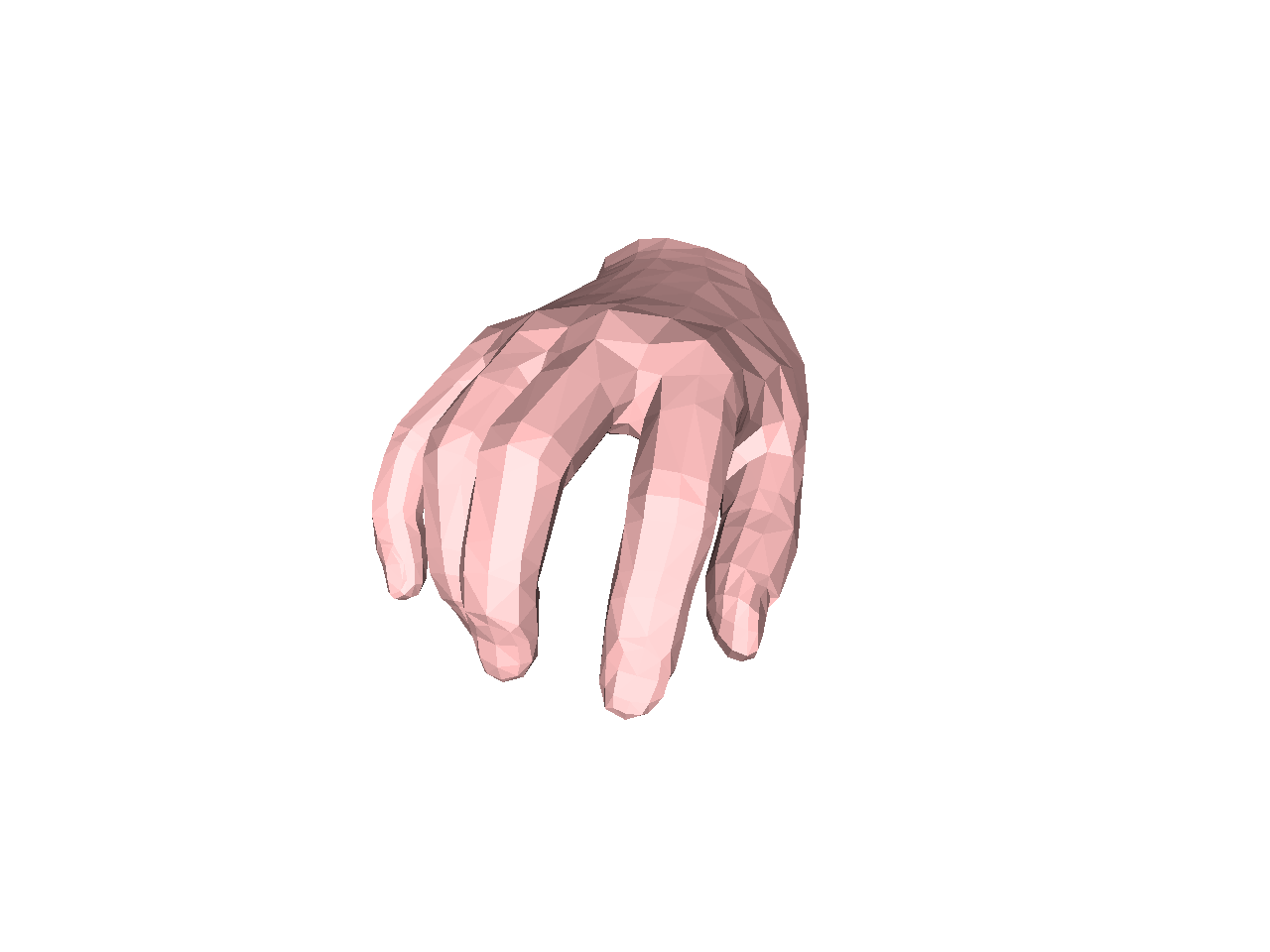}		&		\hspace*{\ImgSquizPcaPoseWHHa}
							\includegraphics[trim=000mm 000mm 000mm 000mm, clip=false, height=\ImgSquizPcaPoseWSS \textwidth]{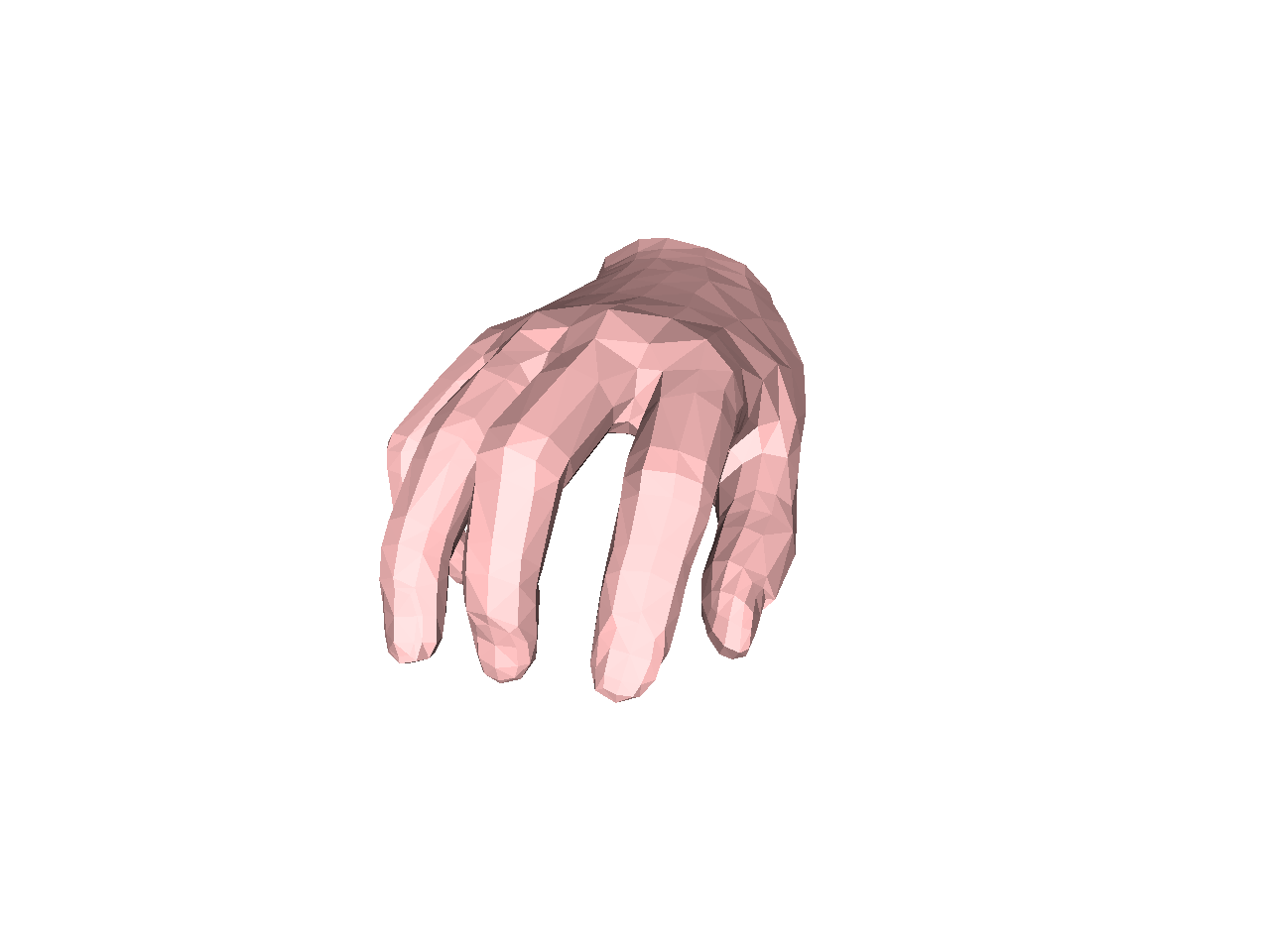}		&		\hspace*{\ImgSquizPcaPoseWHta}
							\multirow{-6}{*}{\parbox{09mm}{mean -\stdNoPoseMIN ~std\\ \\}}																															\\				[\ImgSquizPcaPoseWVva]
							\parbox{10mm}{Mean Pose}																																									&		\hspace*{\ImgSquizPcaPoseWHta}
							{PC 1}																																													&		\hspace*{\ImgSquizPcaPoseWHHa}
							{PC 2}																																													&		\hspace*{\ImgSquizPcaPoseWHHa}
							{PC 3}																																													&		\hspace*{\ImgSquizPcaPoseWHHa}
							{PC 4}																																													&		\hspace*{\ImgSquizPcaPoseWHHa}
							{PC 5}																																													&		\hspace*{\ImgSquizPcaPoseWHHa}
							{PC 6}																																													&		\hspace*{\ImgSquizPcaPoseWHHa}
							{PC 7}																																													&		\hspace*{\ImgSquizPcaPoseWHHa}
							{PC 8}																																													&		\hspace*{\ImgSquizPcaPoseWHHa}
							{PC 9}																																													&		\hspace*{\ImgSquizPcaPoseWHHa}
							{PC 10}																																													&		\hspace*{\ImgSquizPcaPoseWHta}
							{}
	\end{tabular}
	\caption{
				{\bf PCA pose space.}
				The left-most image depicts the \emph{mean pose}, while the rest of the columns depict the effect of the first ten \emph{principal components} (PCs) of the pose space.
				The effect of each PC is shown by adding $\mypm \stdNoPosePLU$ standard deviations (std) to the mean pose, as indicated.
	}
	\label{fig:pcaPCsPose}
\end{figure*}

%% file: tex_main/04_handsBody_4d_alignment.tex
\section{Capturing body and hand motion}\label{sec:captureHandBodiesMotion}

To evaluate \smplH we use 4D sequences of bodies and hands in motion; i.e.~temporal sequences of 3D scans. 
We fit the proposed model to the scans, bringing all of them into correspondence. 

\subsection{Camera system}\label{sec:cameraSystem}

For this purpose we use a 3dMDbody.u~\cite{3dMDbody} multi-camera active stereo system capturing 3D scans at $60$ frames per second (fps). 
The camera system consists of $22$ pairs of stereo cameras, $22$ color cameras, $34$ speckle projectors and arrays of white-light LED panels. 
Here we do not use the color cameras.
For each frame $t$ the system outputs a 3D scan $\scanverts^t$ computed from the stereo cameras and explaining a projected texture pattern.

The camera pairs are positioned in a way to cover multiple overlapping view-points and allow a wide range of full-body motions such as arm movements, running on the spot, throwing or jumping. 
All cameras synergistically capture full bodies and no camera is dedicated to the capturing of hands. 
This enables the use of holistic models like the proposed one, however it poses additional challenges for hands due to low resolution for the hand region, the big size of the projected speckle pattern compared to finger width, and the high velocity of finger motion.
These factors result in large occluded regions without observed points during several frames.
The above suggest the the need for strong hand models that are combined with full body models.

\subsection{Sequences}\label{sec:sequences}

Using the camera system described in Sec.~\ref{sec:cameraSystem} we capture a new set of challenging 4D sequences with a variety of subjects and actions performed. 
The captured sequences are separated in $3$ groups. 
In the first group we capture $11$ sequences of $10$ subjects, $5$ male and $5$ female, of varying size and shape, each performing an unconstrained improvised action. 
For these sequences we perform a comparison between the proposed model and SMPL.
The next group of $28$ sequences are captured for a single subject performing actions designed around hand motion including basketball, finger counting or keyboard typing, among others.
The final set of $2$ sequences captures $2$ professional female actors performing improvised movements that express fear. 
The duration of the sequences (in frames) is $189$ to $1125$ for the first group, $204$ to $890$ for the second and more than $3500$ for the third one. 
In total, our dataset has $41$ sequences with a total of $27,156$ frames and corresponding 3D scans. 
All subjects are dressed  with identical minimal clothing, namely tight fitting swimwear bottoms for both men and women and a sports bra top for women. 
All subjects gave written informed consent to participate in this study.

\subsection{Mesh registration}\label{seq:meshRegistration}

The goal of full body registration is to bring all temporal 3D scans into correspondence by registering a single 3D template for body and hands to all of them. 
The template $\template$ is a 3D watertight triangulated mesh comprised of $6890$ vertices and $13,776$ triangles. %V^t \in 

Similar to Section \ref{sec:handModel_RegistrationBootstraping}, registration is cast as a minimization problem. 
We try to minimize the discrepancy between the model and the data by minimizing the distance between the scan and the registration with respect to the registration vertex locations. 

For registration we use a two stage approach, similar to~\cite{Loper:SIGASIA:2015,Dyna:SIGGRAPH:2015}.
We first compute a specific template $\template^j$ for each subject, $j$, using a subset of the scans, typically the first $50$ frames where the subject is mostly static.  
We then use $\template^j$ to align the proposed model to all the scans (including the subset of the first step). 
By adopting the subject-specific templates the optimization of the second stage becomes significantly faster. 
To keep local optimization tractable, optimization for each frame is initialized with the pose of the previous frame in both stages. 
In the absence of a good initialization for the first frame, we introduce an extra initial stage for that frame in which a stronger pose prior is used.

\subsubsection{Subject-specific template}\label{seq:registration_personalizedTemplate}

For the first frame, where initialization is far away from the optimum, we exploit the fact that subjects were instructed to start with an approximate ``A-pose'' by following a conservative approach with strong pose priors. 
For this frame we solve only for a model fit by optimizing pose and shape
\begin{equation}
	E_0(\shape, \pose; \scanverts, \anneal) = 	\anneal \lambda_{\bar{g}} E_{\bar{g}} + 
																			\frac{\lambda_{\thetab}}{\anneal} E_{\thetab} + 
																			\frac{\lambda_{\thetah}}{\anneal} E_{\thetah} + 
																			\lambda_\beta E_{\beta}
																			\label{eq:registration_personalizedTemplate_modelOnly}
\end{equation}
where
\begin{align}
	E_{\bar{g}}(\shape,\pose; \scanverts)							&=								E_{g^*}(\shape,\pose;\scanverts) + 0.1 E_{g'}(\shape,\pose;\scanverts)								\label{eq:registration_personalizedTemplate_modelOnly_Egbar}		\\
  	E_{g^*}(\shape,\pose; \scanverts)	&= \sum_{\scanvert  \in \scanverts} \rho(\min_{m          \in \modelsurf(\shape,\pose)} \| \scanvert - m          \|) 		\label{eq:registration_personalizedTemplate_modelOnly_Eg}			\\
  	E_{g'}(\shape,\pose; \scanverts)	&= 								\sum_{m \in M(\shape,\pose)}          \rho(\min_{\scanvert \in \mathcal{C}}  \| \scanvert - m          \|)	\label{eq:registration_personalizedTemplate_modelOnly_Egtonos}	\\
	E_{\thetab}(\poseb) 									&=								\| \poseb \|^2 																							\label{eq:registration_personalizedTemplate_modelOnly_Ethetab}	%\\
\end{align}
where $\modelsurf(\shape,\pose)$ refers to the surface entailed by the model vertices $M(\shape,\pose)$, $\mathcal{C}$ to the surface entailed by the scan vertices $\scanverts$, and the \emph{data} term $E_{\bar{g}}$ includes a \emph{scan-to-mesh} energy $E_{g^*}$ similar to Equation~\eqref{eq:initial_registration_Eg} and a \emph{mesh-to-scan} $E_{g'}$. 
We found empirically that $20$ shape parameters $\beta_i \in \shape$ suffice to effectively capture shape identity.  

While in general $E_{g^*}$ suffices for fitting, the penalization of mismatching model vertices imposed by $E_{g'}$ avoids problems like excessively long hands in the templates extracted in this stage.
In that respect, $E_{\bar{g}}$ can be seen as an approximation of the Hausdorff distance, as argued in~\cite{Tkach:SIGGRAPH:2016, tagliasacchi2016modern}; replacing the \emph{max} operators in the Hausdorff formulation by \emph{sum} makes it differentiable and adds extra robustness to spurious scan data.

Furthermore, 
$E_{\beta}$ is a \emph{shape} term 	similar to Equation~\eqref{eq:initial_registration_Ebeta}, 
$E_{\thetab}$ is a pose prior 		similar to Equation~\eqref{eq:initial_registration_Etheta}. 
This prior is applied for the \emph{body} pose parameters $\poseb$ up to (and including) the wrist, excluding the fingers, i.e. $\poseh$. 
We empirically found that a Gaussian pose prior is not suitable for hands since in case of disappearing hand data in scans it traps the hand pose in a local minimum around the mean pose, even when data reappears. 
Since hand pose deviations are not Gaussian, we found Gaussian Mixture Models (GMM) to be more expressive. 
Inspired by \cite{Bogo:ECCV:2016,Olson:2013:IferenceNetworksMixtures}, we employ a GMM by approximating the sum in the GMM by a max operator
\begin{align}
	E_{\thetah}(\poseh) 	&	\equiv 	-\log \sum   \big( 			    g_j \mathcal{N}(\poseh \gmmean_{\thetah, j}, \Sigma_{\thetah, j}) \big)		\\
						    &	\approx	-\log \max_j \big( 			    g_j \mathcal{N}(\poseh \gmmean_{\thetah, j}, \Sigma_{\thetah, j}) \big)		\\
						    &	=			  \min_j \Big( -\log \big(  g_j \mathcal{N}(\poseh \gmmean_{\thetah, j}, \Sigma_{\thetah, j}) \big) \Big)
						\label{eq:gmm_hands}
\end{align}
where $g_j$ are the mixture model weights of $N$ Gaussians, and $\gmmean_{\thetah, j}$, $\Sigma_{\thetah, j}$ are the mean and covariance matrices of each component of the mixture.
The GMM is computed separately for left and right but based on the same data, namely all the left and right poses from the \mano dataset, mirrored as required. 
We empirically found $N = 10$ to work well. 
The objective Function \eqref{eq:registration_personalizedTemplate_modelOnly} is used in an annealing framework based on the annealing weight $\anneal$. 
At the beginning of the process, where the initial solution is far away from the desired optimum, optimization can be easily trapped in a local minimum. 
We therefore start with a smaller weight $\anneal = 100$ to have a higher prior term and a lower data term. 
After a rough alignment with this weight we repeat registration with a higher weight $\anneal = 500$ to allow for better fitting with a weaker prior. 
The objective Function \eqref{eq:registration_personalizedTemplate_modelOnly} is used only for the first frame, as a first step for rough initialization of the following stage. 

Next we optimize shape, pose, and a full coupled aligned mesh, $\algn$, by minimizing 
\begin{equation}
	E(\shape, \pose, \algn; \scanverts, \anneal) = 	\anneal \lambda_{g}                         E_g(\algn; \scanverts) 	+ 
														    \frac{\lambda_{\thetab}}{\anneal} 	E_c 		 + 
														    \frac{\lambda_{\thetah}}{\anneal} 	E_{\thetah}	 + 
														    \lambda_\beta                     	E_{\beta} 	 +
														    \lambda_r                         	E_r
														    \label{eq:registration_personalizedTemplate_coupled}
\end{equation}
where 
\begin{align}
	E_r(\pose) &= \sum_k \| \theta_k - \theta'_k \|^2 
\end{align}
and where $E_g$ and $E_c$ are defined by Equations \eqref{eq:initial_registration_Eg} and \eqref{eq:initial_registration_Ec}, respectively.
Recall that $E_c$ is a ``coupling'' term that regularizes the aligned mesh to be close to the model.

We drop the body pose prior $E_{\thetab}$ to allow more freedom during fitting. 
However, we keep the hand pose prior $E_{\thetah}$ since hand data can be very noisy, even disappearing completely for several frames. 
For this reason the data term is simplified to $E_g$, removing the sum over model points in Equation~\eqref{eq:registration_personalizedTemplate_modelOnly_Egtonos} which is particularly sensitive to holes in the scan.
The term $E_{\beta}$ is defined in Equation~\eqref{eq:initial_registration_Ebeta}. 
As a form of temporal pose smoothness against jittery motion, we introduce a zero-velocity regularizer $E_r$ that penalizes large changes between the current $\theta$ and the previous $\theta'$ pose parameters during optimization, especially in the complete absence of data. 

After aligning all frames in the initial subsequence, we then unpose all of these aligned meshes similar to~\cite{Loper:SIGASIA:2015} and manually select a subset through visual inspection, to filter out unposing artifacts. 
We then average these unposed alignments to obtain the subject-specific template $\template^j$.

\subsubsection{Subject-specific sequence alignment}\label{seq:registration_subjectSpecificRegistration}

After first acquiring a subject-specific template $\template^j$ we avoid optimizing over $\shape$ in the second stage, to gain significant optimization speed up and robustness. 
We therefore perform alignment similar to Section \ref{seq:registration_personalizedTemplate} but we drop the shape term $E_{\beta}$.
Therefore the first frame is initialized with model-based optimization
\begin{equation}
	E_0(\pose; \scanverts, \anneal) = 		\anneal \lambda_{\bar{g}}                   E_g(\shape,\pose; \scanverts)	+ 
										            \frac{\lambda_{\thetab}}{\anneal}   E_{\thetab} 			        + 
											        \frac{\lambda_{\thetah}}{\anneal}   E_{\thetah}
    \label{eq:registration_subjectSpecificRegistration_modelOnly}
\end{equation}
followed by coupled alignment
\begin{equation}
	E(\pose, \algn; \scanverts, \anneal) = 	\anneal \lambda_g                           E_g(\algn; \scanverts)	+ 
	                            			        \frac{\lambda_{c}}{\anneal}         E_c					    + 
	                            					\frac{\lambda_{\thetah}}{\anneal}   E_{\thetah}			    + 
	                            					\lambda_r                           E_r  .
    \label{eq:registration_subjectSpecificRegistration_coupled}
\end{equation}
Subsequent frames are optimized with coupled alignment only. All the terms are defined as in Section~\ref{seq:registration_personalizedTemplate}
By adopting the two stage approach of first acquiring a subject-specific template and then tracking with subject-specific registration, we reduce the dimensionality of the latter by $20$ parameters and, most importantly, make the registrations more accurate and robust.

\subsubsection{Optimization}

The objective functions~\eqref{eq:registration_personalizedTemplate_modelOnly}, \eqref{eq:registration_personalizedTemplate_coupled}, \eqref{eq:registration_subjectSpecificRegistration_modelOnly}, \eqref{eq:registration_subjectSpecificRegistration_coupled} are minimized using a gradient-based dogleg minimization \cite{nocedal2006numerical}. 
Gradients are computed with automatic differentiation using OpenDR \cite{Loper:ECCV:2014}. 
The optimization takes approximately $4$ minutes per frame on a $3,7$ GHz Quad-Core Intel Xeon E$5$ computer using $4$ threads, where only the search of closest-points is multithreaded. 

%% file: tex_main/05_experiments.tex
\input{tex_FIG/FIG_experiments_MANO}

\section{Experiments}

In the following we present quantitative and qualitative evaluation of \mano and qualitative evaluation of the \smplH model. 

\subsection{\mano Evaluation}	\label{sec:evaluation_handOnly}

\paragraph{Compactness}
Figure~\ref{fig:plotCompactness} plots the compactness of the \mano pose and shape space, left and right, respectively. 
These plots show the variance of the training-set captured by a varying number of components. 
The plot for the pose space shows that $6$, $10$ and $15$ components explain $81\%$, $90\%$ and $95\%$ correspondingly of the full space. 
Although more components lead to a lower error, they also lead to over-fitting to noise and higher computational demands.
Therefore, a larger number of components is suitable for clean hand-only datasets while full-body data benefits for the extra regularization provided by a small number of components.

\paragraph{Generalization}
To evaluate the generalization capabilities of the pose space in \mano we create a test-dataset of $50$ right hand scans as shown in Figure~\ref{fig:plotsTestDataset}, in which scans with severe occlusions have been removed.
The subjects in this dataset are not part of the data used for model training.
Figure~\ref{fig:plotGeneralizationPose} shows the generalization plot for the pose space, illustrating how the trained model generalizes to unseen poses. 
The plot depicts the mean and standard deviation of the scan-to-mesh (s2m) error in millimeters (mm) for each number of components for the \mbox{low-D} space. 
The plot shows that \mano generalizes nicely and the error decreases monotonically for an increasing number of components for the \mbox{low-D} pose space. 
As the number of components increases, the curve of the \mbox{low-D} space approximates asymptotically the minimum error of $0.93$ mm achieved by the full space. 

To study the generalization of the shape space, the $6$ subjects of the test-dataset in Figure~\ref{fig:plotsTestDataset} are insufficient. Therefore, we perform a leave-one-out evaluation of the shapes in the {\em training set.}
We report the mean absolute error (mabs) on the template vertices of each test-subject. 
Figure~\ref{fig:plotGeneralizationShape} depicts the generalization plot for the shape space, that shows how the model generalizes to unseen shapes. 
Similarly to Figure~\ref{fig:plotGeneralizationPose} for the pose space, this plot for the shape space decreases monotonically for an increasing number of components. 
	
\paragraph{Training dataset size} 
We then study the performance of our model trained on different datasets. 
First we examine the effect of including mirrored poses in the dataset by comparing a model trained only on the right hand poses, and another trained on the dataset augmented with the mirrored left ones.
While the former approach could potentially exploit the differences between hands by creating two specific \mano models for left and right hands, the latter approach exploits more training data for both hands by fusing the datasets.
For the full-dimensional pose space, the former approach has an error of $1.05$ mm and the latter $0.93$ mm. 
Both approaches were evaluated in the independent test-dataset, as shown in Figure~\ref{fig:plotsTestDataset}. 
Since this test dataset contains bigger variance for the pose space, we choose to use a fixed personalized template for each subject and optimize only over pose. 
For brevity we include in \textbf{supplementary material} a generalization plot that compares the two approaches for a varying number of components for the pose space. 
As previously described, the chosen model for \mano is the one that exploits both right and mirrored left data, due to its lower error and more coherent training procedure.

\paragraph{Effect of pose blend shapes}
Regarding the effect of pose blend shapes, we fit the test-dataset scans with the pose blend shapes activated and deactivated, using again a fixed personalized template for each subject. 
The latter is essentially an LBS approach performing similarly to methods like \cite{MSR_2016_Siggraph_handTrack, Tzionas:IJCV:2016, LucaHands}. 
The error for the full pose space is $0.93$ mm for activated blend shapes, while for deactivated $1.3$ mm. 
This numerical gain is in accordance with the qualitative results shown in Fig. \ref{fig:poseBlendShapesOnOff}. 
We thus conclude that the pose blend shapes facilitate not only improved visual realism, but also better fitting. 
	
\paragraph{Effect of parameter learning}
One remaining question concerns the overall effect of learning the model parameters $(\mathcal{S},\mathcal{P},\mathcal{W},\mathcal{J},\template)$ on the accuracy of fitting new data. 
Since we want to study the overall effect of learning all parameters, for this experiment we change the evaluation protocol, and we optimize over both pose and shape during fitting. 
For the independent data in Figure~\ref{fig:plotsTestDataset}, an untrained model reports for the full pose-space an s2m error of $2.90$ mm, while a model with learned parameters is more than $60\%$ better, with an error of $1.01$ mm. 
For brevity we include in \textbf{supplementary material} another generalization plot that compares a varying number of pose components. 

\input{tex_FIG/FIG_experiments_SMPLH_1}

\subsection{\smplH Evaluation}

Our test sequences for \smplH are captured with a scanner configured to accommodate full bodies and not only hands. 
Therefore the hands are often extremely noisy as shown in the scans (top row) of Figures \ref{fig:teaser}, \ref{fig:final_results_comparison_no_hands}, \ref{fig:final_results_1}. 
Please note that, as shown in Figure~\ref{fig:final_results_1} (top), it is not uncommon that the whole hand or even the arm completely disappears in the input scans for several consecutive frames. 
The above challenges call for a strong \mbox{low-D} model as described in Section \ref{sec:HAND_model} for \mano and Section~\ref{sec:SMPLH_modelIntegration} for \smplH. 
Based on the compactness plot for the pose space shown in Figure~\ref{fig:plotCompactness}, we use $6$ components for the evaluation in this section as a trade-off between accuracy and robustness.

As shown in Figure~\ref{fig:final_results_comparison_no_hands}, the holistic approach of \smplH improves significantly the visual realism compared with SMPL. 
The figure shows that SMPL (Fig. \ref{fig:final_results_comparison_no_hands} green) always keeps flat open hands, which often results in a lack of realism and expressiveness.
In contrast, \smplH (Fig.~\ref{fig:final_results_comparison_no_hands} pink) is able to accurately capture expressive hand poses, significantly increasing realism compared with SMPL.

This joint modeling proves robustness against severe conditions that traditionally challenge performance capture methods. 
Figure~\ref{fig:final_results_1} shows that \smplH performs naturally even under severe missing data, fast motions or measurement noise, which are especially prevalent for the case of hands. 

\input{tex_FIG/FIG_experiments_SMPLH_fail}

\input{tex_FIG/FIG_experiments_SMPLH_2}

Finally, it is worth noting that our method is not perfect. Two typical failure cases are shown in Figure~\ref{fig:failure}: scans of people interacting with objects, and hand poses which are not well covered by the low-dimension pose space.
    
Extensive results can be seen in the \textbf{supplementary video}. 

%% file: tex_FIG/FIG_experiments_MANO.tex
\begin{figure}[ht]
	\includegraphics[trim=000mm 000mm 000mm 000mm, clip=true, width=0.99\linewidth]{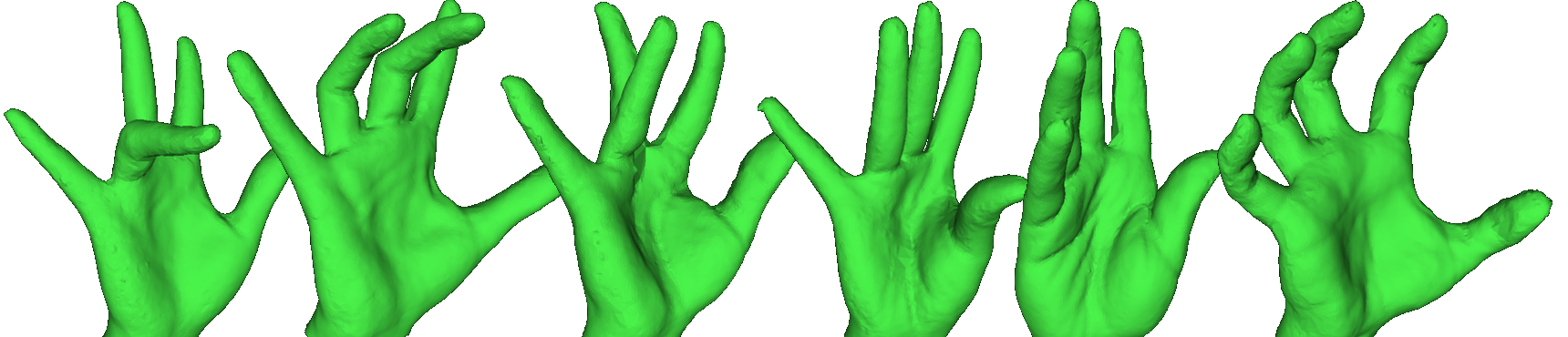}
	\caption{
  				Representative scans of the test-dataset created to evaluate the \mano model.
  				The dataset contains $50$ scans of $6$ subjects performing single- and double-finger articulation, as well as a more coordinated movement of all fingers. 
	}
	\label{fig:plotsTestDataset}
\end{figure}

\begin{figure*}[t]
	\includegraphics[trim=010mm 000mm 016mm 000mm, clip=true, width=0.38\linewidth]{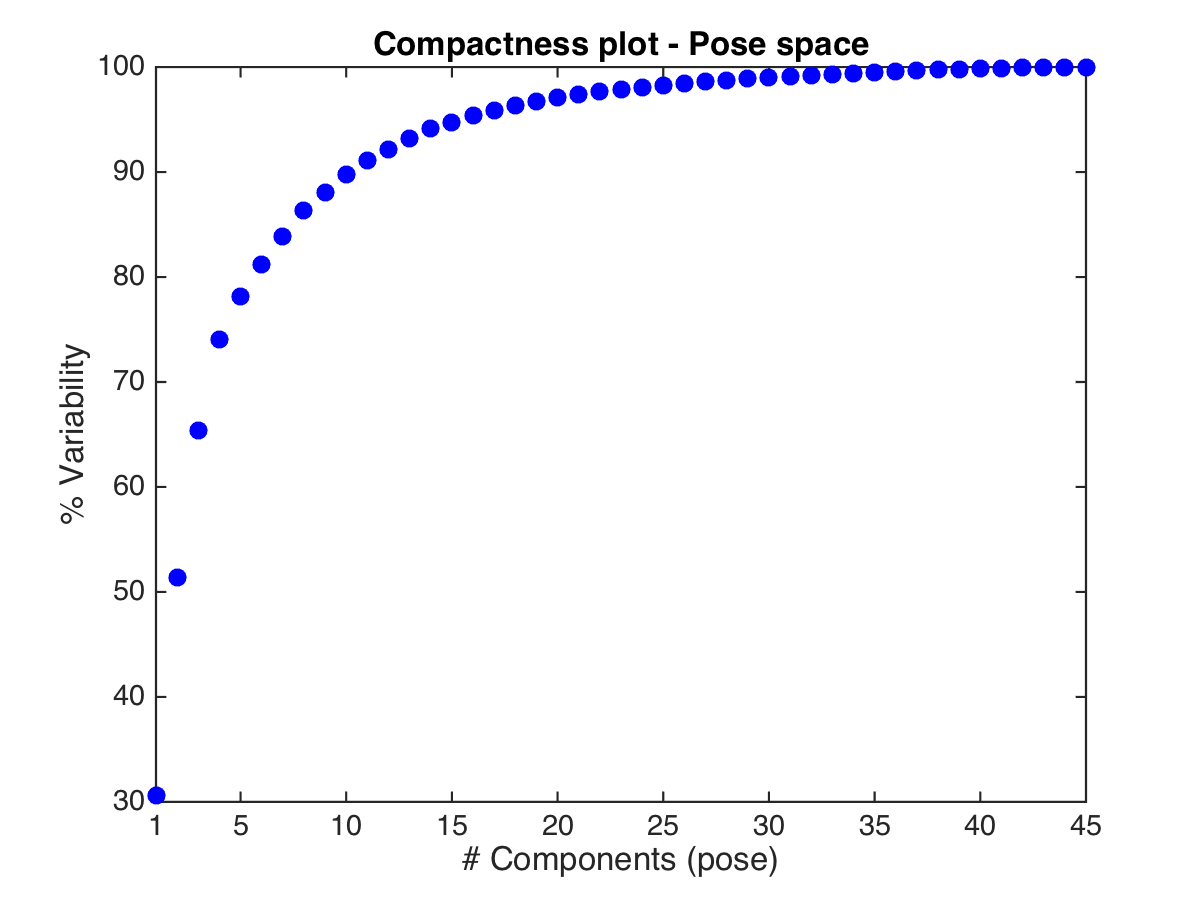}\label{fig:plotCompactnessPose}		\hspace{10mm}
	\includegraphics[trim=010mm 000mm 016mm 000mm, clip=true, width=0.38\linewidth]{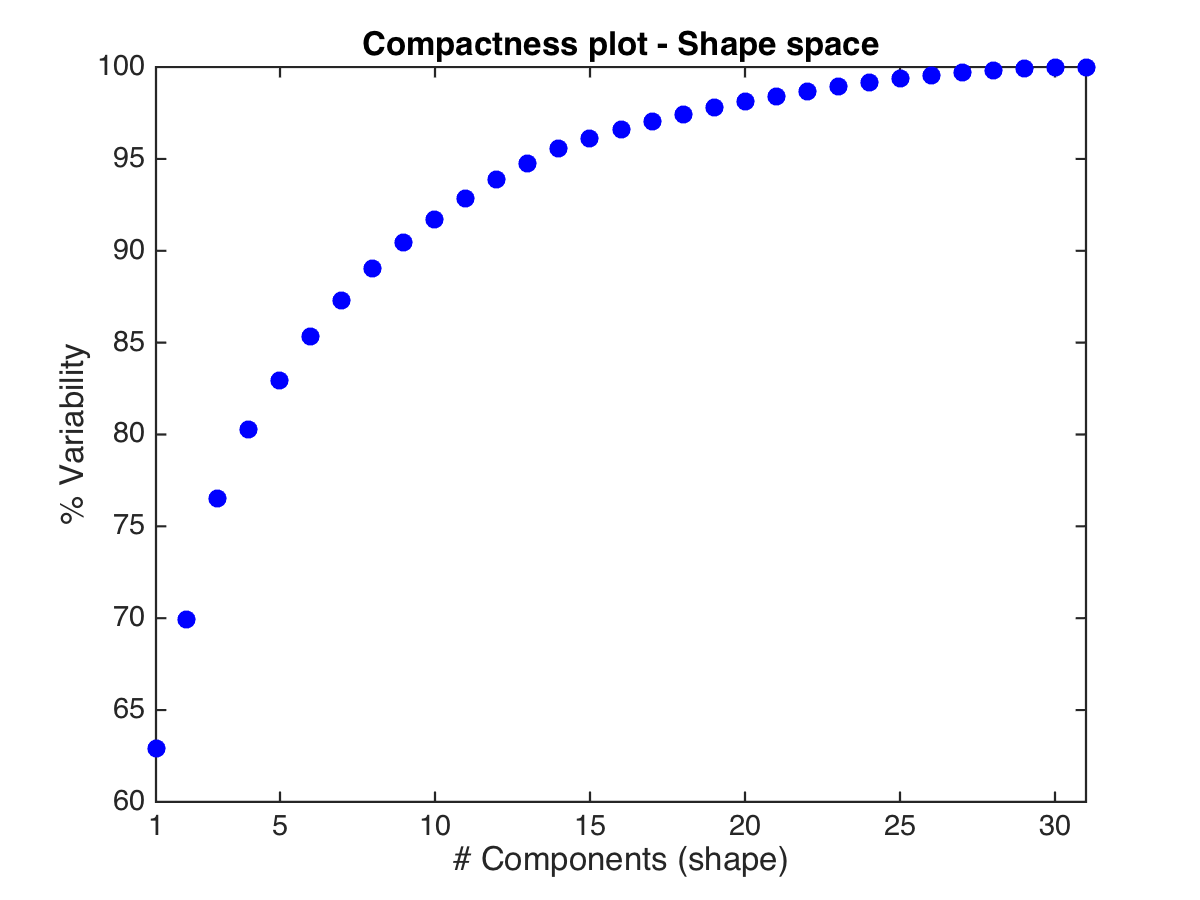}\label{fig:plotCompactnessShape}
	\caption{
  				Compactness plots for the pose (left) and shape (right) space. 
  				For the former, $6$ and $15$ components explain $81\%$ and $95\%$ correspondingly of the space. 
	}
	\label{fig:plotCompactness}
\end{figure*}

\begin{figure}[t]
	\subfloat{\includegraphics[trim=017mm 000mm 017mm 000mm, clip=true, height=0.75\linewidth]{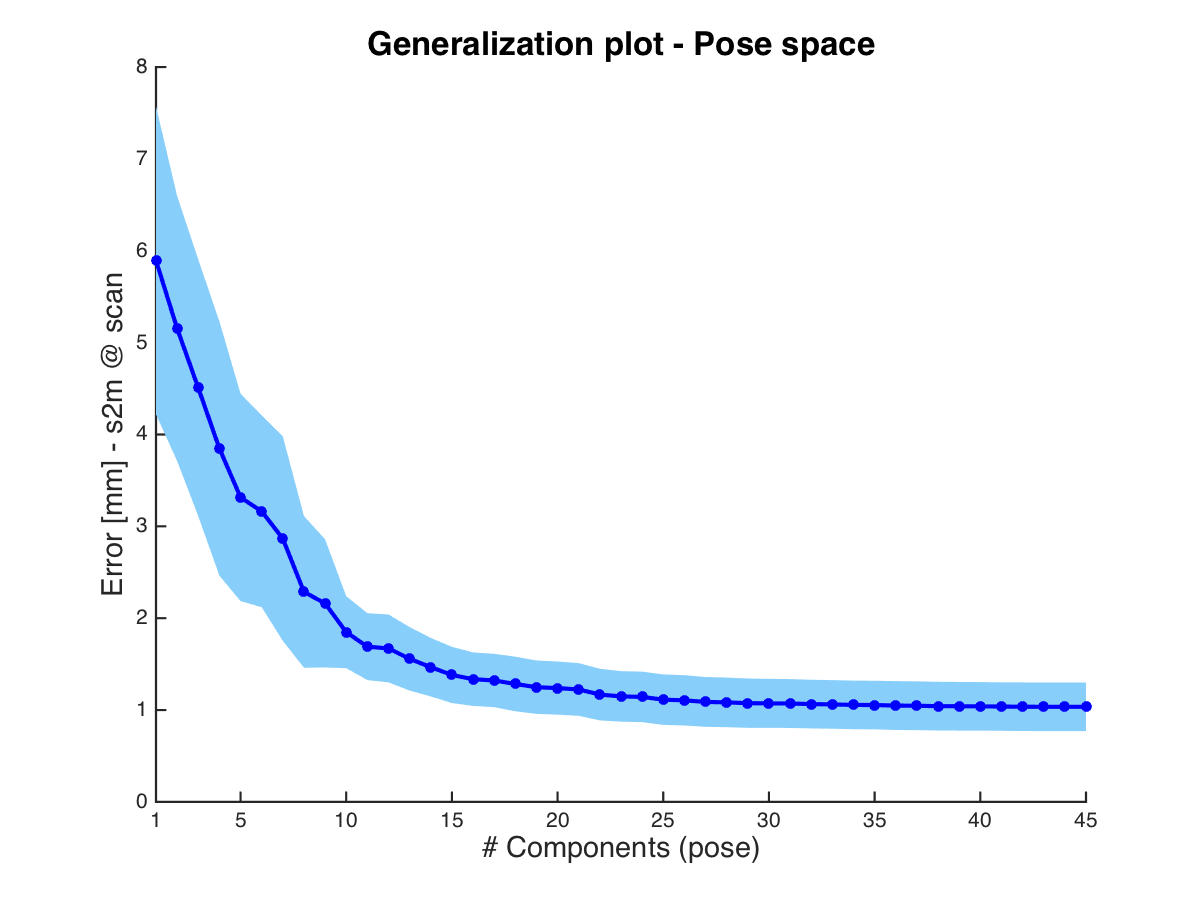}\label{fig:plotGeneralizationPoseS2M}  }
	\caption{
  				Generalization plot for the pose space. 
  				We report the mean scan-to-mesh (s2m) error for a varying number of components for the \mbox{low-D} pose space. 
  				For each number of components we depict the mean and standard deviation of the error. 
  				The error for the full space is $0.93$ mm.
	}
	\label{fig:plotGeneralizationPose}
\end{figure}

\begin{figure}[t]
	\subfloat{\includegraphics[trim=010mm 000mm 010mm 000mm, clip=true, height=0.75\linewidth]{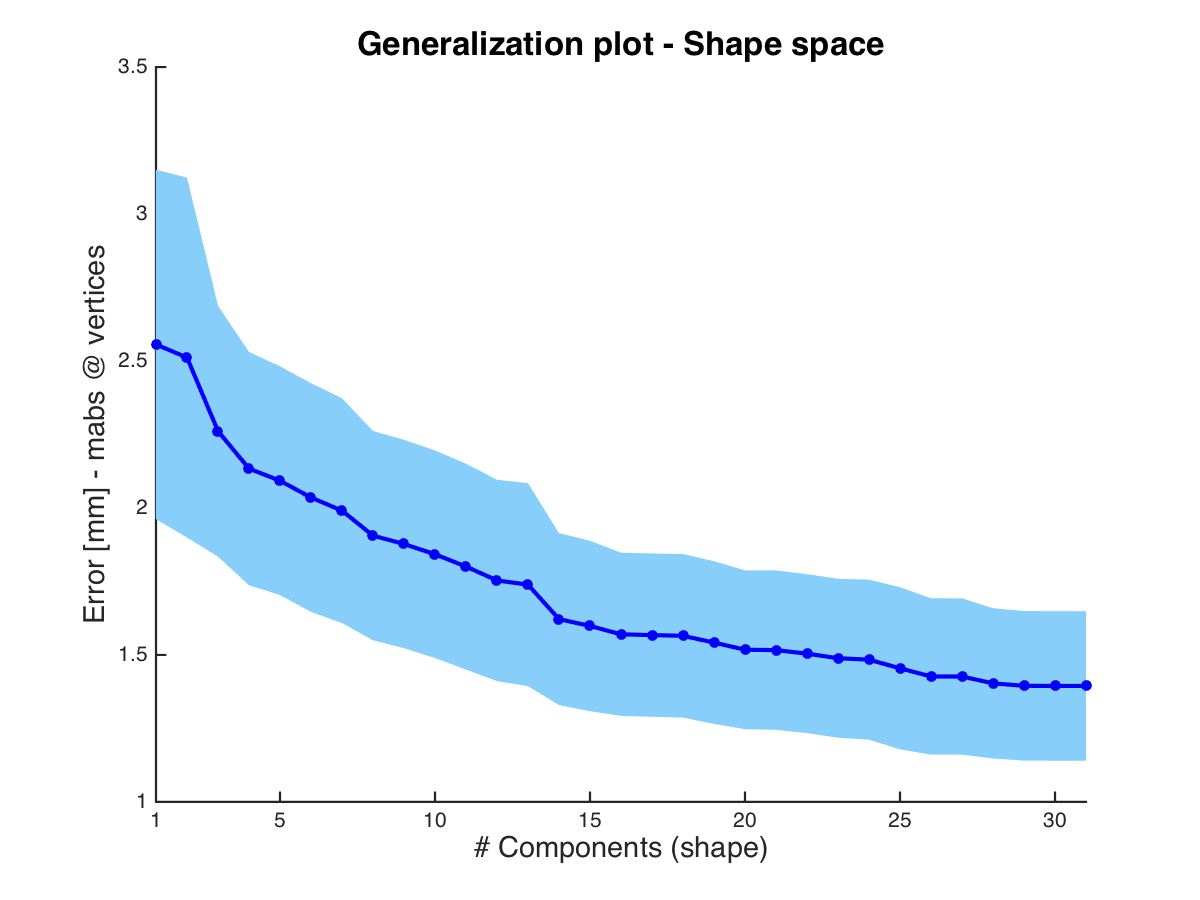}		}
	\caption{
  				Generalization plot for the shape space. 
  				For benchmarking we follow a leave-one-out approach on the training subjects. 
  				We report the mean absolute error (mabs) on the vertices of the template of each test-subject, for a varying number of components for the \mbox{low-D} shape space. 
  				For each number of components we depict the mean and standard deviation of the error. 
	}
	\label{fig:plotGeneralizationShape}
\end{figure}

%% file: tex_FIG/FIG_experiments_SMPLH_1.tex
\begin{figure*}[ht!]
	\begin{tabular}{c}
	\subfloat{	\includegraphics[trim=000mm 000mm 000mm 000mm, clip=true, width=1.0\linewidth]{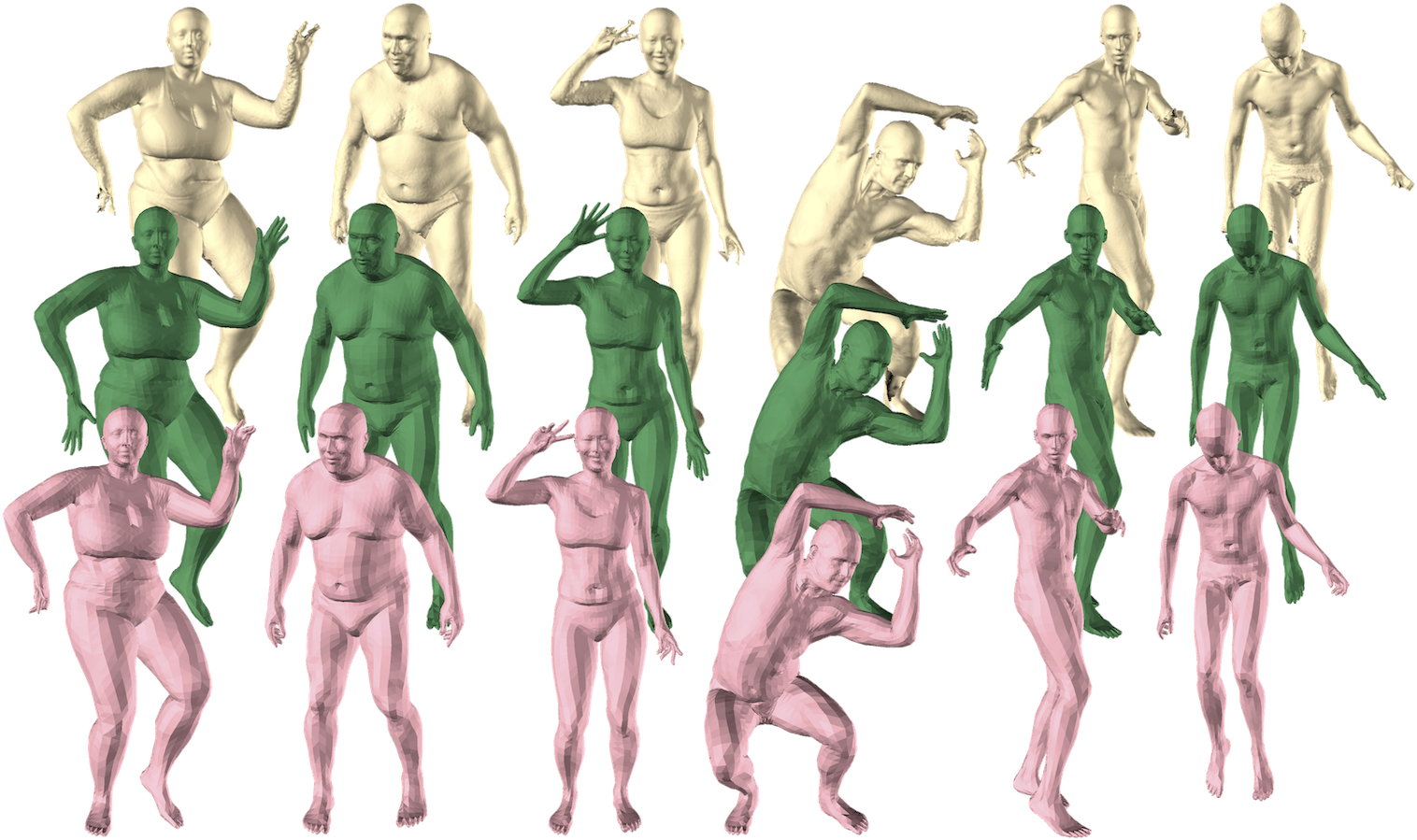}	}
	\end{tabular}
	\caption[]{
		Example 3D scans (\emph{white}) from our 4D sequences and corresponding registrations of \smplH (\emph{pink}). 
		A small discrepancy between (white) scans and corresponding (pink) registrations suggests an expressive model and accurate registration method. 
		Our holistic approach results in natural motion capture even under challenging conditions. 
		We compare \smplH (\emph{pink}) with SMPL \cite{Loper:SIGASIA:2015} (\emph{green}). 
		The latter is representative of existing state-of-the-art methods without joint modeling of body and hands, and always results in open flat hands. 
		On the contrary, \smplH performs significantly more realistically, capturing accurate and expressive hand poses.
   }
\label{fig:final_results_comparison_no_hands}
\end{figure*}

%% file: tex_FIG/FIG_experiments_SMPLH_fail.tex
\begin{figure}
  \begin{center}
  		\includegraphics[width=1.0\linewidth]{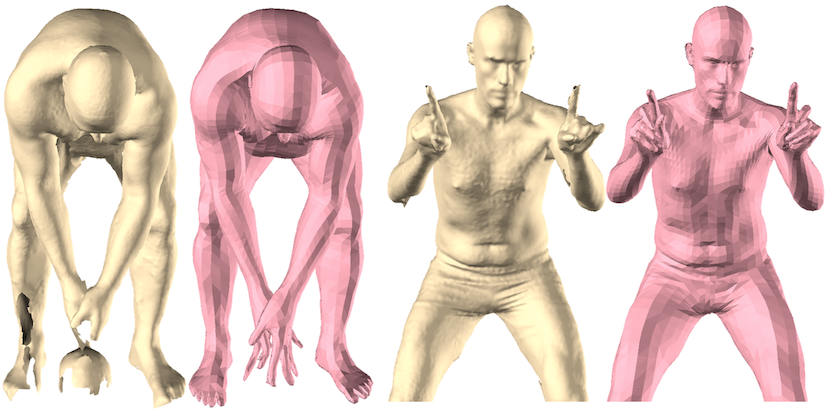}
  \end{center}
  \caption{
  			Typical failure cases for hands. 
  			Each pair shows the scan (white) and the corresponding registration of {SMPL+H} (pink). 
			Left: At the moment hand-object interaction is not explicitly modeled, thus wrong correspondences are established between the hands of the model  and the object scan data. 
			Right: Unusual hand poses may fall outside the low-dimensional pose space. 
			\\
  }
  \label{fig:failure}
\end{figure}

%% file: tex_FIG/FIG_experiments_SMPLH_2.tex
\begin{figure*}[th!]
 \includegraphics[width=.90\linewidth]{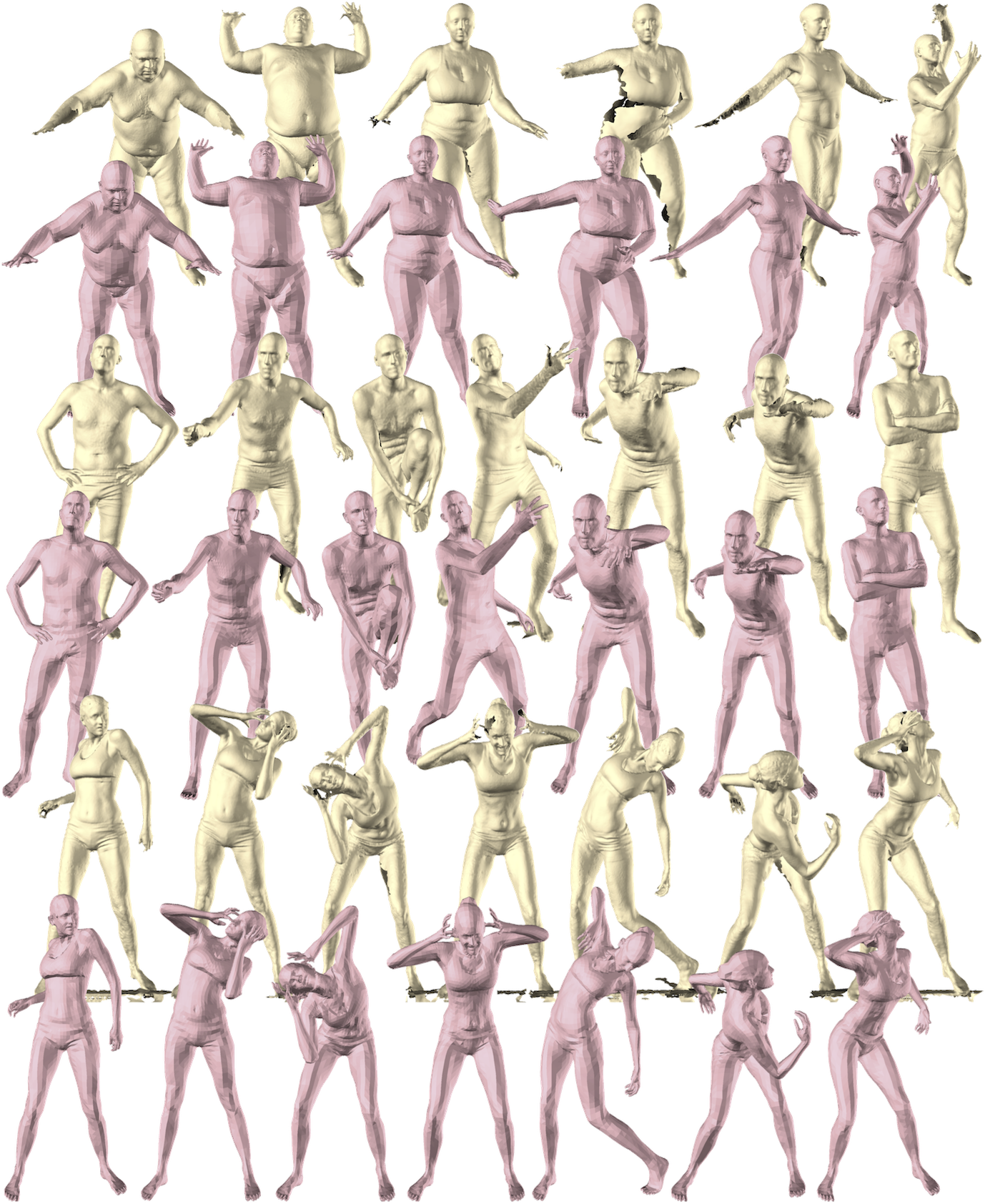} 
 \caption[]{
        Example 3D scans (\emph{white}) from our 4D sequences and corresponding registrations of \smplH (\emph{pink}). 
		A small discrepancy between (white) scans and corresponding (pink) registrations suggests an expressive model and accurate registration method. 
		Our holistic approach results in natural motion capture even under challenging conditions, like the illustrated cases of severe missing data due to fast motion, occlusion, finger-webbing or measurement noise. 
		Please note that it is not uncommon that the whole hand or even the arm completely disappear for several frames, posing increased challenges. 
}
\label{fig:final_results_1}
\end{figure*}

%% file: tex_main/06_discussion.tex
\section{Conclusions}\label{sec:discussion} 

We propose \mano, a new model of the human hand that is learned from examples, is low-dimensional, is easy to pose and fit to data, and is compatible with graphic engines because it is built on linear blend skinning with blend shapes.
The realism of the shape model and its simple formulation makes it appealing for many applications.

We combine \mano with a model of the body to produce \mbox{SMPL+\-H}, the first model of human body shape that is learned from examples and includes full hand articulation.
The expressiveness and robustness of this model is demonstrated by fitting it to 4D scan data of people performing natural movements.
We show that the optimization is stable even when the raw data is noisy, low resolution, and partially missing.
By capturing the hand and body together we extract avatars that appear natural in their movements. 

\subsection{Discussion}

\paragraph{Model:}
While here we build on top of SMPL \cite{Loper:SIGASIA:2015}, our dataset, design choices, and insights can be applied in training related models of the hand using representations like SCAPE \cite{anguelov2005scape}. 

\paragraph{Training dataset:}
The quantitative evaluation of \mano in Section \ref{sec:evaluation_handOnly} shows that our model benefits from augmenting the dataset through mirroring. 
This points to the need for richer training datasets in the future, potentially using dynamic 4D scan sequences instead of static 3D scans. 

\paragraph{Real-time applications:}
Blend shapes encode vertex offsets (per-vertex displacement), as described in Section \ref{sec:model}, thus their memory footprint is comparable to the footprint of the mesh. 
Furthermore, they are linear as described with Equations \eqref{eq:blendShapes_POSE} and \eqref{eq:blendShapes_SHAPE}, thus their runtime is comparable to LBS. 
As a result our model with linear blend shapes is both memory and runtime efficient, therefore it is suitable also for real-time applications. 

\paragraph{Low-dimensional pose space efficiency:}
Computational efficiency is further aided by the faster optimization in the low-dimensional pose space. 
However a low-dimensional space can not model the full space by definition. 
When subjects use their hands in a natural manner, it explains their poses well, while complex unnatural poses may fall outside this space. 
In the latter case the low-dimensional pose space can be used as an computationally efficient initialization for a subsequent optimization in the full pose space. 
Alternatively, we plan to train non-linear latent space models using deep learning but this will require more training data. 

\paragraph{Failure cases:}
Our \mbox{SPML+H} sequences (Sec. \ref{sec:sequences}) include failure cases with unusual hand poses that fall outside the low-dimensional pose space, as described above. 
Example cases are included in our \textbf{supplementary video}. 
Furthermore, we capture a sequence that includes hand-object interaction (a person picks up a heavy ball, see Figure~\ref{fig:failure}), a typical failure case since at the moment we do not explicitly model hand-object interaction. 
However such cases can be tackled in future work inspired by works like \cite{LucaHands, Tzionas:IJCV:2016, Tzionas:ICCV:2015, egohandsDetection_2015_iccv, Rogez:CVPR:2015, srinath_eccv2016_handObject, Oikonomidis_1hand_object}. 
Please note that all failure cases are included in our dataset \cite{dataset_SMPLH_PSHAND} to spur future work. 

\paragraph{Industrial applications:}
Our approach is relevant for industrial applications in several different ways. 
First, while our capture environment is complex, the ability to capture fully body and hand performance without markers may make this appropriate for high-end applications. 
Second, the body model with hand articulation is directly usable by animators as it is designed to be compatible with existing animation systems. 
We believe this is the first learned model of bodies and hands for which this is true. 
Third, this provides a practical step towards the simultaneous capture of hands and bodies, which will enable animators to learn how body pose and hand pose are correlated; this is useful both as reference material and for training machine learning models relating the two. 

\subsection{Limitations and Future Work}

There are several avenues for future work.
We showed tracking results with \smplH that involved small amounts of self contact but we currently do not explicitly reason about this. 
A major limitation of previous 3D body shape tracking systems is that they do not allow self contact, while many natural motions and poses require it. 
We argue that a good hand model is valuable towards this direction, as we should be better able to fit it to scans where it is in contact with other body parts.

Contact with other objects is also important. 
Our hand training set includes such cases but we remove the objects and do not explicitly reason about hand-object contact during fitting.  
Of course, reasoning about such interaction with objects requires one to also reason about the object's surface.
This is beyond our scope here but the \mano model should help facilitate research in this direction.
The extension with interaction-specific blend shapes could be useful to model non-rigid skin deformation during manipulation. 

We believe that high-quality data such as our dataset \cite{dataset_SMPLH_PSHAND} is crucial to build a strong model like \mano. 
However, after creating a strong model with such costly data, \mano is valuable for fitting lower quality data of other modalities like RGB-D or RGB, or for training methods to deal with such data. 

The same applies for the combination of body and hands. 
Here we fit \smplH to 4D capture data. 
It would be interesting to fit the model to lower-quality RGB-D sequences.
Since body models like SMPL have been fit to such data \cite{Bogo:ICCV:2015} with high-quality results, doing so for the body and hands together may be possible.
Again the challenge will be estimating the fingers from low-resolution data.

To address that challenge, we plan to use \mano and \smplH for a very different purpose -- to generate realistic training data for deep learning, a topic gaining increasing research focus \cite{simon2017hand,varol17}.
The \mano training set includes high-quality texture maps so we can generate hands with realistic textures, pose them, light them, and then put them in scenes with varying backgrounds. 
We then plan to train deep networks to estimate the shape and pose of the hand from images.

Finally we plan to use \smplH registrations to train models capturing how hand motion is correlated with body motion. 
Existing approaches \cite{jorg2012data} are based on synthetic data from animations. 
Inspired by them, we would like to capture hand and body correlations based on real markerless mocap data, while using realistic pose and shape spaces. 
Existing mocap datasets \cite{CMU_mocap_db, h36m_pami} do not contain hand information and \smplH could fill this gap. 
Given sufficient \smplH registrations, learning a mapping from body motions to hand poses would be possible. 
This would help towards more realistic and robust motion capture, and towards enriching existing animations or mocap sequences by adding correlated and plausible hand poses. 

%% file: tex_main/07_acknowledgements.tex
\begin{acks}
	We thank T. Alexiadis, A. Keller, J. Marquez, S. Polikovsky, E. Holderness and J. Purmort for help with trial coordination, scanner setup and data acquisition. 
	We thank Lilla LoCurto and Bill Outcault for the artistic sequences, captured as part of the ``art in residence'' program.
	We thank T. Zaman for the video voice-over, as well as T. Ntouni and A. Quiros Ramirez for help with the project website. 
	Finally, we thank the co-authors of \cite{Loper:SIGASIA:2015} M. Loper, N. Mahmood and G. Pons-Moll. 
\end{acks}

%% file: tex_main/08_appendix.tex
\begin{appendices}

\setcounter{figure}{0}
\setcounter{table}{0}
\setcounter{equation}{0}
\renewcommand{\thefigure}{A.\arabic{figure}}
\renewcommand{\thetable}{A.\arabic{table}}
\renewcommand{\theequation}{A.\arabic{equation}}

\input{tex_FIG/FIG_SUPMAT_1_pca}

\section*{Model/Scan mirroring}

	For the creation of the \mano hand model, we first collect a large number of scans of hands in isolation. 
	These scans are obtained with a scanner configured specifically to capture hands with a fixed wrist position. 
	This allows us to capture the nuances of hand deformation.
	After capturing this data for both right and left hands, we create a single augmented dataset by mirroring the left hand scans to appear as right ones. 
	This approach increases the size of the training data and removes the bias introduced by the handedness of the subjects.
    In practical terms, it results in a performance improvement as shown in the experiments section. 
	The augmented dataset enables us to train a single consistent hand model for both hands, i.e. we train the right hand model and generate the left one by mirroring. 
    Model components which depend on the global coordinate frame, like the mesh template $\template$, the shape blend shapes $B_S$ and the pose blend shapes $B_P$, require mirroring. 
	The rest of the components (e.g. the blend weights $\bweights$ and joint regressor $\mathcal{J}$) remain untouched.

    We define the sagittal plane in SMPL, $x$, as our mirroring plane. This entails the following mirroring transformation
    \begin{align}
		M &= 
			\Bigg[
			\begin{array}{ccc}
				-1 & 0 & 0 \\
			 	 0 & 1 & 0 \\
			 	 0 & 0 & 1
			\end{array}
			\Bigg]
    \end{align}
    Mirroring the scan points and the mesh template is trivial, i.e. each scan point $p$ and each vertex $v$ of the template $\template$ are mirrored as $p' = p M$ and $v' = v M$. 
    Each of the shape blendshapes $\mathbf{S}_n$ (Eq. (8) in \cite{Loper:SIGASIA:2015})
    follow a similar procedure
    \begin{align}
          B_S' =  M B_S &= \sum_{n=1}^{|\shape|}\beta_n M \mathbf{S}_n \\
          \mathbf{S}_n' &= M \mathbf{S}_n
    \end{align}
    since they map a (coordinate frame independent) shape coefficients $\beta_n$ to vertex displacements.

	The pose blend shapes are slightly different, since they are multiplied by hand pose rotations, which depend on the global coordinate frame.
    Therefore, the output-mirrored pose blend shapes $M \mathbf{P}_n$ should be modified to account for the mirror input transformation required to transform right poses into left ones.
        
    To understand how to achieve this, consider first that a pose is mirrored by applying its rotation in a mirrored coordinate frame.
    This corresponds to pre- and post- multiplying its corresponding rotation matrix $R$ by the commutative transformation $M$, $R' = M R M$.

    Considering Eq. (9) \cite{Loper:SIGASIA:2015}, we want to obtain the mirrored pose blendshapes $\mathbf{P}_n'$ such that
    \begin{align}
         B_P' &\equiv \sum_{n=1}^{|\shape|}(R_n'(\pose) - R_n'(\pose^*)) \mathbf{P}_n' \\
               &=  M B_P = \sum_{n=1}^{|\shape|}(R_n(\pose) - R_n(\pose^*)) M \mathbf{P}_n
    \end{align}
    where  $R_n'$ are the scalar elements of the mirrored rotations $R'$.
    Therefore, one can obtain the mirrored pose blendshapes $\mathbf{P}_n'$ by applying the rotation un-mirroring transformation (pre- and post-multiplying by $M$) to each of the $3 \times 3$ input blocks in $M \mathbf{P}_n$.

\section*{Evaluation - \mano hand model}

	In the next figures we present variations of the generalization plot for several comparisons. 
	
	Figure~\ref{fig:plotCompareModels} shows the performance of our model trained on different datasets. 
	The red curve shows the \mano model trained only on the right hand poses (i.e. no mirroring augmentation). 
	This approach requires to train two separate left and right \mano models, making the models biased by the handedness of the subjects. 
	The blue curve shows the \mano model trained on an augmented dataset, for which left hand poses are mirrored and combined with the right hand ones. 
	This approach leads to a single hand model from a single training procedure (the left model is generated by mirroring the resulting right hand model) with a larger training dataset and no handedness bias. 
	It is our chosen approach for \mano.
	The plot shows that the latter approach performs favorably, pointing to the need for richer training datasets in the future, potentially using dynamic 4D scans instead of static 3D scans. 
	For the full pose space the former approach has an error of $1.05$ mm and the latter $0.93$ mm. 

	Figure~\ref{fig:plotTrainingError} shows a comparison of the error before and after training. 
	For the full-space the initial model has an error of $2.90$ mm, while the trained model has an error of $1.01$ mm.

    \pagebreak
    \input{tex_FIG/FIG_SUPMAT_3_generalization_BeforeAfter}
    \input{tex_FIG/FIG_SUPMAT_2_generalization_Datasets}

\end{appendices}

%% file: tex_FIG/FIG_SUPMAT_1_pca.tex
\begin{figure*}
    \centering
  	\includegraphics[width=1.0 \linewidth]{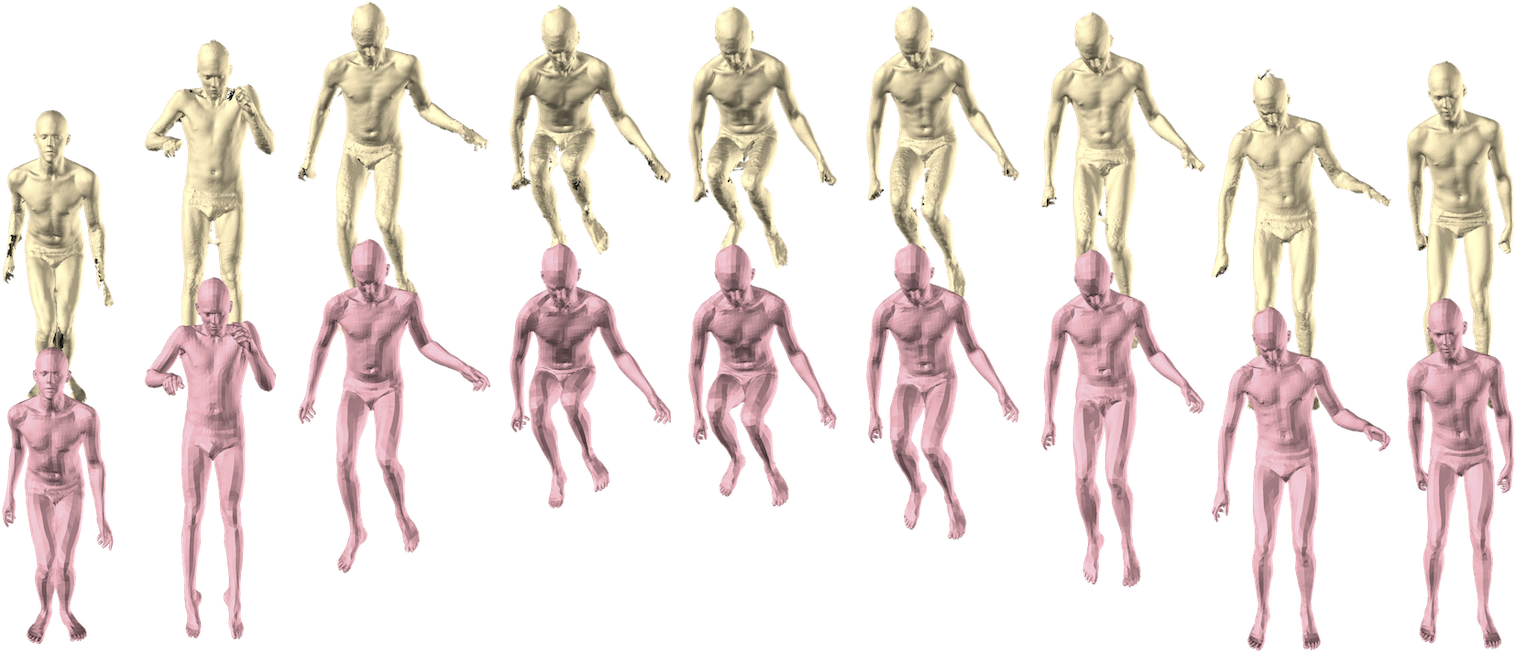}
  \vspace{-4mm}
  \caption{
  			Sample frames (non-sequential) from a jumping action. 
  			Fast motions result in missing hand data in the 3D scans, posing challenges to existing motion capture methods. 
	  		The 3D scans are shown with \emph{beige} color, while the resulting registration of SMPL+H is shown with \emph{pink}. 
  			The proposed model and registration method result in natural motion capture even under fast motions and missing visual data for several frames, e.g. in the hand region. 
  }
  \label{fig:pcaPCsPoseSUP}
\end{figure*}

%% file: tex_FIG/FIG_SUPMAT_3_generalization_BeforeAfter.tex
\begin{figure}[ht]
	\centering
	\includegraphics[trim=017mm 000mm 017mm 000mm, clip=true, width=.99\linewidth]{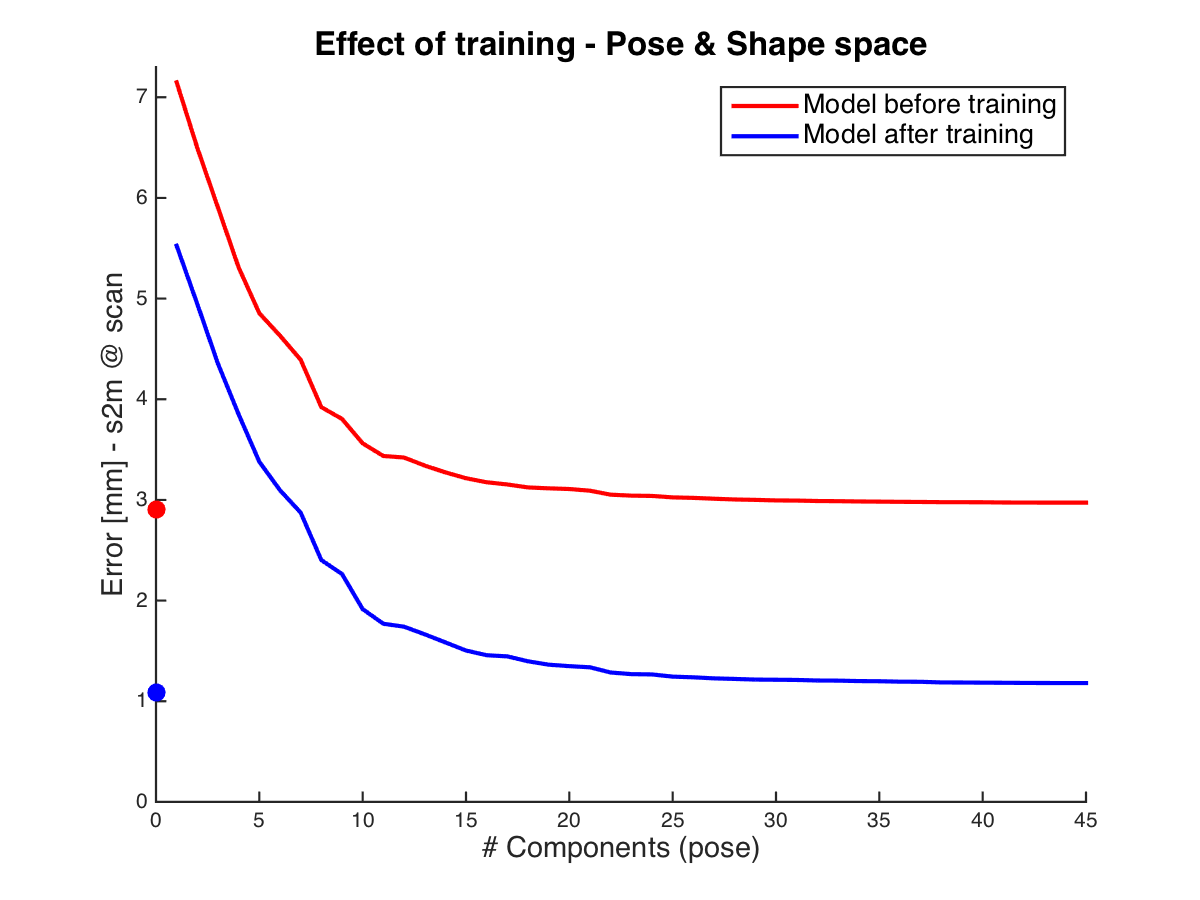}
	\vspace{-5mm}
	\caption{
  				Generalization plot for the model before and after training. 
  				We report the mean scan-to-mesh (s2m) error for a varying number of pose-space components. 
  				For $x=0$ we show the error for the full space. 
	}
	\label{fig:plotTrainingError}
\end{figure}

%% file: tex_FIG/FIG_SUPMAT_2_generalization_Datasets.tex
\begin{figure}[ht]
    \centering
    \includegraphics[trim=017mm 000mm 017mm 000mm, clip=true, width=.99\linewidth]{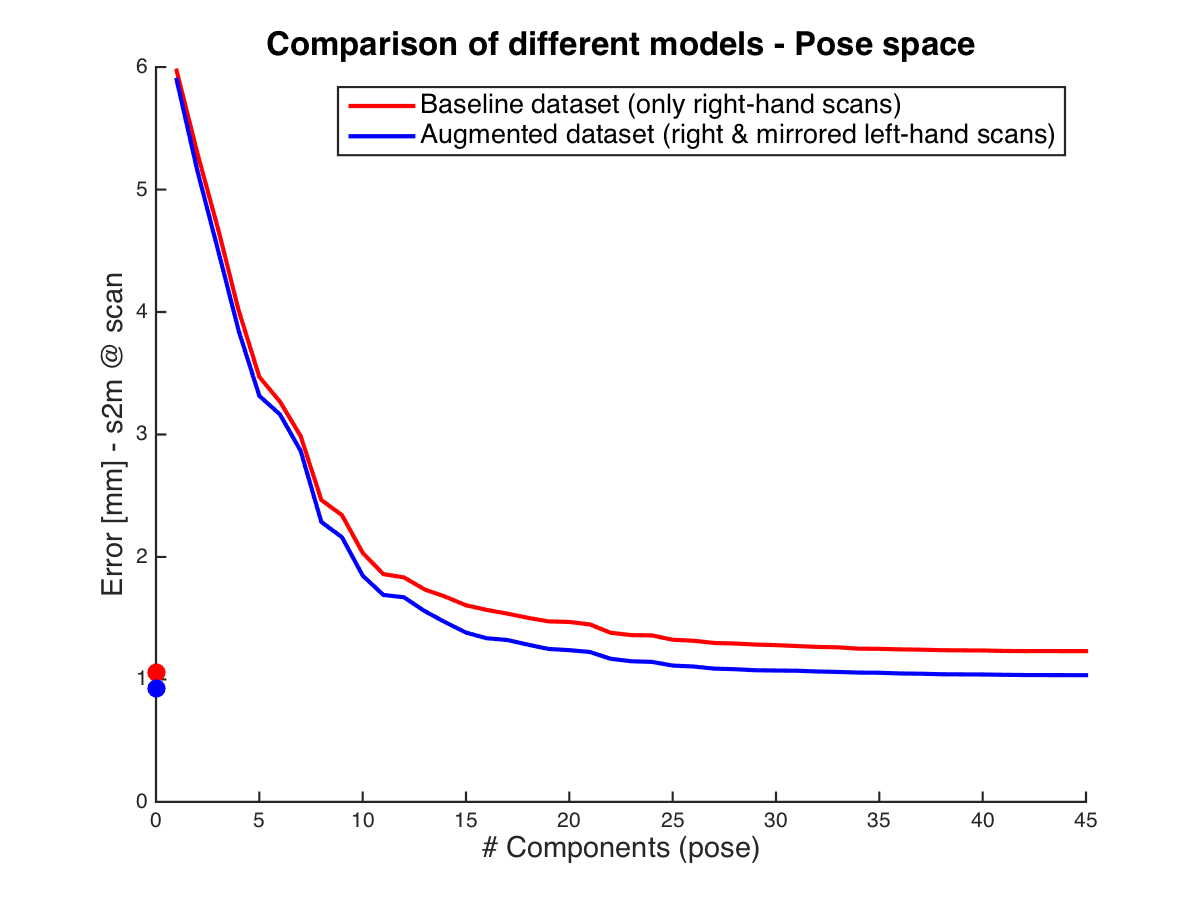}
    \vspace{-5mm}
    \caption{
        Generalization plot for the model trained on different datasets. 
        The \emph{baseline dataset} is comprised of only right-hand scans, while the \emph{augmented dataset} is comprised of right- and mirrored left-hand scans. 
        We report the mean scan-to-mesh (s2m) error for a varying number of pose-space components. 
        For $x=0$ we show the error for the full space. 
        The plot shows that an augmented dataset leads to a lower fitting error of the model, pointing to the need for bigger future training datasets.
    }
    \label{fig:plotCompareModels}
\end{figure}